\documentclass[12pt]{amsart}
\usepackage{amsmath,amssymb,amsfonts,amscd,epsfig}

\theoremstyle{plain}
\newtheorem{thm}{Theorem}
\newtheorem{lem}[thm]{Lemma}

\newtheorem{prop}[thm]{Proposition}
\newtheorem{rem}[thm]{Remark}
\newtheorem{defn}[thm]{Definition}
\newtheorem{ex}[thm]{Example}

\newtheorem{dt}[thm]{Definition and Theorem}
\newtheorem{con}[thm]{Conjecture}
\numberwithin{thm}{section}
\numberwithin{equation}{section}

\newcommand{\sQ}{{\mathcal Q}}

\newcommand{\A}{{\mathbb A}}

\newcommand{\CC}{{\mathbb C}}

\newcommand{\FF}{{\mathbb F}}

\newcommand{\NN}{{\mathbb N}}

\newcommand{\PP}{{\mathbb P}}
\newcommand{\QQ}{{\mathbb Q}}
\newcommand{\RR}{{\mathbb R}}

\newcommand{\ZZ}{{\mathbb Z}}

\title{Quantum periods:\\A census of $\phi^4$-transcendentals}
\author{Oliver Schnetz}\address{
Department Mathematik\\
Bismarkstra\ss e 1$\frac{1}{2}$\\
91054 Erlangen\\
Germany\\
E-mail address: schnetz@mi.uni-erlangen.de}
\begin{document}
\begin{abstract}
Perturbative quantum field theories frequently feature rational linear combinations of multiple zeta values (periods).
In massless $\phi^4$-theory we show that the periods originate from certain `primitive' vacuum graphs.
Graphs with vertex connectivity 3 are reducible in the sense that they lead to products of periods with lower loop order.
A new `twist' identity amongst periods is proved and a list of graphs (the census) with their periods, if available,
is given up to loop order 8.
\end{abstract}
\maketitle
\tableofcontents

\section{Introduction}
The last decade has seen a renewed interest in perturbative quantum field theory (pQFT).
On the one hand, progress has been achieved on amplitudes with many legs and a low number of loops (zero or one)
\cite{W1}, \cite{W2}, \cite{B1} (and the references therein).
From an experimentalist point of view these results will be vital in the analysis of upcoming LHC-data.
On the other hand the study of many loops with a low number of external legs is important for the
understanding of high precision experiments like the measurement of the anomalous magnetic moment of
the electron \cite{O1}, \cite{G1}, \cite{H1}. Huge theoretical efforts on the numerical \cite{A1}, \cite{A2} as well as on the analytical side \cite{L1},
\cite{P1} (see below) are accompanied by new insights from conjectured relations between pQFT, number theory and knot theory \cite{BK}, \cite{B3},
\cite{B4}, \cite{K1}, Hopf algebras \cite{C1}, \cite{C2}, \cite{E1}, and algebraic geometry \cite{BEK}, \cite{BLO}, \cite{C3}, \cite{BROW}.

This article focuses on the second aspect of pQFT. Since the basic concepts are motivated by physical examples
let us look at the magnetic moment of the electron which is a benchmark problem of perturbative Quantum Electrodynamics (pQED).
Fifty years of computations provide us with 3 orders of radiative corrections to the `classical' value $g=2$.
The coefficient of the first order was derived in 1948 \cite{S1}, the second order in 1957 \cite{S2}, \cite{P2}.
The calculation of the third order was finished in 1996 \cite{L1}. We give the result in a slightly unconventional way by
introducing Euler sums
\begin{eqnarray}\label{01}
U_n&=&\sum_{k=1}^\infty \frac{(-1)^k}{k^n}\;=\;(2^{1-n}-1)\zeta(n),\quad (\hbox{if }n\ge2),\nonumber\\
U_{3,1}&=&\sum_{k>l\geq1} \frac{(-1)^k}{k^3}\frac{(-1)^l}{l}\;=\; -0.117~875~999~650\ldots.
\end{eqnarray}
Now, we can give the result for $g-2$ in terms of Euler sums and rational numbers (the coupling $\alpha$ is measured to $\alpha/\pi=0.002~322~819~455\ldots$),
\begin{eqnarray}\label{02}
\frac{g-2}{2}&=&\frac{1}{2}\frac{\alpha}{\pi}\;+\;\left(-\,U_3-6U_2U_1-U_2+\frac{197}{2^4 3^2}\right)\!\left(\frac{\alpha}{\pi}\right)^2\nonumber\\
&&\hspace*{-33pt}+\left(\frac{86}{3^2}U_5\!+\!\frac{166}{3^2}U_3U_2\!-\!\frac{50}{3}U_{3,1}\!-\!\frac{13}{5}U_2^2\!-\!\frac{278}{3^3}U_3
\!-\!\frac{1192}{3}U_2U_1\!-\!\frac{34202}{3^3 5}U_2\!+\!\frac{28259}{2^6 3^4}\right)\!\left(\frac{\alpha}{\pi}\right)^3\!.\nonumber\\
&&
\end{eqnarray}
This result stands out from other multi-loop calculations because it is very likely correct:
The above number can actually be measured to a precision that controls the calculation.

We see that the first order is given by a rational number, whereas the second order is provided
by a sum of 4 terms: a rational number plus 3 transcendentals. (We do not distinguish between transcendentals and very-likely-transcendentals here.)
We may consider the sum as an element in a 4-dimensional vector space over $\QQ$. This picture, however, may be premature: If we give the Euler sums
a grading (a weight) by adding the indices in a product (rational numbers have weight 0) we see that the first two transcendentals are of weight 3. Maybe we should
combine the two numbers to provide a sole transcendental (written as $U_3+6U_2U_1$) resulting in a 3-dimensional
vector space over $\QQ$ for the second order. How can we tell? We have to look at all other sorts of QED-experiments and check if we can write the
second order in terms of $U_2$ and $U_3+6U_2U_1$. The Lamb shift e.g.\ is of this type. The second order coefficient
reads $(U_3+6U_2U_1)+49/(2^2 3^2)U_2-4819/(2^6 3^4)$ \cite{L2}. Moreover, we see that we actually
need (a minimum of) two transcendentals at two loops because the ratio between the weight 2 and the weight 3 transcendentals differs from
Eq.\ (\ref{02}). It seems to be a general fact that transcendentals of different weight cannot be combined.
On the other hand, the full photon propagator features a $U_3$ not paired by a $U_2U_1$ \cite{R1}. However, the photon propagator
is gauge-dependent and hence not an observable quantity.

Looking at the third order contribution in Eq.\ (\ref{02}) we see transcendentals up to weight 5. The grade grows in steps of
2 with every loop order. Moreover, the third order coefficient features all lower order transcendentals and some of their products.
(It cannot contain $U_3^2$ because this has weight 6, but $U_2^2U_1$ is absent for some unknown reason.)
Both are generic features: The coefficients lie in a graded $\QQ$-algebra and the grade grows in steps of 2 with the loop order.
The new numbers at order 3 are one (at least) weight 5 transcendental $U_5$,
one weight 4 transcendental $U_{3,1}$, and---in the case that a sole weight 3 transcendental suffices
at second order---one weight 3 transcendental to account for the new ratio between $U_3$ and $U_2U_1$.

In this paper we focus on perturbative massless $\phi^4$-theory which is technically less intricate than pQED
but still shows the structure we are interested in. It is known (up to 6 loops) that the $\phi^4$ beta-function expands into a power series in
the coupling $g$ with coefficients that are rational linear combinations of multiple zeta values (MZVs).
The transcendentals are periods in the sense of \cite{K3}. Such periods were found to be generic for pQFTs \cite{B8}.
In \cite{BK} the $\phi^4$-theory periods were reported up to loop order 7 (all rational linear combinations of MZVs with 3 numbers missing).
Here, we want to extend this list to loop order 8 (the `census'). Another objective of the paper
is to simplify the graph theoretical side of the problem by lifting it to primitive 4-regular (vacuum) graphs.
This lift uses a well-known `conformal' symmetry of primitive graphs in massless renormalizable QFTs \cite{BK}.
Primitive vacuum graphs are relatively sparse at low loop order (e.g.\ 2 at 5 loops or 14 at 7 loops), however they become quite
abundant at higher loops (7~635~677 at 14 loops, see Table 1, Sect.\ \ref{sect3}).

As a side effect of the approach we recognize that primitive vacuum graphs with vertex connectivity 3 evaluate to products of lower order periods
(Thm.\ \ref{thm1}). Graphically the product is described as gluing along triangles (see Fig.\ 4).
Among the 73 primitive vacuum graphs up to loop order 8 we have 13 products.

In the following we concentrate on irreducible (non-product, vertex connectivity $\ge 4$) primitive graphs and implement another
two reductions: The new twist identity which is quite ubiquitous at high loop order (Thm.\ \ref{thm2}) and the well known
but rather sparse Fourier identity \cite{BK}, \cite{B7} (Thm.\ \ref{thm3}) which is slightly extended in Remark \ref{rem2}.
Both identities together reduce the number of irreducible periods up to loop order 8 from 60 to 48.

We use `exact numerical methods' \cite{BK} in Sect.\ \ref{sect3} to identify 31 of the remaining 48 irreducible periods.
All of these periods are found to be {\it integer} linear combinations of MZVs (as suggested for some periods in \cite{BROW}).
The missing 17 periods (2 at loop order 7 and 15 at loop order 8) are inaccessible by the method available today.

From a physical point of view one may doubt the value of these considerations because the periods considered here are not directly linked to
observables. They are rather a kind of QFT-concentrate originating from the most complicated Feynman graphs of a given order.
They hence may serve as a test for calculational techniques. If one is able to calculate all periods of a certain order one has
a good method to calculate all amplitudes in this order.
Regretfully the today's analytical methods last only for the first few loop orders (five, maybe six, in massless $\phi^4$-theory).

From a mathematical point of view the appearance of MZV periods hints towards (algebraic) geometries of mixed Tate type \cite{BEK}, \cite{BLO}, \cite{BROW}.
Every period in $\phi^4$-theory that is a rational linear combinations of MZVs reveals a connection between quantum field theory and
mixed Tate motives. However, it cannot be conjectured (by what we know today) that the entire $\phi^4$-theory (all periods to all orders)
stays in the realm of MZVs \cite{BROW}, \cite{QFTFq}.
Moreover, it is unclear if $\phi^4$-periods (in the sense of this article) exhaust the number contents of $\phi^4$-theory (as suggested by
the spirit of the Hopf-algebra approach to renormalization \cite{C1}, \cite{C2}).

Although much of the material spawned from a conjectured connection to knot theory \cite{BK}, \cite{B3}, \cite{K1} this link stayed somewhat vague
such that the author decided not to include it into this paper. However, it may well be possible that this connection will reappear as the knowledge
on QFT-periods develops.

{\bf Acknowledgements.} The author is very grateful for discussions with S. Bloch, D. Broadhurst, F.C.S. Brown, and K. Yeats.
H. Frydrych contributed a C$++$ program and G. Hager was kindly helping the author to use the RRZE computing cluster.

\section{$\phi^4$-periods}
\subsection{Background}

We consider massless euclidean $\phi^4$-theory (see e.g.\ \cite{IZ}) in 4 (space-time) dimensions with interaction term normalized to
\begin{equation}\label{1}
L_{\rm int}=-\frac{16\pi^2g}{4!}\int_{\RR^4}{\rm d}^4x\,\phi(x)^4.
\end{equation}
It is convenient to `irrationalize' the coupling by a factor of $16\pi^2$ to eliminate unwanted factors of $\pi$.
We focus on the 4-point-function and obtain for the amplitude of a Feynman-graph $\Gamma$ (for examples see Fig.\ 1)
\begin{equation}\label{2a}
A_\Gamma=(2\pi)^4\delta^4(q_1+q_2+q_3+q_4)\frac{16\pi^2 g}{|q_1|^2\cdots |q_4|^2}\cdot \left(\frac{g}{\pi^2}\right)^{h_1}\int_{\RR^{4h_1}}{\rm d}^4p_1\cdots{\rm d}^4p_{h_1}\frac{1}{\prod_{i=1}^nQ_i(p,q)}
\end{equation}
where we introduced the following notation: The momentum-conserving 4-dimensional $\delta$-function $\delta^4$ with `external' momenta $q_1,\ldots,q_4$,
the `loop order' $h_1$ giving the number of independent cycles in $\Gamma$.
The graphs we consider are `one-particle irreducible' meaning that (except for the four external edges) the graph has edge-connectivity $\geq2$.
The $n$ `propagators' $1/Q_i$ (associated to `interior' edges) are inverted rank 4 quadrics in the coordinates of momentum vectors.
Each quadric is the square $|\bullet|^2$ of a 4-dimensional euclidean vector which is a (signed) sum of (some of) the external momenta
$q_1,\ldots q_4$ and internal momenta $p_1,\ldots,p_{h_1}$.

The first half of the right hand side (up to the $\cdot$) is the amplitude of the tree graph (with 4 edges) whereas
the second half (for $h_1>0$) is a divergent integral:
Graphs that contribute to the $\phi^4$ 4-point function have $n=2h_1$. Thus, the differential form on the right hand side of Eq.\ (\ref{2a}) has total degree 0.
The integral diverges logarithmically (like $\int_1^\infty{\rm d}p/p$) for large $p_i$. Since for large $p_i$ the value of the external momenta becomes
irrelevant we may nullify the $q_i$ to characterize the divergence by a mere number (if it exists) given by the projective integral
\begin{equation}\label{3}
P_\Gamma=\pi^{-2h_1}\int_{\PP\RR^{4h_1-1}}\frac{\Omega(p)}{\prod_{i=1}^nQ_i(p,0)}.
\end{equation}
Here we have introduced the projective volume measure which is defined in $\PP^m$ with coordinates $x_0,\ldots,x_m$ as
\begin{equation}\label{3a}
\Omega(x)=\sum_{i=0}^m(-1)^i{\rm d}x_0\cdots\hat{{\rm d}x_i}\cdots{\rm d}x_m.
\end{equation}
We assume an orientation on $\PP\RR^{4h_1-1}$ (which is an orientable space) is chosen such that $P_\Gamma>0$.
Readers not familiar with projective integrals may prefer to set one of the coordinates of one of the internal momenta to 1 and interpret
the integral in Eq.\ (\ref{3}) as volume integral over the remaining $4h_1-1$ coordinates.

In the following we consider the differential form $\Omega(p)/\prod Q_i(p,0)$ in Eq.\ (\ref{3}) as degree 0 meromorphic $4h_1-1$ form in complex projective
space $\PP\CC^{4h_1-1}$. It is of top degree as meromorphic form and hence closed in the complement of $\prod Q_i(p,0)=0$.
As odd dimensional real projective space the domain of integration is orientable and compact without boundary and thus a cycle in $\PP\CC^{4h_1-1}$.
However, the cycle of integration meets the singularities of the differential form which in general leads to an ill-defined integral. To ensure that the integral
converges we need an extra condition on the graph $\Gamma$.
\begin{dt}\label{dt1}
A graph $\Gamma$ is primitive if it has $n(\Gamma)=2h_1(\Gamma)$ edges and every proper subgraph $\gamma<\Gamma$ has $n(\gamma)>2h_1(\gamma)$.
The period $P_\Gamma$, Eq.\ (\ref{3}), is well-defined if and only if $\Gamma$ is primitive.
\end{dt}
\begin{proof}
This is Prop.\ 5.2 in \cite{BEK}.
\end{proof}
Algebraically, primitive means primitive for the coproduct in the Connes-Kreimer Hopf algebra of renormalization \cite{C1}.
Geometrically, the subgraph-condition in Def.\ \ref{dt1} means that every sub-cycle of $\PP\RR^{4h_1-1}$ meets the polar divisor of the differential form
with a codimension (in $\PP\CC^{4h_1-1}$) that is strictly larger than in the case of a transversal intersection. This suggests that the integration cycle only
`touches' the singularities of the differential form and that it is hence possible to deform the cycle in a way that it entirely lies in the complement $\prod Q_i(p,0)\neq0$
without altering the value of the integral (although it is not obvious how to do this). In this sense $\pi^{2h_1}P_\Gamma$ becomes a period in
$\{p\in\PP\CC^{4h_1-1}:\prod Q_i(p,0)\neq0\}$. In any case, the parametric representation, Eq.\ (\ref{6e}) makes $P_\Gamma$ an algebraic period in the sense of
M. Kontsevich and D. Zagier \cite{K3}. We call it a $\phi^4$-(quantum)-period. One finds these quantum periods in all sorts of perturbative calculations (like the beta-function
or the anomalous dimension) within the quantum field theory considered. In fact, the role that quantum periods play in the Hopf-algebra of renormalization suggests
that there might exist a clever renormalization scheme such that they form a complete $\QQ$-base for the coefficients of the perturbative expansion of scalar functions.
This gives quantum periods a prominent role within quantum field theory.

\begin{figure}[t]
\epsfig{file=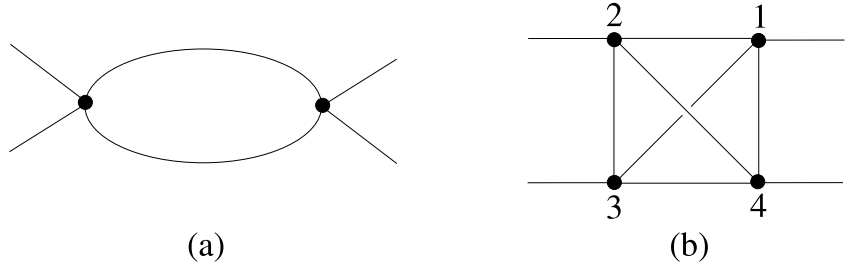,width=14cm}
\caption{A primitive graph with one loop (a) and one with three loops (b).}
\end{figure}

We postpone formal definitions and close this subsection with the first calculation of a $\phi^4$-period.
\begin{ex}\label{ex1}
Consider the graph plotted in Fig.\ 1(a). For the period we find
\begin{eqnarray*}
P_1&=&\pi^{-2}\int_{\PP\RR^3}\frac{\Omega(p)}{|p|^2\cdot|p|^2}\\
&=&\pi^{-2}\int_{\RR^3}\frac{{\rm d}^3{\bf p}}{({\bf p}^2+1)^2}\\
&=&\pi^{-2}4\pi\int_0^\infty\frac{p^2{\rm d}p}{(p^2+1)^2}\\
&=&1.
\end{eqnarray*}
In the second line we used $p=(1,{\bf p})$ to make the integral affine and in the third line we introduced polar coordinates to
transform the integral to a standard one-dimensional integral.
Notice, that graph 1(a) is the only $\phi^4$-period known to evaluate to a rational number. Most likely, it is the only rational $\phi^4$-period.
\end{ex}

\subsection{Feynman rules}

Feynman rules are prescriptions how to translate a Feynman-graph $\Gamma$ into an analytical expression, the amplitude $A_\Gamma$.
In our setup---primitive 4-point functions without external momenta in massless 4-dimensional $\phi^4$-theory---these expressions evaluate to positive
numbers.

We have four different ways to use Feynman rules: Position and momentum space where integrands are
products of inverted quadrics and the variables are 4 dimensional vectors assigned to vertices and cycles, respectively.
Alternatively we may use Feynman's parametric space either in its original form or in a dual version with variables attached
to edges of the graph. Although the transition from position or momentum space to parametric space is due to Feynman it is known in the mathematical literature
as `Schwinger-trick'. To avoid confusion we stick to this name in the following diagram that summarizes the interconnection between the different approaches.
\begin{equation}
\begin{array}{ccccc}
&&&&\hbox{dual}\\
\hbox{position}&&\hbox{Schwinger}&&\hbox{parametric}\\
\hbox{space}&&\longleftrightarrow&&\hbox{space}\\
\hbox{(vertices)}&&\hbox{trick}&&\hbox{(edges)}\\[1ex]
\uparrow&&&&\uparrow\\
\hbox{Fourier}&&&&\hbox{Cremona}\\
\hbox{transformation}&&&&\hbox{transformation}\\
\downarrow&&&&\downarrow\\[1ex]
\hbox{momentum}&&\hbox{Schwinger}&&\hbox{parametric}\\
\hbox{space}&&\longleftrightarrow&&\hbox{space}\\
\hbox{(cycles)}&&\hbox{trick}&&\hbox{(edges)}
\end{array}
\end{equation}

\begin{figure}[t]
\epsfig{file=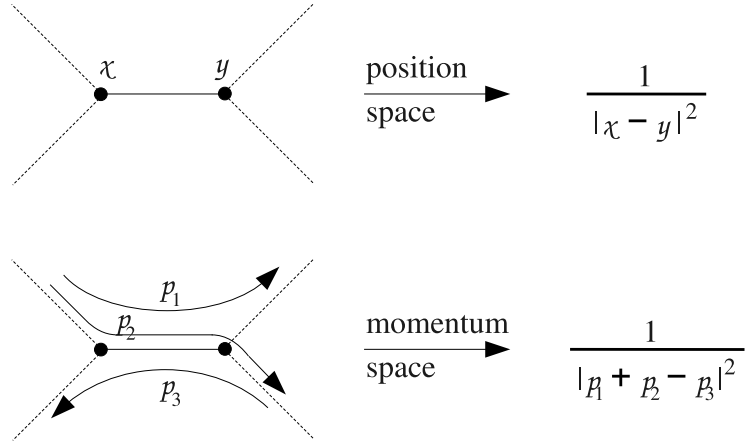,width=\textwidth}
\caption{Propagators for a massless bosonic quantum field theory.}
\end{figure}

In position space every edge joining vertices with variables $x,y\in\RR^4$ contributes by a factor $1/|x-y|^2$ to the Feynman integrand (see Fig.\ 2).
In momentum space every edge contributes by a factor $1/|\sum\pm p_i|^2$ with variables $p_i\in\RR^4$ associated to cycles $P_i$ (choose a basis) that run through the
edge in one ($+$ sign) or opposite ($-$ sign) direction. The integration ranges over the whole real space. The similarity between the propagator in position and
in momentum space is due to the Fourier-symmetry (see Subsect.\ \ref{Fourier})
\begin{equation}\label{2}
\int\frac{{\rm d}^4x}{(2\pi)^2}\frac{{\rm e}^{{\rm i}px}}{x^2}=\frac{1}{p^2},
\quad\int\frac{{\rm d}^4p}{(2\pi)^2}\frac{{\rm e}^{-{\rm i}px}}{p^2}=\frac{1}{x^2}.
\end{equation}

In (dual) parametric space, the integrand is the inverse square of the (dual $\bar\Psi_\Gamma$)
graph polynomial $\Psi_\Gamma$ defined by a sum over all spanning trees of $\Gamma$.
\begin{eqnarray}\label{3b}
\Psi_\Gamma(\alpha)&=&\sum_{T\,\rm span.\,tree}\;\prod_{e\not \in T}\alpha_e,\\
\bar \Psi_\Gamma(\alpha)&=&\sum_{T\,\rm span.\,tree}\;\prod_{e \in T}\alpha_e\;=\;\Psi_\Gamma(\alpha^{-1}) \prod_e\alpha_e.\nonumber
\end{eqnarray}
The integration ranges over positive values of $\alpha_e$. Feynman parameters roughly halve the dimension of the integral (at the expense
of having a boundary). They are particularly useful for calculations at low loop order (for which they were invented) and for studying the algebraic
geometry of the periods \cite{BEK}, \cite{BLO}, \cite{DOR}, \cite{BROW}.

Formal definitions of momentum space and parametric space Feynman rules can be found in \cite{BEK} and \cite{BLO}.
Here, let us explain the rules by way of example.
\begin{ex}\label{ex2}
Consider the graph plotted in Fig.\ 1 (b). We delete the external edges and find for the amplitude in the
four possible settings,
\begin{enumerate}
\item
momentum space. We attach variables $p_1$, $p_2$, $p_3\in\RR^4$ to the cycles $(123)$, $(243)$, $(341)$, resp., and obtain (edges
$(12)\cdot(23)\cdot(24)\cdot(34)\cdot(14)\cdot(13)$)
\begin{equation}\label{4b}
A_{(1)\,b}^{\rm mom}=\int_{\PP\RR^{11}}\frac{\Omega(p)}{|p_1|^2|p_1-p_2|^2|p_2|^2|p_2-p_3|^2|p_3|^2|p_3-p_1|^2}.
\end{equation}
\item
position space. We attach the variable $x_i\in\RR^4$ to vertex $i$ and set $x_4=0$ (to `break' translational invariance, see Thm.\ \ref{dt2}) and obtain
(using a projective setup)
\begin{equation}\label{4a}
A_{(1)\,b}^{\rm pos}=\int_{\PP\RR^{11}}\frac{\Omega(x)}{|x_1|^2|x_1-x_2|^2|x_2|^2|x_2-x_3|^2|x_3|^2|x_3-x_1|^2}.
\end{equation}
This integral trivially evaluates to the same number as $A_{(1)\,b}^{\rm mom}$.
\item
parametric space. We attach variables $\alpha_{ij}\in\RR$ to the edges $(ij)$ and obtain the projective integral with boundary
\begin{equation}\label{4d}
A_{(1)\,b}^{\rm par}=\int_\Delta\frac{\Omega(\alpha)}{\Psi_{(1)\,b}(\alpha)^2},
\end{equation}
where $\Delta$ is the 5-dimensional projective simplex $\alpha_{ij}>0$ and
\begin{eqnarray}\label{4d1}
\Psi_{(1)\,b}(\alpha)&=&
\alpha_{24}\alpha_{34}\alpha_{12}+\alpha_{24}\alpha_{34}\alpha_{13}+\alpha_{34}\alpha_{23}\alpha_{24}
+\alpha_{13}\alpha_{24}\alpha_{12}+\alpha_{14}\alpha_{24}\alpha_{12}+\alpha_{24}\alpha_{13}\alpha_{23}\nonumber\\
&&+\;\alpha_{24}\alpha_{14}\alpha_{23}+\alpha_{24}\alpha_{14}\alpha_{13}+\alpha_{34}\alpha_{12}\alpha_{13}
+\alpha_{23}\alpha_{34}\alpha_{12}+\alpha_{14}\alpha_{34}\alpha_{12}\\
&&+\;\alpha_{34}\alpha_{14}\alpha_{13}+\alpha_{34}\alpha_{14}\alpha_{23}+\alpha_{14}\alpha_{12}\alpha_{23}+\alpha_{13}\alpha_{12}\alpha_{23}+\alpha_{14}\alpha_{23}\alpha_{13}.\nonumber
\end{eqnarray}
\item
dual parametric space. Similarly we obtain in dual parametric space
\begin{equation}\label{4c}
A_{(1)\,b}^{\rm dual\, par}=\int_\Delta\frac{\Omega(\alpha)}{\bar\Psi_{(1)\,b}(\alpha)^2}
\end{equation}
with
\begin{eqnarray}\label{4c1}
\bar\Psi_{(1)\,b}(\alpha)&=&
\alpha_{14}\alpha_{23}\alpha_{13}+\alpha_{14}\alpha_{12}\alpha_{23}+\alpha_{14}\alpha_{12}\alpha_{13}
+\alpha_{34}\alpha_{14}\alpha_{23}+\alpha_{23}\alpha_{34}\alpha_{13}+\alpha_{14}\alpha_{34}\alpha_{12}\nonumber\\
&&+\;\alpha_{34}\alpha_{12}\alpha_{13}+\alpha_{23}\alpha_{34}\alpha_{12}+\alpha_{24}\alpha_{14}\alpha_{23}
+\alpha_{24}\alpha_{14}\alpha_{13}+\alpha_{24}\alpha_{13}\alpha_{23}\\
&&+\;\alpha_{23}\alpha_{24}\alpha_{12}+\alpha_{13}\alpha_{24}\alpha_{12}+\alpha_{24}\alpha_{34}\alpha_{13}+\alpha_{24}\alpha_{14}\alpha_{34}+\alpha_{24}\alpha_{34}\alpha_{12}.\nonumber
\end{eqnarray}
\end{enumerate}
\end{ex}
Algebraically, these projective integrals may be considered as residues.
We will show how to evaluate them in the case of the above example in Ex.\ \ref{ex3}.

For the purpose of this paper position space Feynman rules are best suited. We will mainly use these in the following.

\subsection{Vacuum graphs}

\begin{figure}[t]
\epsfig{file=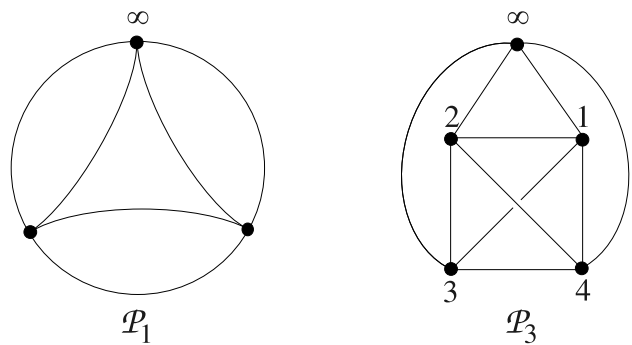,width=\textwidth}
\caption{The completions of the graphs Fig.\ 1 (a) and Fig.\ 1 (b) give $P_1$ and $P_3$ in the census Table 4 of Sect.\ \ref{sect3}.}
\end{figure}

Every 4-point graph in $\phi^4$-theory can be uniquely completed to a 4-regular graph by attaching one extra vertex to the external edges (see Fig.\ 3).
However, the converse is obviously not true: By deleting a vertex from a 4-regular graph one can in general obtain quite different 4-point graphs.
The power of the completion to 4-regular graphs lies in the fact that all these 4-point functions give the same $\phi^4$-period (if any, see Thm.\ \ref{dt2}).
This property is specific to massless renormalizable quantum field theories. It is a consequence of conformal symmetry which is broken on the quantum level but
retained in the residues of primitive graphs.

It is considerably more economical (and more symmetrical) to work with completed graphs because they are (for high loop order) much less in number.
In $\phi^4$-theory 4-regular graphs are vacuum graphs: they have no external edges to be associated to incoming or outgoing particles.
With no external momenta it is natural to assign a pure number to them.
However, we do not give them a physical interpretation. In quantum field theory vacuum amplitudes cancel by normalization.
Here, we consider them as equivalence classes of 4-point graphs that evaluate to the same period.

To obtain a well-defined period we need a criterion for a 4-regular graph to provide primitive 4-point graphs after the removal of a vertex.

\begin{defn}\label{def1}
A 4-regular graph $\Gamma$ with $\geq3$ vertices is (completed) primitive if and only if the only way to split $\Gamma$ with four edge-cuts is
to separate off a vertex.
\end{defn}

Completed primitive graphs may be considered as having `almost' edge-connectivity 6. The completed graphs in Fig.\ 3 are primitive.
We reserve the letter $\ell$ for the `loop order' of the completed graph which is the number of independent cycles $h_1$ of the graph minus any vertex,
\begin{equation}\label{5x}
\ell(\Gamma)=h_1(\Gamma-v)\quad\hbox{ (for any vertex $v$ in $\Gamma$)}.
\end{equation}
The examples in Fig.\ 3 have loop order $\ell=1$ and $\ell=3$, respectively.

It is easy (using B.D. McKay's {\it nauty} \cite{NAU} and writing a little C$++$ program) to list primitive 4-regular graphs. In Table 1 (Sect.\ \ref{sect3})
we count the number of completed primitive graphs up to $\ell=14$. We find that up to $\ell=8$ we have a mere 73 graphs while at $\ell=14$ they are
more than seven million in number.

We need the following elementary lemma:

\begin{lem}\label{lem1}
\begin{enumerate}
\item
For every subgraph $\gamma$ of a 4-regular graph $\Gamma$ with $n(\gamma)$ edges, $N(\gamma)$ vertices,
and `edge-deficiency' $d(\gamma)=\sum_{{\rm vertices}\, v\,{\rm in}\,\gamma} [4-\deg(v)]$ we have
\begin{equation}\label{5z}
d(\gamma)=4N(\gamma)-2n(\gamma).
\end{equation}
\item
If $\Gamma$ is completed primitive with $N$ vertices it has vertex-connectivity $\geq3$ and
\begin{equation}\label{5y}
\ell=N-2.
\end{equation}
\item
The only non-simple completed primitive graph is $P_1$ [see Fig.\ 3].
\end{enumerate}
\end{lem}
\begin{proof}
The number of half edges in $\gamma$ is $2n$ and also $\sum\deg(v)=4N-d$ yielding Eq.\ (\ref{5z}).

By graph homology any connected graph with $N$ vertices has
\begin{equation}\label{5a}
n=h_1+N-1
\end{equation}
edges. Let $\gamma=\Gamma-v$ be a 4-regular graph minus one vertex. We have $d(\gamma)=4$ and from  Eqs.\ (\ref{5z}), (\ref{5a})
$\ell=h_1(\gamma)=n(\gamma)-N(\gamma)+1=N(\gamma)-3=N(\Gamma)-2$ proving Eq.\ (\ref{5y}).

If $\Gamma$ has vertex-connectivity 2 one may cut the two `right' edges of the first cut-vertex and the two `left' edges of the second cut-vertex
to obtain a non-trivial 4-edge cut rendering $\Gamma$ non-primitive. Hence primitive graphs have vertex-connectivity $\geq3$.

Graphs with loops are never primitive. If $\Gamma$ is non-simple and primitive it has a double edge connecting vertices $a$ and $b$ (say).
Cutting the other 4 edges connected to $a$ and $b$ splits the graph. Because $\Gamma$ is primitive these 4 edges have to connect to a single vertex yielding the
graph $P_1$ of Fig.\ 3.
\end{proof}

The following proposition assures that 4-regular graphs lead to well-defined periods if and only if they are completed primitive.

\begin{prop}\label{prop1}
Let $\Gamma$ be a 4-regular graph and $v$ a vertex in $\Gamma$. Then $\Gamma-v$ is primitive if and only if $\Gamma$ is completed primitive.
\end{prop}

\begin{proof}
If $\Gamma$ is completed primitive then $\Gamma$ is connected and $n(\Gamma-v)=2h_1(\Gamma-v)$.
Assume $\Gamma-v$ is not primitive. Then there exists a proper subgraph $\gamma$ of $\Gamma-v$ with $n(\gamma)\leq2h_1(\gamma)$.
Because $\gamma$ is a proper subgraph the complement of $\gamma$ in $\Gamma$ has at least two vertices.
Since $d(\gamma)=4N(\gamma)-2n(\gamma)$ [Eq.\ (\ref{5z})] $=2n(\gamma)-4h_1(\gamma)+4$ [Eq.\ (\ref{5a})] $\leq4$ the subgraph $\gamma$
connects to its complement by not more than 4 edges. This makes $\Gamma$ non-primitive.

If, on the other hand, $\Gamma-v$ is primitive then $\Gamma$ cannot have a nontrivial split by four cuts because every part of the split would have
$n=2h_1$. The part of the split that does not contain $v$ is a proper subgraph of $\Gamma-v$ hence rendering $\Gamma-v$ non-primitive.
\end{proof}

\subsection{The period}

In this subsection we give six equivalent definitions for a $\phi^4$-period.

\begin{dt}\label{dt2}
Let $\Gamma$ be a 4-regular graph with loop order $\ell$. If $\Gamma$ is completed primitive then the following equations define the same number $P_\Gamma$, otherwise
all equations are ill-defined. In the first case $P_\Gamma$ is the $\phi^4$-period of $\Gamma$.
\begin{enumerate}
\item
Projective momentum space.
Choose one vertex in $\Gamma$ with label `$\infty$'. With projective momentum space Feynman rules for $\Gamma-\infty$ (see Eq.\ (\ref{4b}) for an example) we have
\begin{equation}\label{6a}
P_\Gamma=\pi^{-2\ell}A_{\Gamma-\infty}^{\rm mom}=\pi^{-2\ell}\int_{\PP\RR^{4\ell-1}}\frac{\Omega(p)}{\prod_1^{2\ell}|\sum\pm p_i|^2}.
\end{equation}
\item
Affine momentum space.
Choose one vertex in $\Gamma$ with label `$\infty$'. Use standard momentum space Feynman rules for $\Gamma-\infty$ and set one momentum vector
(say $p_1$) to any unit-vector in $\RR^4$. Name this unit-vector `$1$' to obtain
\begin{equation}\label{6b}
P_\Gamma=\pi^{-2(\ell-1)}\int_{\RR^{4(\ell-1)}}\frac{{\rm d}^4p_2\cdots{\rm d}^4p_\ell}{\left.\prod_1^{2\ell}|\sum\pm p_i|^2\right|_{p_1\rightarrow1}}.
\end{equation}
\item
Projective position space.
Choose two vertices in $\Gamma$ with labels `$\infty$' and `$0$'. With projective position space Feynman rules for $\Gamma-\infty$ (see Eq.\ (\ref{4a}) for an example) and
$x_0=0$ where $x_0$ is the position vector associated to the vertex $`0'$
\begin{equation}\label{6c}
P_\Gamma=\pi^{-2\ell}A_{\Gamma-\infty}^{\rm pos}=\pi^{-2\ell}\int_{\PP\RR^{4\ell-1}}\frac{\Omega(x)}{\left.\prod_1^{2\ell}|x_i-x_j|^2\right|_{x_0\rightarrow0}}.
\end{equation}
\item
Affine position space.
Choose three vertices in $\Gamma$ with labels `$\infty$', `$0$', and `$1$'. Use standard position space Feynman rules for $\Gamma-\infty$, set $x_0=0$, and set the position vector
$x_1$ to any unit-vector `$1$' in $\RR^4$ to obtain
\begin{equation}\label{6d}
P_\Gamma=\pi^{-2(\ell-1)}\int_{\RR^{4(\ell-1)}}\frac{{\rm d}^4x_2\cdots{\rm d}^4x_\ell}{\left.\prod_1^{2\ell}|x_i-x_j|^2\right|_{x_0\rightarrow0,\, x_1\rightarrow1}}.
\end{equation}
\item
Parametric space. Choose one vertex in $\Gamma$ with label `$\infty$'. Parametric Feynman rules for $\Gamma-\infty$ (see Eq.\ (\ref{4d}) for an example) give
($\Delta=\{\alpha_i>0\}$)
\begin{equation}\label{6e}
P_\Gamma=A_{\Gamma-\infty}^{\rm par}=\int_\Delta\frac{\Omega(\alpha)}{\Psi_{\Gamma-\infty}(\alpha)^2}.
\end{equation}
\item
Dual parametric space. Choose one vertex in $\Gamma$ with label `$\infty$'. Dual parametric Feynman rules for $\Gamma-\infty$ (see Eq.\ (\ref{4c}) for an example) give
\begin{equation}\label{6f}
P_\Gamma=A_{\Gamma-\infty}^{\rm dual\,par}=\int_\Delta\frac{\Omega(\alpha)}{\bar\Psi_{\Gamma-\infty}(\alpha)^2}.
\end{equation}
\end{enumerate}
\end{dt}
\begin{proof}
It was proved in Thm.\ \ref{dt1} and Prop.\ \ref{prop1} that the existence of the integral in Eq.\ (\ref{6a}) is equivalent to $\Gamma$ being primitive.
The validity and equivalence of the list of equations is proved in six steps.

First, we show that Eq.\ (\ref{6a}) is equivalent to Eq.\ (\ref{6b}) for an identical choice of `$\infty$'. From Eq.\ (\ref{6a}) we go to affine space
by setting the 1-component of $p_1$ to $1$, hence $p_1=(1,{\bf p}_1)$ for ${\bf p}_1\in\RR^3$. Next we rescale all $p_i$, $i\geq2$ by $p_i\mapsto |p_1|p_i$.
Because $\deg(\Omega)=4\ell$ we obtain with `1'$\,=p_1/|p_1|$
\begin{equation*}
\pi^{-2\ell}\int_{\RR^3}\frac{{\rm d}^3{\bf p}_1}{|p_1|^4}\cdot\int_{\RR^{4(\ell-1)}}
\frac{{\rm d}^4p_2\cdots{\rm d}^4p_\ell}{\left.\prod_1^{2\ell}|\sum\pm p_i|^2\right|_{p_1\rightarrow1}}.
\end{equation*}
The first factor evaluates to $\pi^2$ by Ex.\ \ref{ex1} whereas the second factor is independent of the direction of $p_1/|p_1|$ by rotational symmetry.
We also see that the period in Eq.\ (\ref{6b}) does not depend on the choice of $p_1$.

Second, we prove that Eq.\ (\ref{6a}) is equivalent to Eq.\ (\ref{6e}) for an identical choice of `$\infty$'.
A series of elementary integrations leads to
\begin{equation}\label{7a}
\frac{1}{Q_1Q_2\cdots Q_{2\ell}}=(2\ell-1)!\int_0^\infty\cdots\int_0^\infty
\frac{{\rm d}\alpha_2\cdots{\rm d}\alpha_{2\ell}}{(Q_1+\alpha_2Q_2+\ldots+\alpha_{2\ell}Q_{2\ell})^{2\ell}}.
\end{equation}
Here $\sQ=Q_1+\alpha_2Q_2+\ldots+\alpha_{2\ell}Q_{2\ell}$ is the `universal quadric' \cite{BLO}.
For $\alpha_i>0$ it is given by a positive definite $4\ell\times4\ell$ matrix $M$ which is block diagonal
with 4 identical blocks of $\ell\times\ell$ matrices $N$, one for each space-time dimension.
By a real linear transformation $S$ we bring $M$ into its normal form which is a unit-matrix,
$S^{T}MS=1$\hspace{-3.1pt}I. The projective volume form transforms by the determinant of $S$. Note that $\det(S)=\det(M)^{-1/2}=\det(N)^{-2}$ and
\begin{equation}\label{7b}
P_\Gamma=\pi^{-2\ell}(2\ell-1)!\int_{\PP\RR^{4\ell-1}}\frac{\Omega(p)}{(\sum_{i=1}^{2\ell}|p_i|^2)^{2\ell}}
\cdot\int_0^\infty\cdots\int_0^\infty\frac{{\rm d}\alpha_2\cdots{\rm d}\alpha_{2\ell}}{\det(N)^2}.
\end{equation}
We translate the first integral on the right hand side into an affine integral over the unit sphere $S^{4\ell-1}=\{\sum_{i=1}^{2\ell}|p_i|^2=1\}$.
The projective volume form induces the standard measure on $S^{4\ell-1}$.
Because the sphere is a double cover of the real projective space (and the integrand is 1) we obtain vol$(S^{4\ell-1})/2=\pi^{2\ell}/\Gamma(2\ell)$
for the first integral. After transition to projective space in the second integral
we finally have to show than $\det(N)=\Psi_\Gamma$ which is the result of the Matrix-Tree Theorem Prop.\ 2.2 in \cite{BEK}.

Third, we prove that Eq.\ (\ref{6e}) is equivalent to Eq.\ (\ref{6f}) for an identical choice of `$\infty$'. This is obvious from a Cremona transformation which
in affine space $\alpha_1=1$ amounts to a series of one-dimensional inversions $\alpha_i\mapsto 1/\alpha_i$, $i=2,\ldots,2\ell$.

Fourth, we show that Eq.\ (\ref{6c}) is equivalent to Eq.\ (\ref{6f}) for an identical choice of `$\infty$'. This is achieved by literally the same method as in the second step.
Starting from position space leads to a matrix $N$ that is the (`0',`0') minor of the `graph Laplacian'. Another Matrix-Tree Theorem (see e.g.\ \S4 in \cite{LOV}) assures
that the determinant of any ($i,i$)-minor of the graph Laplacian is given by the dual graph polynomial. As a side-effect we see that the period in
Eq.\ (\ref{6c}) does not depend on the choice of `0'.

Fifth, we show that Eq.\ (\ref{6c}) is equivalent to Eq.\ (\ref{6d}) for an identical choice of `$\infty$'. This is exactly the same proof as in the first step showing the
same equivalence in momentum space. As a consequence the period in Eq.\ (\ref{6d}) cannot depend on the choices of `0' and `1'.

Sixth, we have to prove that $P_\Gamma$ does not depend on the choice of `$\infty$'. This is done in affine position space using Eq.\ (\ref{6d}).
An inversion $x_i\mapsto x_i/|x_i|^2$, $i\neq$ `0', `1', `$\infty$' transforms propagators $|x_i-x_j|^{-2}$ to $|x_i|^2|x_j|^2|x_i-x_j|^{-2}$ and $|x_i|^{-2}$ to $|x_i|^2$.
Together with the change in the integration measures ${\rm d}^4x_i\mapsto {\rm d}^4x_i|x_i|^{-8}$ we observe that 3-valent vertices in $\Gamma-\infty$
become connected to `0' whereas vertices connected to `0' in $\Gamma-\infty$ become 3-valent. Keeping in mind that 3-valent vertices are connected to `$\infty$' in $\Gamma$,
inversion interchanges the choices for `0' and `$\infty$'. Because the choice of `0' is arbitrary before and after the inversion the period cannot depend on
the choice of `$\infty$'. Going backwards the same has to be true for any of the formulae we gave for the period.
\end{proof}

We close this subsection with the first calculation of a non-trivial $\phi^4$-period.

\begin{ex}\label{ex3}
Consider the graph $P_3$ plotted in Fig.\ 3. With any choice for `0', `1', `$\infty$' we obtain from Eq.\ (\ref{6d}) [compare Eq.\ (\ref{4a})]
\begin{equation*}
P_3=\pi^{-4}\int_{\RR^8}\frac{{\rm d}^4x_2{\rm d}^4x_3}{|1-x_2|^2|x_2|^2|x_2-x_3|^2|x_3|^2|x_3-1|^2}.
\end{equation*}
The best way to evaluate this integral is by using Gegenbauer-Techniques \cite{C4}. Quite efficiently one may use
\begin{equation}
\frac{1}{|x-y|^2}=\frac{1}{|xy|}\int_{-\infty}^\infty\frac{{\rm d}p}{\pi}\sum_{n=1}^\infty C_{n-1}(\cos\theta_{xy})\left|\frac{x}{y}\right|^{ip}\frac{n}{n^2+p^2}
\end{equation}
where $\theta_{xy}$ is the angle between $x$ and $y$. Orthogonality of the Gegenbauer polynomials ($\hat y=y/|y|$)
\begin{equation}
 \int_{S^3}\frac{{\rm d}\hat y}{2\pi^2}C_{n-1}(\cos\theta_{xy})C_{m-1}(\cos\theta_{yz})=\frac{\delta_{n,m}}{n} C_{n-1}(\cos\theta_{xz})
\end{equation}
and $\int_0^\infty{\rm d}x|x|^{{\rm i}p-1}=2\pi\delta(p)$ leads to ($C_{n-1}(1)=n$)
\begin{eqnarray*}
P_3&=&16\pi^{-1}\sum_{n=1}^\infty\int_{-\infty}^\infty{\rm d}p\frac{n^2}{(n^2+p^2)^3}\\
&=&16\pi^{-1}\zeta(3)\int_{-\infty}^\infty{\rm d}p\frac{1}{(1+p^2)^3}\\
&=&6\zeta(3).
\end{eqnarray*}
\end{ex}
The only other period that can be calculated that easily is $P_4$ if one chooses for `0' and `$\infty$' opposite vertices of the octahedron graph.
The result is $20\zeta(5)$, see Table 4, Sect.\ \ref{sect3}. The periods $P_3$ and $P_4$ are the first two members of the zig-zag family that conjecturally evaluates
to a rational multiples of $\zeta(2\ell-3)$, see Subsect.\ \ref{sym} and Eq.\ (\ref{8d}).

A calculation of $P_3$ using parametric space can be found in \S7 of \cite{BRO1}.

\subsection{Vertex-connectivity 3: The product identity}

The periods of primitive 3-vertex-connected graphs reduce to products of periods of smaller graphs.
\begin{defn}\label{def2}
A completed primitive graph is reducible if it has vertex-connectivity 3, otherwise it is irreducible.
\end{defn}

\begin{figure}[t]
\epsfig{file=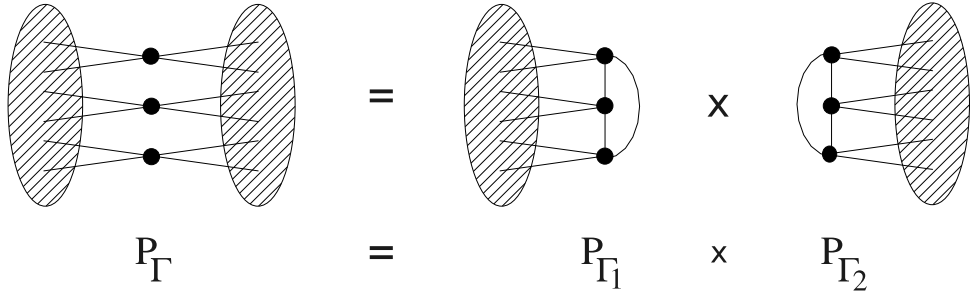,width=\textwidth}
\caption{Vertex connectivity 3 leads to products of periods.}
\end{figure}

With this definition we obtain the following theorem (see Fig. 4).
\begin{thm}\label{thm1}
A reducible completed primitive graph $\Gamma$ is the gluing of two completed primitive graphs $\Gamma_1$ and $\Gamma_2$ on triangle faces
followed by the removal of the triangle edges. The period of $\Gamma$ is the product of the periods of $\Gamma_1$ and $\Gamma_2$,
\begin{equation}\label{7}
P_{\Gamma}=P_{\Gamma_1}P_{\Gamma_2}.
\end{equation}
\end{thm}
\begin{proof}
The gluing of 4-valent graphs $\Gamma_1$, $\Gamma_2$ along triangles with vertices $v_1$, $v_2$, $v_3$ leads to a graph with 6-valent $v_1$, $v_2$, $v_3$
whereas all other vertices remain 4-valent. After the removal of the triangle edges we obtain a 4-regular graph $\Gamma$. This graph has vertex-connectivity $\leq3$
because it splits with the removal of $v_1$, $v_2$, $v_3$. If $\Gamma_1$ and $\Gamma_2$ are primitive then they have well-defined periods.
By Eq.\ (\ref{7}) (independently proved below) $\Gamma$ has a well-defined period and it is primitive by Thm.\ \ref{dt2}. By Lemma \ref{lem1} it has vertex-connectivity $\geq3$
(hence $=3$) making $\Gamma$ reducible.

If, on the other hand, a primitive graph $\Gamma$ has vertex-connectivity 3 it splits into $\gamma_1$ and $\gamma_2$ by the removal of $v_1$, $v_2$, $v_3$.
We attach $v_1$, $v_2$, $v_3$ to $\gamma_1$ and $\gamma_2$ in the same way they were attached in $\Gamma$ and define $d_{i,j}$ as the degree of $v_i$ in $\gamma_j$.
We have (1) $d_{i,1}+d_{i,2}=4$ because $v_i$ had degree 4 in $\Gamma$, (2) $d_{1,j}+d_{2,j}+d_{3,j}$ is even by Eq.\ (\ref{5z}),
and (3) $d_{1,j}+d_{2,j}+d_{3,j}>4$ because $\Gamma$ is primitive. The only solution for (1), (2), (3) is all $d_{i,j}=2$ making the split graphs 4-regular after
the addition of the triangles $(v_1,v_2,v_3)$. Again, they are primitive by Eq.\ (\ref{7}).

To prove Eq.\ (\ref{7}) we use Eq.\ (\ref{6d}) and choose $v_1=$`0', $v_2=$`1', $v_3=$`$\infty$'. In this case the integral on the right hand side becomes a product
of two integrals, according to the vertex sets of $\Gamma_1$ and $\Gamma_2$ respectively. The triangle $(v_1,v_2,v_3)$ gives an extra propagator connecting
0 and 1. This propagator is $|1|^{-2}=1$  by definition and does not change the integrand. Hence we are free to add or remove the triangle.
Because, by Eq.\ (\ref{5y}), $\ell(\Gamma)-1=N(\Gamma)-3=N(\Gamma_1)+N(\Gamma_2)-6=\ell(\Gamma_1)-1+\ell(\Gamma_2)-1$ the right hand side of Eq.\ (\ref{6d})
factors into $P_{\Gamma_1}P_{\Gamma_2}$. We obtain Eq.\ (\ref{7}) if $P_\Gamma$ exists
and going backwards we also have Eq.\ (\ref{7}) if $P_{\Gamma_1}$ and $P_{\Gamma_1}$ exist.
\end{proof}

Thm.\ \ref{thm1} gives a `multiplication' on graphs. However, graphs with no triangles (like $P_{6,4}$ in Table 4) do not `multiply'.
Moreover the `multiplication' depends on the way the triangles are chosen. If $\Gamma_1$ has $n_1$ triangles and $\Gamma_2$ has $n_2$ triangles there are
$6n_1n_2$ ways to glue (for small graphs many of these will give isomorphic results). See Table 2 for the number of non-isomorphic gluings of irreducible graphs.
In the special case that `$\infty$' is one of the split vertices Thm.\ \ref{thm1} follows form Prop.\ 39 in \cite{BROW}.

In Table 1, Sect.\ \ref{sect3}, we see that 13 of the 73 primitive graphs up to loop order 8 are reducible. Because their periods can be
derived from periods of smaller graphs we did not include them in Table 4. At 14 loops 93.7\% of the primitive graphs are irreducible.

Note that irreducible graphs have vertex-connectivity 4 because it is always possible to separate off a vertex by removing its neighbors.

\subsection{Vertex-connectivity 4: The twist identity}

\begin{figure}[t]
\epsfig{file=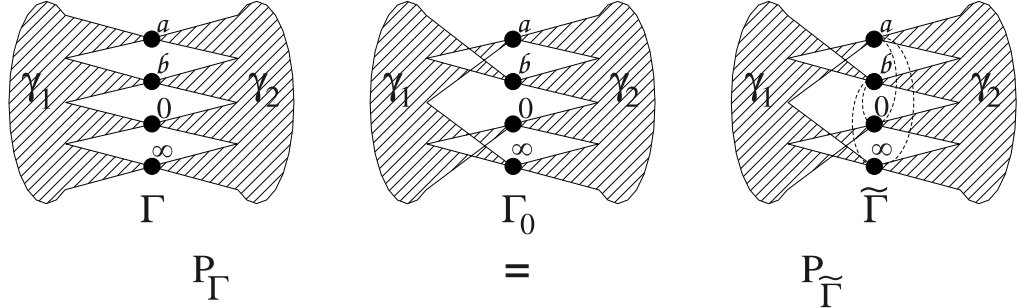,width=\textwidth}
\caption{Twist transformation: Twist the `left' graph $\gamma_1$ (or equivalently the `right' graph $\gamma_2$) to obtain $\Gamma_0$. Try to move edges to opposite sides
of the dashed 4-cycle (if necessary) to obtain a 4-regular graph. If successful the new graph is the twisted graph $\tilde\Gamma$ with $P_\Gamma=P_{\tilde\Gamma}$.}
\end{figure}

Vertex-connectivity 4 leads to an identity between periods of graphs depicted in Fig.\ 5.

\begin{thm}\label{thm2}
Let $\Gamma$ be a completed primitive graph with vertex-connectivity 4 realized by the vertices $a$, $b$, $c$, $d$. Let $\gamma_1$ and $\gamma_2$ and be the
split graphs with the vertices $a$, $b$, $c$, $d$ added in the way they were attached in $\Gamma$. Connect vertices $a,b\in\gamma_1$ to vertices $b,a\in\gamma_2$
and vertices $c,d\in\gamma_1$ to vertices $d,c\in\gamma_2$ (resp.) to obtain $\Gamma_0$.
If $\Gamma_0$ is 4-regular then $\tilde\Gamma=\Gamma_0$. Otherwise assume it is possible to swap
edges $ac\leftrightarrow bd$ or $ad\leftrightarrow bc$ to (uniquely) obtain a 4-regular graph $\tilde\Gamma$. Then $\tilde\Gamma$ is primitive and
\begin{equation}\label{7c}
P_\Gamma=P_{\tilde\Gamma}.
\end{equation}
\end{thm}
\begin{proof}
We start from Eq.\ (\ref{6c}) with `0'$\,=c$ and `$\infty$'$\,=d$. For simplicity we use $a$ and $b$ as variables associated to vertices $a$ and $b$.
The other vertices of $\gamma_1$ have the variables $x_i$ whereas the variables located at the vertices of $\gamma_2$ are $y_j$.
We use quaternions to define the following projective degree 1 coordinate transformation
\begin{equation}\label{7d}
\sigma:\quad x_i\mapsto ax_i^{-1}b,\quad a\mapsto a,\quad b\mapsto b,\quad y_j\mapsto y_j.
\end{equation}
The transformation $\sigma$ is the identity on $\gamma_2$ while the propagators in $\gamma_1$ are transformed as $|x_i-x_j|^{-2}\mapsto |x_ix_j/(ab)|^2|x_i-x_j|^{-2}$,
$|x_i-a|^{-2}\mapsto |x_i/a|^2|b-x_i|^{-2}$, $|x_i-b|^{-2}\mapsto |x_i/b|^2|a-x_i|^{-2}$ (interchanging $a$ and $b$),
and $|x_i|^{-2}\mapsto |x_i/(ab)|^2$. The integration measure transforms under
$\sigma$ by a Jacobian determinant which can be calculated by a sequence of inversions $x_i\mapsto x_i^{-1}$ and rescalings $x_i\mapsto ax_i$, $x_i\mapsto x_ib$ (choose
one of the $y_j$-components $=1$ to make the measure affine). The inversions reproduce the propagators $|x_i-x_j|^{-2}$ and interchange $0$ and $\infty$ (see
step 6 in the proof of Thm.\ \ref{dt2}).

If $\gamma_1-\{a,b,0,\infty\}$ (minus the edges attached to $a,b,0,\infty$) has $N$ vertices and $n$ edges
and $d_a$, $d_b$, $d_0$, $d_\infty$ are the degrees of $a$, $b$, $0$, $\infty$ in
$\gamma_1$ (respectively) then $\sigma$ generates a factor $|a|^{4N-2n-2d_a-2d_0}$. Because $\gamma_1-\{a,b,0,\infty\}$ has edge-deficiency (see Lemma \ref{lem1})
$d_a+d_b+d_0+d_\infty$ we can rewrite the factor using Eq.\ (\ref{5z}) as $|a|^{-d_a+d_b-d_0+d_\infty}$.
On the other hand, in $\Gamma_0$ the vertices $a$, $b$, $0$, $\infty$ have degrees
$D_a=d_b+4-d_a$, $D_b=d_a+4-d_b$, $D_0=d_\infty+4-d_0$, $D_\infty=d_0+4-d_\infty$, respectively. Making the vertices 4-regular by moving $s$-times edge $a0$ to edge $b\infty$
($s=-1,0,1$ with $s=-1$ meaning moving edge $b\infty$ to edge $a0$) and $t$-times edge $a\infty$ to edge $b0$ amounts to a factor $|a|^{2s}$. Because after moving edges
the vertices $a$ and $0$ have degree 4 we obtain the conditions $D_a-s-t=4$ and $D_0-s+t=4$. This determines $s$ to $s=(D_a+D_0)/2-4=(d_b-d_a+d_\infty-d_0)/2$ and hence
the factor from moving the edges equals the factor from the transformation $\sigma$. By symmetry the same holds for the powers of $|b|$. Thus $\sigma$ transforms Eq.\ (\ref{6c})
into a period-integral for $\tilde\Gamma$. This proves Eq.\ (\ref{7c}) and because $P_{\tilde\Gamma}$ is finite the graph $\tilde\Gamma$ is (completed) primitive
by Thm.\ \ref{dt2}.
\end{proof}

Note that the twist is symmetric under exchanging $\gamma_1$ and $\gamma_2$ although the proof is not.
Moreover, a double twist with respect to the same vertices is the identity.
Although defined for all primitive graphs, the twist transformation operates on irreducible graphs.

\begin{rem}\label{rem1}
The twisted graph $\tilde\Gamma$ is irreducible if and only if $\Gamma$ is irreducible.
\end{rem}
\begin{proof}
Assume $\Gamma$ splits into $\Gamma_1$ and $\Gamma_2$ by removing the vertices $a,b,c$.

If $a,b$, and $c$ lie in $\gamma_2$ (or in $\gamma_1$) then $\Gamma_1$ or
$\Gamma_2$ is a subgraph of $\gamma_2$ (otherwise $\Gamma_1$ and $\Gamma_2$ would be connected via $\gamma_1$ and could not split with the removal of $a,b,c$).
Because the transformation $\sigma$ in  Eq.\ (\ref{7d}) is the identity on $\gamma_2$ the twisted graph $\tilde\Gamma$ is reducible.

If $a,b$, and $c$ do not all lie in $\gamma_1$ or $\gamma_2$ then it is easy to see that there exists another set of three vertices
that splits $\Gamma$ and fully lies in $\gamma_1$ or in $\gamma_2$.

Hence $\Gamma$ is irreducible if $\tilde\Gamma$ is irreducible. The converse is true because a twist of $\tilde\Gamma$ is isomorphic to $\Gamma$.
\end{proof}

Note that a 4-vertex cut of $\Gamma$ does not necessarily lead to twist graphs. If twist graphs exist for vertices $(abcd)$, $(acbd)$, and $(adbc)$
(changing the labeling)  then the transformations form a Klein four-group $C_2\times C_2$ and this is the largest set of transformations
one can get from one split. There may be more twist identities for other ways to split $\Gamma$ by removing 4 vertices.

In many cases the graphs $\Gamma$ and $\tilde\Gamma$ are isomorphic. This is always the case when the twist transformation is applied to the neighbors of a vertex.
But still non-trivial splits of primitive graphs are quite common. Amongst the 60 irreducible graphs up to loop order 8 we have 10 non-trivial twist identities.
In particular at high loop order the twist-identity appears to be quite frequent. By applying the twist in different ways to the same graph one obtains larger
equivalence classes of graphs. At loop order 11 the twist identity reduces the number of potentially different periods from 8~687 irreducible
graphs to 6~300 by forming equivalence classes of up to 12 graphs.

\subsection{Planar graphs: The Fourier identity}\label{Fourier}

An identity that can already be found in \cite{BK} is the re-interpretation of momentum vectors as position vectors (introduced in \cite{B7}).
Because the momentum space Feynman rules are derived from position space by a Fourier-transform we call it a Fourier identity.

To allow for the re-interpretation of momentum vectors as position vectors the Feynman graph has to have a planar embedding. When starting
from a completed graph $\Gamma$ we first have to identify a vertex $v=$`$\infty$' (if possible) such that $\Gamma-v$ has a planar embedding.
Once we have a planar embedding we may determine the dual graph and try to complete it to a 4-regular graph by adding a vertex (see Fig.\ 6).
If $\Gamma$ was irreducible then $\Gamma-v$ has vertex-connectivity 3 and the dual graph is unique.

\begin{figure}[t]
\epsfig{file=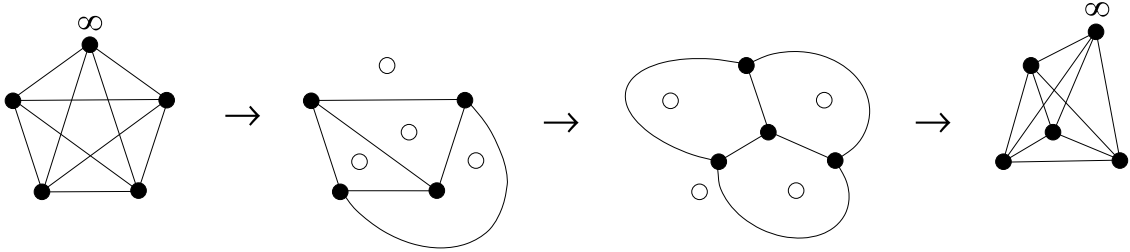,width=\textwidth}
\caption{Fourier transformation: remove one vertex, draw the dual (if possible), add one vertex and connect it to all 3-valent vertices.
If the result is 4-regular then it is the Fourier transformed graph $\hat\Gamma$ with $P_\Gamma=P_{\hat\Gamma}$.
In the above example Fourier transformation leads back to the original graph. In some (rare) cases it leads to new graphs.}
\end{figure}

\begin{thm}\label{thm3}
Let $\Gamma$ be a completed primitive graph.
If $\Gamma-v$ has a planar embedding for some vertex $v$ and if the dual graph can be completed to a 4-regular graph $\hat\Gamma$ by adding one vertex $\hat v$
then $\hat\Gamma$ is primitive and
\begin{equation}\label{8a}
P_\Gamma=P_{\hat\Gamma}.
\end{equation}
\end{thm}
\begin{proof}
The momentum space Feynman rules of $\Gamma-v$ are identical to the position space Feynman rules of $\hat\Gamma-\hat v$ if one sets the outside-vertex to 0
and uses the other vertices of $\hat\Gamma-\hat v$ as cycle base of $\Gamma-v$. Eq.\ (\ref{8a}) is thus a consequence of the equivalence of Eqs.\ (\ref{6a}) and (\ref{6c}).
The graph $\hat\Gamma$ is primitive by Thm.\ \ref{dt2} because it has a finite period.
\end{proof}

Similar to the twist identity the Fourier identity establishes an equivalence relation between graphs with equivalence classes that can have more than two elements
(one may be able to choose different vertices $v$). However, Fourier identities are rare. We have five identities up to loop order 8. Three of these can also
be obtained by the twist identity. The two new identities reduce the number of irreducible periods up to loop order 8 to 48 (see Table 1, Sect.\ \ref{sect3},
31 of these periods have been identified as MZVs).
The first instance of a Fourier equivalence class with three elements is found at loop order 11 where the number of independent identities is 43
(as compared to 2~387 independent twist identities).

\begin{figure}[t]
\epsfig{file=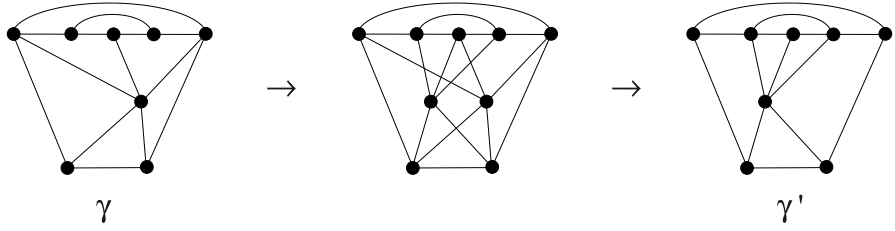,width=\textwidth}
\caption{Extended Fourier transformation: Go from $\gamma$ to $\gamma'$ by adding one vertex, connect it to all 3-valent vertices, and remove the vertex with degree $\geq5$
in $\gamma$.}
\end{figure}

It is possible to slightly extended the Fourier identity (see Fig.\ 7).

\begin{rem}\label{rem2}
One can extend the Fourier identity to some graphs that have a dual $\gamma$ (after the removal of a vertex) which does not complete to a 4-regular graph.
Assume $\gamma$ fails to complete to a 4-regular graph because is has one vertex $w$ with degree $\geq5$ whereas all other vertices have degree 3 or 4.
Define a graph $\gamma'$ by adding a vertex connected to all 3-valent vertices of $\gamma$ followed by the removal of $w$.
If $\gamma'$ is planar determine a dual $\hat\gamma'$. If $\hat\gamma'$ fails to complete to a 4-regular graph for the same reason as $\gamma$ then continue
to perform the above transformation (if possible) until $\hat\gamma^{\prime\cdots\prime}$ completes to a 4-regular graph  $\hat\Gamma^{\prime\cdots\prime}$ or
$\hat\gamma^{\prime\cdots\prime}$ has more than one vertex with degree $\geq5$. In the first case one has
\begin{equation}\label{8b}
P_\Gamma=P_{\hat\Gamma^{\prime\cdots\prime}}.
\end{equation}
\end{rem}
\begin{proof}
In position space (Eq.\ (\ref{6c}) or Eq.\ (\ref{6d}) with $v=\infty$ and $w=0$) the transformation between $\gamma$ and $\gamma'$ is an inversion, step 6 in the
proof of Thm.\ \ref{dt2}.
\end{proof}

As in the case of the twist identity a double Fourier transformation (not changing the deleted vertex) is the identity.
Moreover, the Fourier transformation operates on irreducible graphs.

\begin{rem}\label{rem3}
The Fourier transformed graph $\hat\Gamma$ is irreducible if and only if $\Gamma$ is irreducible.
\end{rem}
\begin{proof}
Assume $\Gamma$ splits into $\Gamma_1$ and $\Gamma_2$ by removing the vertices $a$, $b$, $c$. Then $a$, $b$, and $c$ connects to both split graphs with two edges each (see Fig.\ 4).

If $v$ is one of the vertices $a$, $b$, or $c$ then $\Gamma-v$ has vertex-connectivity 2. Its dual, too, has vertex-connectivity 2, hence $\hat\Gamma$ is reducible.

If $v$ is none of the vertices $a$, $b$, or $c$ then the dual graph has vertex-connectivity 3 realized by vertices that lie `between' $ab$, $bc$, $ca$
on the two-sphere $S^2$. In the same way as $\Gamma$ each of these vertices connects to the split graphs with two edges (draw the dual of $\Gamma$ in Fig.\ 4).
Let $\Gamma_1$ and $\Gamma_2$ be the split graphs with reattached vertices $a$, $b$, $c$ (the graphs $\Gamma_1$ and $\Gamma_2$ in Fig.\ 4 without the triangle $(01\infty)$).
If $v$ lies in the `left' graph $\Gamma_1$ (without restriction) then $\Gamma_1-v$ has edge-deficiency $d_1=10$ (see Lemma \ref{lem1}) whereas $\Gamma_2$ has
deficiency $d_2=6$. By application of Eqs.\ (\ref{5z}) and (\ref{5a}) we can calculate the deficiencies of the split graphs of $\hat\Gamma-\hat v$ to $\hat d_1=6$ (the `left' part)
and $\hat d_2=10$. Because $\hat\Gamma$ is 4-regular the extra vertex $\hat v$ connects only to the `right' part. Thus $\hat\Gamma$ retains vertex-connectivity 3.

Hence $\Gamma$ is irreducible if $\hat\Gamma$ is irreducible. The converse is true because $\hat\Gamma$ Fourier-transforms to $\Gamma$.
\end{proof}

The author did not find an extended Fourier symmetry that leads out of the subset of irreducible graphs although the above proof does not apply to this case.

It is well possible that there exist more transformations that leave the period invariant although up to date all identities found numerically
are of twist of Fourier type.

\subsection{Weight and the double triangle reduction}

In general, a period is an integral of an algebraic differential form over a simplex bounded by algebraic inequalities \cite{K3}.
Sums and products of periods are periods which makes the set of periods a $\bar\QQ$-algebra.

Special periods are multiple zeta values (MZVs) (for data and a recent overview see \cite{MZDM}).
\begin{equation}\label{9a}
\zeta(n_1,\ldots,n_r)=\sum_{k_1>\ldots>k_r\geq1}\frac{1}{k_1^{n_1}\cdots k_r^{n_r}}\quad n_1\geq2,\;n_i\in\NN.
\end{equation}
The sum over the exponents on the right hand side $n=n_1+\ldots+n_r$ is the weight of the MZV. It is invariant under regularized shuffle and quasi shuffle (stuffle) relations
which are conjectured to generate all relations between MZVs. Restricting oneself to these two sets of identities (defining formal MZVs) the weight of an MZV is
a well-defined concept. When one considers MZVs as real numbers it seems hopeless trying to prove that there exist no weight-violating relations amongst them.
We keep this in mind although we do not always stress the difference.

In this paper we encounter $\QQ$-linear combinations of MZVs. The vector space $\QQ[MZV]$ of such numbers forms a $\QQ$-algebra (conjecturally) graded by the weight.
A number in $\QQ[MZV]$ has pure weight $n$ if it has contributions from the weight $n$ sector only, otherwise it mixes weights.
The dimensions of the pure weight $n$ subspaces are conjectured to follow a Fibonacci type sequence $d_n=d_{n-2}+d_{n-3}$, (see \cite{Z1} and \cite{B3}).
All MZVs are periods because there exists a representation of $\zeta(n_1,\ldots,n_r)$ as an integral of a rational $n$-form over a simplex.

In quantum field theory we do not have a standard representation for periods: The integrals given in Eqs.\ (\ref{6a})--(\ref{6f}) are much too complicated for
that purpose. In this case we do best to consider a period as a real number and call it an element in $\QQ[MZV]$ if a rational linear combination
of MZVs evaluates to it. In such a situation clean statements can still be made on upper bounds of its maximum weight according to the following definition
(which is not in general consistent with the definition of weight in Hodge theory).

\begin{defn}\label{def3}
The maximum weight of a period $P$ is the smallest integer $n$ such that $P$ is the integral of an algebraic $n$ form over a simplex bounded by algebraic inequalities.
\end{defn}

The maximum weight of a number given by a rational linear combination of (formal) MZVs is smaller or equal to the maximum of its weights.
If, e.g.\ a number evaluates to $\zeta(3)+2\zeta(2,2)-\zeta(3,2)/2$ its maximum weight is $\leq5$.
(A similar concept with sums replacing integrals leads to a `maximum depth' in the case of MZVs.) For $\phi^4$-periods we have the following statement

\begin{lem}\label{lem2}
The maximum weight of a $\phi^4$-period of loop order $\ell$ is $\leq2\ell-3$.
\end{lem}
\begin{proof}
After three integrations starting from Eq.\ (\ref{6e}) with $\alpha_{2\ell}=1$ we are left with $2\ell-4$ integrals over rational linear combinations of logarithms.
For details see \S10.3 in \cite{BROW}.
\end{proof}

In Table 4, Sect.\ \ref{sect3}, we observe weights $2\ell-3$, $2\ell-4$, and in one case ($P_{8,16}$) the mixing of weights $2\ell-5$, and  $2\ell-6$.
We would like to have a graph-theoretical criterion that predicts the maximum weight of a $\phi^ 4$-period.
Such a criterion is still missing (see \cite{BY} for recent results).
What we have is the partially proved conjecture that a `double triangle reduction' does not alter the maximum weight of the period (see Fig.\ 8).
The double triangle reduction is the completed version of the construction in Thm.\ 130 of \cite{BROW} (see also \cite{BY}).

\begin{figure}[t]
\epsfig{file=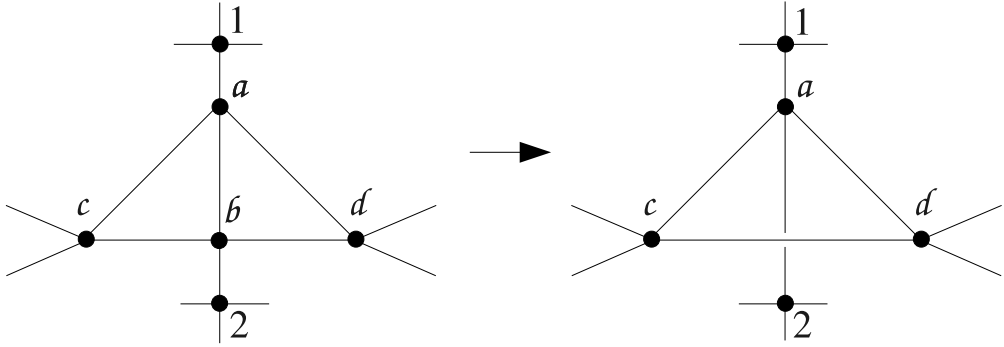,width=\textwidth}
\caption{Double triangle reduction: Replace a joint vertex of two attached triangles by a crossing.}
\end{figure}

\begin{defn}\label{def4}
Assume a graph has an edge $ab$ that is the common edge of (exactly) two triangles $(abc)$ and $(abd)$, $c\neq d$.
In the (double triangle) reduced graph one of the vertices of the edge $ab$ is replaced by a crossing with edge $cd$.
\end{defn}

It is obvious that the reduced graph does not depend on which vertex of the common edge is replaced by the crossing. Moreover we have the following proposition.

\begin{prop}\label{prop2}
Any double triangle reduction of a completed primitive graph is completed primitive.
\end{prop}
\begin{proof}
First, we observe that $P_1$ (see Fig.\ 3) does not have a double triangle reduction. All other primitive graphs are simple by Lemma \ref{lem1}.

Second, we see that the reduction of a simple graph can only be non-simple in the cases (a) and (b) of Fig.\ 9.
While Fig.\ 9 (a) is ruled out by definition (because 3 triangles meet in an edge) we find that Fig.\ 9 (b) can only be a subgraph of a primitive
graph if all four `external' edges connect to the same vertex (otherwise one obtains a non-trivial cut by these 4 edges rendering the graph (completed) non-primitive).
In this case we obtain $P_3$ (see Fig.\ 3) which does not have a double triangle reduction because 3 triangles meet in every edge.

Third, assume the reduced graph of a simple primitive graph $\Gamma$ has a non-trivial split by cutting 4 edges $a$, $b$, $c$, $d$.
The reduction of the double triangle gives a single triangle. If none of its edges is in $\{a,b,c,d\}$ then the original graph $\Gamma$ has the same non-trivial split.
Otherwise two edges of the triangle (say $a$ and $b$) have to be in $\{a,b,c,d\}$.
In this case the split cuts two edges of a vertex. Cutting the other two edges of the vertex (say $e$ and $f$) together with $c$ and $d$ gives another split.
This split is non-trivial because otherwise $e$ and $f$ had to connect to the same vertex forming a double edge (making $\Gamma$ non-simple).
Moreover, none of the edges of the reduced triangle is in $\{c,d,e,f\}$ providing $\Gamma$ with a non-trivial split.
\end{proof}

\begin{figure}[t]
\epsfig{file=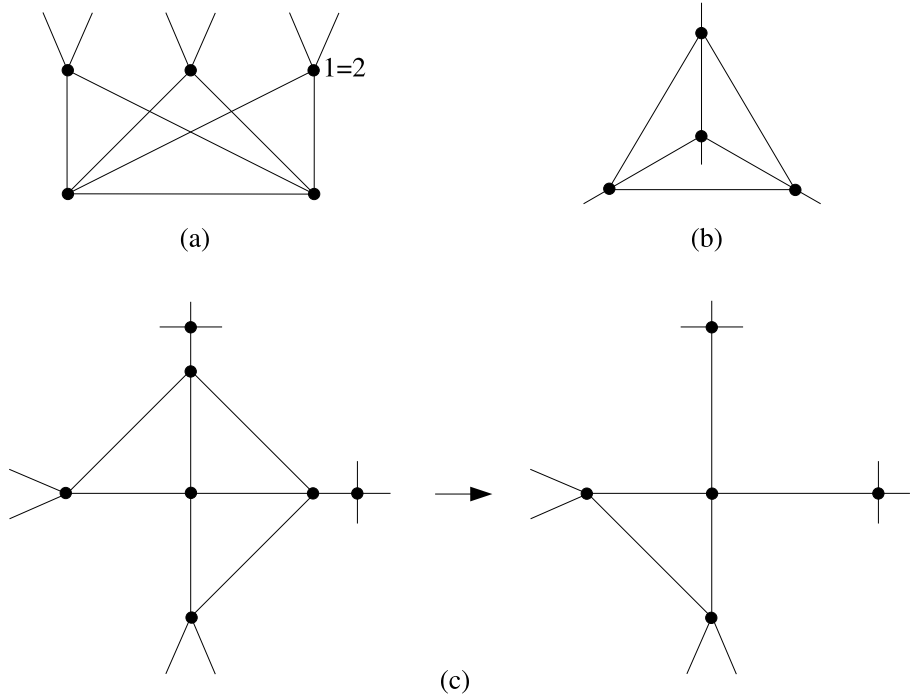,width=12cm}
\caption{Double triangle reductions of completed primitive graphs commute. Cases (a) and (b) are not allowed. In case (c) any application
of two double triangle reductions gives the same result.}
\end{figure}

Because of Prop.\ \ref{prop2} it is possible to compare periods of graphs with the periods of their reduced graphs.

\begin{con}\label{con1}
Double triangle reduction does not alter the maximum weight of the period.
\end{con}

A closely related statement is proved as Thm.\ 36 in \cite{BY}.

To conjecture the maximum weight of an unknown period one would like to apply as many double-triangle reductions as possible. Afterwards, with some luck,
the period of the completely reduced graph is known and one can read of (an upper bound for) the maximum weight. Because of the following proposition
the completely reduced graph is uniquely determined by the original graph.

\begin{prop}\label{prop3}
Double triangle reductions commute.
\end{prop}
\begin{proof}
The statement is obvious unless 3 triangles allow for different reductions. All possible cases are given in Fig.\ 9.
The situations (a) and (b) are ruled out in the proof of Prop.\ \ref{prop2}. In Fig.\ 9 (c) any application of two reductions
leads to the depicted result.
\end{proof}

Double triangle reductions may transform irreducible graphs to reducible graphs. In this case we can use the product identity Eq.\ (\ref{7}) to further
simplify the graph. By the following proposition any sequence of reductions and product splits (splitting the graph on vertex sets which lead to vertex-connectivity 3,
Thm.\ \ref{thm1} and Fig.\ 4) lead to the same result, the `ancestor' of the graph.

\begin{prop}\label{prop4}
Double triangle reductions and product splits commute.
\end{prop}
\begin{proof}
The product identity only applies if 3 vertices $a$, $b$, $c$ connect two sub-graphs in the way depicted in Fig.\ 4.
Any double triangle has to be contained in left or in the right sub-graph (including the edges connected to $a$, $b$, and $c$).
It cannot be teared apart by the product split and it hence does not matter if one splits before or after the reduction.
\end{proof}

\begin{defn}\label{def5}
Let $\Gamma$ be a completed primitive graph. Any sequence of reductions and product splits (Thm.\ \ref{thm1}) terminates at a product
of double-triangle free irreducible graphs, the ancestor of $\Gamma$. We call $\Gamma$ a descendant of its ancestor.
The set of descendants of an ancestor is the family of the ancestor.
\end{defn}

\begin{rem}\label{rem4}
Conjecture \ref{con1} is equivalent to the statement that all periods of graphs in a family have the same maximum weight.
\end{rem}

A family has either one member only, a triangle-free (girth $\geq4$) ancestor (like $P_{6,4}$ in Table 4, Sect.\ \ref{sect3}), or it has infinite
cardinality. In the latter case the number of descendants is finite for every loop order $\ell$ and it grows with $\ell$.
For finite loop order $\ell$ the $P_3$-family is the largest. At $\ell=11$, out of 8687 irreducible graphs (see Table 1, Sect.\ \ref{sect3}) 1286 are
descendants of $P_3$, 920 are descendants of $P_3^2$, 132 are descendants of $P_3^3$, and 6 are descendants of $P_3^4$.
The other graphs have different ancestors.

Note that families may be linked by twist identities (Thm.\ \ref{thm2}) or by (extended, Rem.\ \ref{rem2}) Fourier identities (Thm.\ \ref{thm3}).
The first example is the weight 10 ancestor $P_{7,10}$ (see Table 4, Sect.\ \ref{sect3}) which is linked by a Fourier identity to $P_{7,5}$ with ancestor
$P_3^2$. Likewise weight 12 ancestors $P_{8,32}$ and $P_{8,34}$ are linked by a twist. Thus twist and Fourier identities group families to clusters of extended families
(all conjectured to have the same maximum weight). In general it is not easy to see if two families are linked by identities because this link may occur at high
loop order. For example the $P_{7,11}$-family is linked to the $P_{8,36}$-family via a Fourier transformation on descendants with loop order 9 (not included in Table 4).
The author is grateful to K. Yeats for providing this example.

Another way to conjecture the weight of yet unknown periods is by counting the number of zeros of the graph polynomial over finite fields and relies on
\'etale cohomology combined with the Lefschetz fixed-point formula (and on empirical data, see Remark 2.10 in \cite{QFTFq}).

\subsection{Integer multiple zeta values and the index}

Multiple zeta values span a $\ZZ$ module $\ZZ[MZV]$ (the integer MZVs) provided with a ring structure.
The set of integer MZVs is a lattice in the vector space $\QQ[MZV]$. All periods that have been identified up to date are found in this lattice.
For certain classes of graphs (`vertex-width $\leq3$' and all positive sign `Dodgsons') this is proved in \cite{BROW}.

For every $x\in\ZZ[MZV]$ there exists a maximum number $k$ such that $x/k$ is still an integer MZV.

\begin{defn}\label{def6}
For $x\in\ZZ[MZV]$ let the index of $x$ be the maximum $k\in\NN$ such that $x\in k\ZZ[MZV]$.
\end{defn}

For example, Open question 12.8 (10) in \cite{BROW} asks for the index of $\zeta(n)$. Thanks to the database \cite{MZDM} which provides a (conjecturally) complete
set of MZV-relations up to weight 22 we were able to answer this question for all $n\leq19$ (sufficient to loop order $\ell=11$).
\begin{eqnarray}\label{8c}
&n&\hbox{index }[\zeta(n)]\\
&2&1\nonumber\\
&3&1\nonumber\\
&4&4=2^2\nonumber\\
&5&2=2^1\nonumber\\
&6&48=2^43\nonumber\\
&7&16=2^4\nonumber\\
&8&576=2^63^2\nonumber\\
&9&144=2^43^2\nonumber\\
&10&3840=2^83\cdot5\nonumber\\
&11&768=2^83\nonumber\\
&12&6368256=2^{10}3^2691\nonumber\\
&13&1536=2^93\nonumber\\
&14&3870720=2^{12}3^35\cdot7\nonumber\\
&15&552960=2^{12}3^35\nonumber\\
&16&1600045056=2^{14}3^33617\nonumber\\
&17&55296=2^{11}3^3\nonumber\\
&18&1164321423360=2^{16}3^45\cdot43867\nonumber\\
&19&2949120=2^{16}3^25.\nonumber
\end{eqnarray}

The index of the identified $\phi^ 4$-periods can be found in Table 4, Sect.\ \ref{sect3}.

\subsection{Symmetric graphs}\label{sym}

We call a completed $\phi^ 4$-graph symmetric if all vertices are equal.

\begin{defn}\label{def7}
A completed $\phi^ 4$-graph is symmetric if its symmetry group acts transitively on the vertices.
\end{defn}

Amongst the symmetric graphs are the simplest as well as the most complicated graphs (according to their numerical and analytical accessibility).

Up to loop order 8 there are two types of symmetric graphs: The `cyclic' graphs and one cartesian product of cycles.

\begin{defn}\label{def8}
The cyclic graph $C^N_{m,n}$ with $1\leq m,n\leq N-1$ has vertices $1,2,\ldots,N$ and edges connecting $i$ with $i+m\mod N$ and with $i+n\mod N$.
\end{defn}

The graph $P_1$ (Fig.\ 3) is $C^3_{1,1}$ (or $C^3_{1,2}$) while $P_3$ (Fig.\ 3) is $C^5_{1,2}$.

\begin{lem}\label{lem3}
The cyclic graphs have the following properties
\begin{enumerate}
\item $C^N_{m,n}\sim C^N_{n,m}\sim C^N_{-m,n}\sim C^N_{am,an}$ for $a\in\ZZ_{/N\ZZ}^{\;\times}$ are isomorphic.
\item $C^N_{m,n}$ is 4-regular if $m,n\neq N/2$.
\item $C^N_{m,n}$ is simple if $m\neq\pm n\mod N$.
\item $C^N_{m,n}$ is connected if $\gcd(m,n,N)=1$.
\item $C^N_{m,n}$ is completed primitive (with loop order $\ell=N-2$) if $N=3$ or if it is 4-regular, simple, and connected.
\end{enumerate}
\end{lem}
\begin{proof}
Straight forward.
\end{proof}

The series $C^N_{1,2}$ is the completion of the zig-zag series introduced in \cite{BK}. They are descendants of $P_3$ (see Def.\ \ref{def5}).
Their periods are known to be integer MZVs \cite{BROW}. In fact, there exists a strikingly simple conjecture for their periods \cite{BK}.
\begin{con}\label{con2}
\begin{equation}\label{8d}
P_{C^{\ell+2}_{1,2}}=\frac{4(2\ell-2)!}{\ell!(\ell-1)!}\sum_{k=1}^\infty\frac{(-1)^{\ell(k-1)}}{k^{2\ell-3}}\in\QQ\zeta(2\ell-3)\cap\ZZ[MZV].
\end{equation}
\end{con}
The conjecture is proved for $\ell=3$ in \cite{C4}, $\ell=4$ in \cite{C5}, $\ell=5$ in \cite{K2}, and $\ell=6$ in \cite{U1}.

A second series of symmetric primitive graphs arises from cartesian products of cycles with the smallest member the $K_3$\raisebox{.7ex}{\fbox{}}$K_3$ graph $P_{7,10}$ which
is linked by a Fourier transform to $P_{7,5}$ (see Table 4, Sect.\ \ref{sect3}) and was determined by `exact numerical methods' in \cite{BK}.

\begin{con}\label{con3}
\begin{equation}\label{8e}
P_{K_3\raisebox{.5ex}{\fbox{}}K_3}=-189\zeta(3)\zeta(7)+450\zeta(5)^2.
\end{equation}
\end{con}

\section{Tables}\label{sect3}
This section presents a collection of explicit results. In Table 1 we list the number of completed primitive graphs (see Def.\ \ref{def1}) up to loop order $\ell=14$.
Next, we list the number of irreducible graphs (see Def.\ \ref{def2}), (an upper bound for) the number of different periods (see Def.\ \ref{dt2}),
the number of periods that were successfully determined, followed by the number of independent MZVs introduced by these periods (see Table 3).
The author used B.D. McKay's powerful {\it nauty} \cite{NAU} to generate the first column.
\vskip5mm

\begin{center}
\begin{tabular}{rrrrrc}
$\ell$&graphs&irreducible&periods&results&indep.~MZVs\\\hline
1&1&1&1&1&1\\
2&0&0&0&0&0\\
3&1&1&1&1&1\\
4&1&1&1&1&1\\
5&2&1&1&1&1\\
6&5&4&4&4&2\\
7&14&11&9&7&2\\
8&49&41&$\leq 31$&16&7\\
9&227&190&$\leq 136$&1&1\\
10&1~354&1~182&$\leq 846$&1&1\\
11&9~722&8~687&$\leq 6300$&1&1\\
12&81~305&74~204&?&1&1\\
13&755~643&700~242&?&1&1\\
14&7~635~677&7~160~643&?&1&1
\end{tabular}

Table 1: Completed primitive graphs and $\phi^4$-periods.
\end{center}

In Table 2 we summarize the results for reducible graphs in terms of $\ZZ$-linear combination of products of irreducible graphs which are listed
in Table 4. A term $nP_A P_B$ means there exist $n$ non-isomorphic reducible graphs that factorize into $P_A$ times $P_B$ by the product identity (Thm.\ \ref{thm1}).
The sum of the coefficients equals the number of reducible graphs (column 3 minus column 2 in Table 1).
The table does not include symmetry factors (see Table 4) and hence it cannot be interpreted as a contribution to e.g.\ a physical beta-function.
\vskip5mm

\begin{center}
\begin{tabular}{r|l}
$\ell$&sum of reducible graphs\\\hline
5&$P_3^2$\\\hline
6&$P_3P_4$\\\hline
7&$P_3^3+P_3P_5+P_4^2$\\\hline
8&$3P_3^2P_4+P_3(P_{6,1}+2P_{6,2}+P_{6,3})+P_4P_5$\\\hline
9&$2P_3^4+4P_3^2P_5+P_3(3P_4^2+P_{7,1}+4P_{7,2}+3P_{7,3}+3P_{7,4}+P_{7,5}+3P_{7,6}$\\
&$+\,2P_{7,7}+P_{7,8}+2P_{7,9}+P_{7,10}+P_{7,11})+P_4(P_{6,1}+2P_{6,2}+P_{6,3})+2P_5^2$\\\hline
10&\begin{minipage}[t]{14cm}
$6P_3^3P_4+P_3^2(5P_{6,1}+10P_{6,2}+3P_{6,3})$\\
$+\,P_3(7P_4P_5+P_{8,1}+5P_{8,2}+2P_{8,3}+7P_{8,4}+3P_{8,5}+2P_{8,6}+3P_{8,7}+3P_{8,8}+4P_{8,9}$\\
$+\,4P_{8,10}+2P_{8,11}+3P_{8,12}+6P_{8,13}+6P_{8,14}+4P_{8,15}+P_{8,16}+2P_{8,17}+4P_{8,18}$\\
$+\,P_{8,19}+5P_{8,20}+3P_{8,21}+3P_{8,22}+3P_{8,23}+P_{8,24}+4P_{8,25}+2P_{8,26}+3P_{8,27}$\\
$+\,2P_{8,28}+4P_{8,29}+2P_{8,30}+3P_{8,31}+P_{8,32}+2P_{8,33}+P_{8,34}+P_{8,35}+P_{8,36}+2P_{8,37}$\\
$+\,P_{8,38}+P_{8,39})+2P_4^3+P_4(P_{7,1}+4P_{7,2}+3P_{7,3}+3P_{7,4}+P_{7,5}+3P_{7,6}$\\
$+\,2P_{7,7}+P_{7,8}+2P_{7,9}+P_{7,10}+P_{7,11})+P_5(2P_{6,1}+5P_{6,2}+2P_{6,3})$
\end{minipage}
\end{tabular}

Table 2: Results for reducible graphs using the product identity. Note that $P_{6,4}$ is absent because it has no triangle.
\end{center}

Table 3 contains a list of $\phi^4$-transcendentals (except for $Q_0=1$) needed to read Table 4.
The appearance of the `knot numbers' (see Table 3a for their conversion into MZVs)
\begin{equation}\label{9b}
N_{a,b}=\sum_{j>k\geq1}\left(\frac{(-1)^j}{j^ak^b}-\frac{(-1)^j}{j^bk^a}\right)
\end{equation}
introduced by D.J. Broadhurst in \cite{B3} indicates that there might be something like a canonical base for $\phi^4$-periods.
A link to A. Goncharov's coproduct on MZVs \cite{GON} seems possible.
Except for using the $N_{a,b}$s the author did not make an attempt to find a more canonical choice than the shortest possible
presented in the table.
\vskip5mm

\begin{center}
\begin{tabular}{rr|ll}
$\ell$&$\!$wt&number&value\\\hline
1&0&$Q_0=1$&1\\\hline
3&3&$Q_3=\zeta(3)$&1.202~056~903~159\\\hline
4&5&$Q_5=\zeta(5)$&1.036~927~755~143\\\hline
5&7&$Q_7=\zeta(7)$&1.008~349~277~381\\\hline
6&8&$Q_8=N_{3,5}$&0.070~183~206~556\\
&9&$Q_9=\zeta(9)$&1.002~008~392~826\\\hline
7&10&$Q_{10}=N_{3,7}$&0.090~897~338~299\\
&11&$Q_{11,1}=\zeta(11)$&1.000~494~188~604\\
&&$Q_{11,2}=-\zeta(3,5,3)\!+\!\zeta(3)\zeta(5,3)$&0.042~696~696~025\\\hline
8&12&$Q_{12,1}=N_{3,9}$&0.096~506~102~637\\
&&$Q_{12,2}=N_{5,7}$&0.002~046~054~793\\
&&$Q_{12,3}=\pi^{12}/10!$&0.254~703~808~841\\
&13&$Q_{13,1}=\zeta(13)$&1.000~122~713~347\\
&&$Q_{13,2}=-\zeta(5,3,5)\!+\!11\zeta(5)\zeta(5,3)\!+\!5\zeta(5)\zeta(8)$&5.635~097~688~692\\
&&$Q_{13,3}=-\zeta(3,7,3)\!+\!\zeta(3)\zeta(7,3)\!+\!12\zeta(5)\zeta(5,3)\!+\!6\zeta(5)\zeta(8)\hspace*{-5pt}$&6.725~631~947~085
\end{tabular}

Table 3: (Possibly incomplete) list of $\phi^4$-transcendentals up to loop order 8.
\end{center}

\begin{center}
\begin{tabular}{rr|l}
$\ell$&$\!$wt&base\\[1ex]\hline
6&8&$N_{3,5}=\frac{27}{80}\zeta(5,3)+\frac{45}{64}\zeta(5)\zeta(3)-\frac{261}{320}\zeta(8)$\\[1ex]
7&10&$N_{3,7}=\frac{423}{3584}\zeta(7,3)+\frac{189}{256}\zeta(7)\zeta(3)+\frac{639}{3584}\zeta(5)^2-\frac{7137}{7168}\zeta(10)$\\[1ex]
8&12&$N_{3,9}=\frac{27}{512}\zeta(4,4,2,2)+\frac{55}{1024}\zeta(9,3)+\frac{231}{256}\zeta(9)\zeta(3)+\frac{447}{256}\zeta(7)\zeta(5)-\frac{9}{512}\zeta(3)^4$\\[1ex]
&&\hspace{13mm}$-\frac{27}{448}\zeta(7,3)\zeta(2)-\frac{189}{128}\zeta(7)\zeta(3)\zeta(2)-\frac{1269}{1792}\zeta(5)^2\zeta(2)+\frac{189}{512}\zeta(5,3)\zeta(4)$\\[1ex]
&&\hspace{13mm}$+\frac{945}{512}\zeta(5)\zeta(3)\zeta(4)+\frac{9}{64}\zeta(3)^2\zeta(6)-\frac{7322453}{5660672}\zeta(12)$\\[1ex]
&&$N_{5,7}=-\frac{81}{512}\zeta(4,4,2,2)+\frac{19}{1024}\zeta(9,3)-\frac{477}{1024}\zeta(9)\zeta(3)-\frac{4449}{1024}\zeta(7)\zeta(5)+\frac{27}{512}\zeta(3)^4$\\[1ex]
&&\hspace{13mm}$+\frac{81}{448}\zeta(7,3)\zeta(2)+\frac{567}{128}\zeta(7)\zeta(3)\zeta(2)+\frac{3807}{1792}\zeta(5)^2\zeta(2)-\frac{567}{512}\zeta(5,3)\zeta(4)$\\[1ex]
&&\hspace{13mm}$-\frac{2835}{512}\zeta(5)\zeta(3)\zeta(4)-\frac{27}{64}\zeta(3)^2\zeta(6)+\frac{3155095}{5660672}\zeta(12)$\\[1ex]
&&$\pi^{12}/10!=\frac{45045}{176896}\zeta(12)$
\end{tabular}

Table 3a: Conversion or the $N_{a,b}$s (and $\pi^{12}/10!$) into MZVs.
\end{center}

To obtain a complete $\QQ$-base of weight $n$ one has to include all products of $\phi^4$-trans\-cen\-den\-tals with total weight $n$.
Note that the $\QQ$-dimension of the base (including products) is much smaller than the number
of $\QQ$-independent MZVs. The sufficiency of the base has thus predictive power beyond the fact that the known periods are (integer) MZVs.

Because of the absence of lower powers of $\pi$ the appearance of $\pi^{12}/10!$ as $Q_{12,3}$ was not expected by the author.
Notice that at weight 12 we also have the first `push-down' (an MZV that reduces to lower depth Euler sums) \cite{B3}, \cite{MZDM} and
the first `exceptional' relation between odd double sums [between $\zeta(9,3)$ and $\zeta(7,5)$] which was shown in \cite{GAN} to be connected
to the existence of the weight 12 cusp modular form.

Table 4 is the census. We list all 60 irreducible completed primitive graphs of $\phi^4$-theory up to loop order $\ell=8$.

Each row in the table contains the name and a plot of the graph, the first digits of its numerical value (if available),
the size of its automorphism group (due to B.D. McKay's {\it nauty}
\cite{NAU}), its index (if available; see Def.\ \ref{def6}) and ancestor (see Def.\ \ref{def5}), remarks,
the (conjectured) weight, and the exact value (if available).

Except for loop order, the periods are not ordered in a particular way. The name $P_{\ell,\, \#}$ is indexed by the loop order and a number
that represents the order in which it was produced by the generating program.

Analytic results are available for $P_3$ and $P_4$ where simple Gegenbauer techniques suffice \cite{C4}, \cite{C5}, see Ex.\ \ref{ex3}.
Moreover, the zig-zag periods $P_5$ and $P_{6,1}$ have been calculated using the uniqueness relation in \cite{K2} and \cite{U1},
respectively. The only {\it multiple} zeta period that has been calculated is $P_{6,4}$ in \cite{S3}.
All other periods have been determined by a method developed in \cite{BK}: Expand the propagators into Gegenbauer polynomials, evaluate the integrals,
simplify the result, convert the multiple sum into a sequence by introducing some kind of `cutoff' $\Lambda$, accelerate convergence by fitting a power series
in negative exponents of $\Lambda$. It turned out to be very useful to include logarithmic terms in the series up to a certain power leading
to terms of the form $a_{j,k}\Lambda^{-j}\ln^k(\Lambda)$. The desired approximation is then recovered as $a_{0,0}$ (see \cite{B3}).
With a high precision result for the period we use PSLQ to search for a $\QQ$-linear combination of MZVs that reproduces the number.

The method is quite efficient if the expansion into Gegenbauer polynomials does not lead to `multi-$j$-symbols'. We used twist and (extended) Fourier identities
(see Thms.\ \ref{thm2}, \ref{thm3}, and Remark \ref{rem2}) trying to convert the original graph into a (possibly non-$\phi^4$) graph
that delivers a multi-$j$-symbol free expansion. Whenever this was possible we found a highly trustworthy MZV-fit for the period.

The remarks include the number of significant figures (sf) that are available. Note that we needed very high precision results only for some
periods of a given weight to determine the sub-base of $\phi^4$-periods in $\QQ[MZV]$.
Further, we included bibliographic references [Lit], alternative names, and known identities for the graph into the remarks.

A code for quick access to the graph together with a list of all 190 completed primitive irreducible graphs at loop order 9 is included
in the first version of this paper \cite{CEN1}.

Most of the numerical calculations were performed on the Erlanger RRZE cluster.

\begin{tabular}{lllllll}
name&graph&numerical value&$|$Aut$|$&index&ancestor&rem, (sf), [Lit]\\[1ex]
\multicolumn{2}{l}{weight}&\multicolumn{5}{l}{exact value}\\[1ex]\hline\hline
$P_1$&\hspace*{-2mm}\raisebox{-9mm}{\includegraphics[width=12mm]{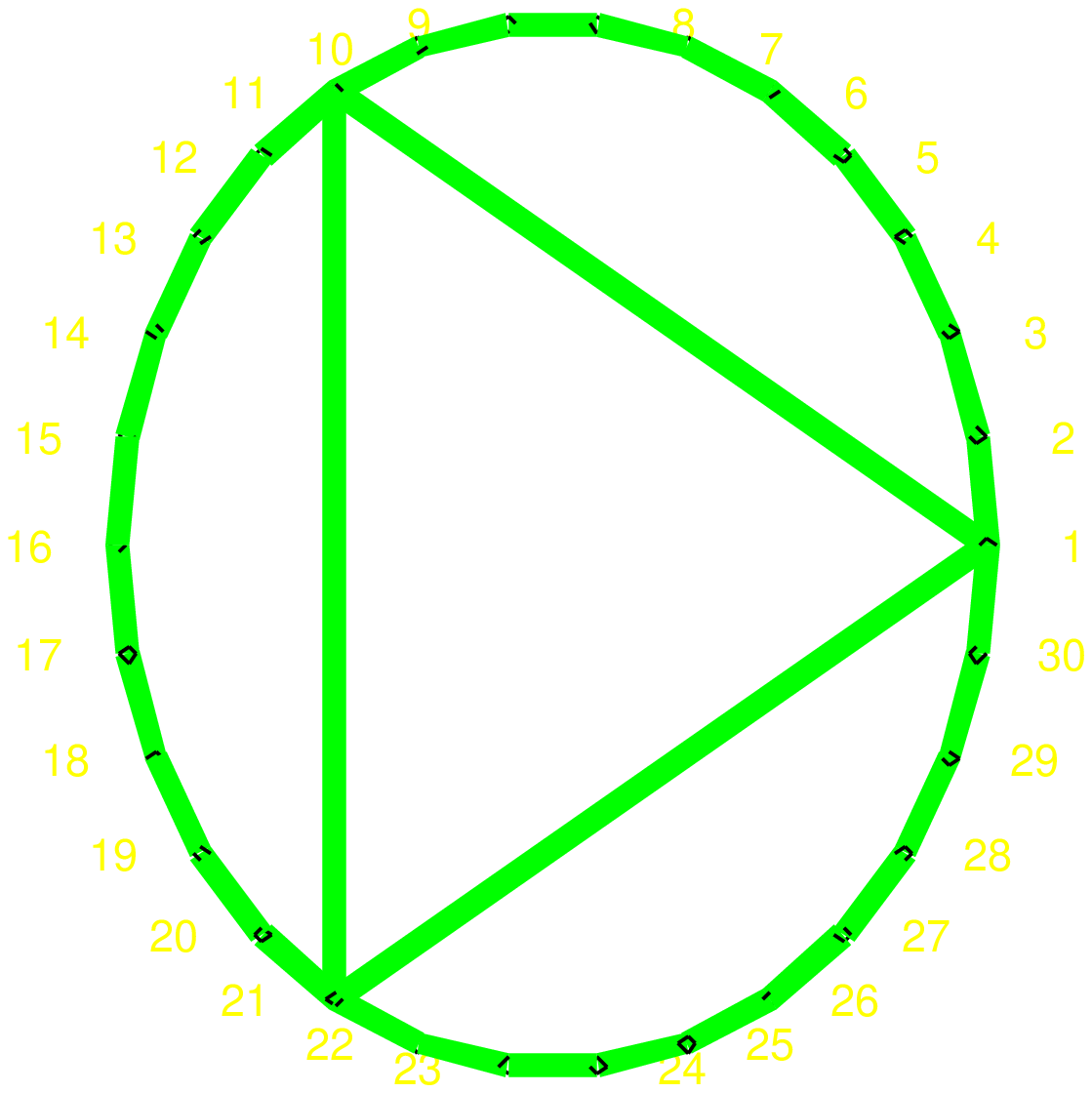}}&1&48&---&$P_1$&$C^3_{1,1}$\\[-6mm]
0&&\multicolumn{5}{l}{1}\\[1ex]\hline\hline
$P_3$&\hspace*{-2mm}\raisebox{-9mm}{\includegraphics[width=12mm]{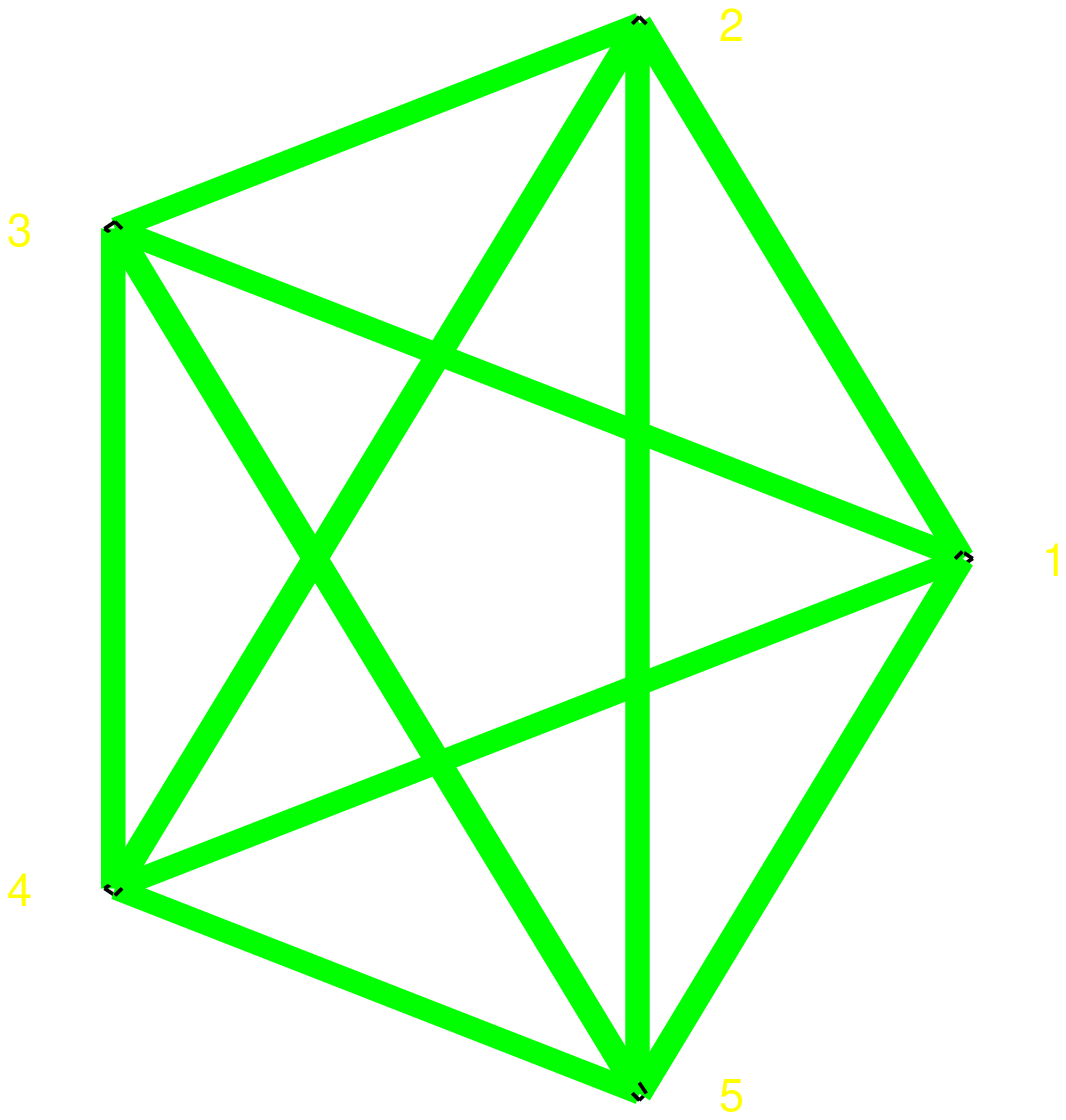}}&7.212~341~418&120&6&$P_3$&$C^5_{1,2}$, $K_5$, \cite{C4}\\[-6mm]
3&&\multicolumn{5}{l}{$6Q_3$}\\[1ex]\hline\hline
$P_4$&\hspace*{-2mm}\raisebox{-9mm}{\includegraphics[width=12mm]{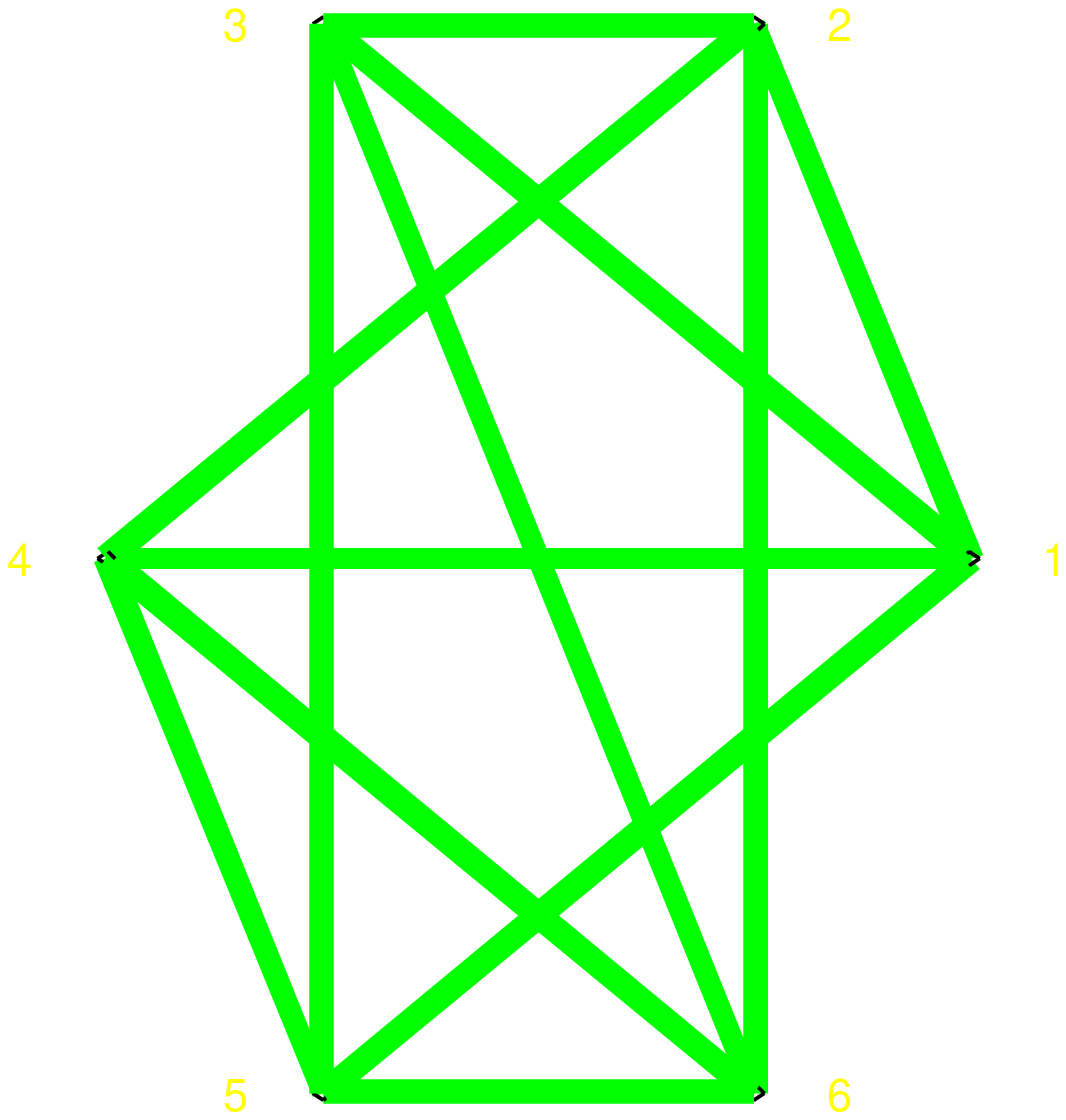}}&20.738~555~102&48&40&$P_3$&$C^6_{1,2}$, $O_3$, \cite{C5}\\[-6mm]
5&&\multicolumn{5}{l}{$20Q_5$}\\[1ex]\hline\hline
$P_5$&\hspace*{-2mm}\raisebox{-9mm}{\includegraphics[width=12mm]{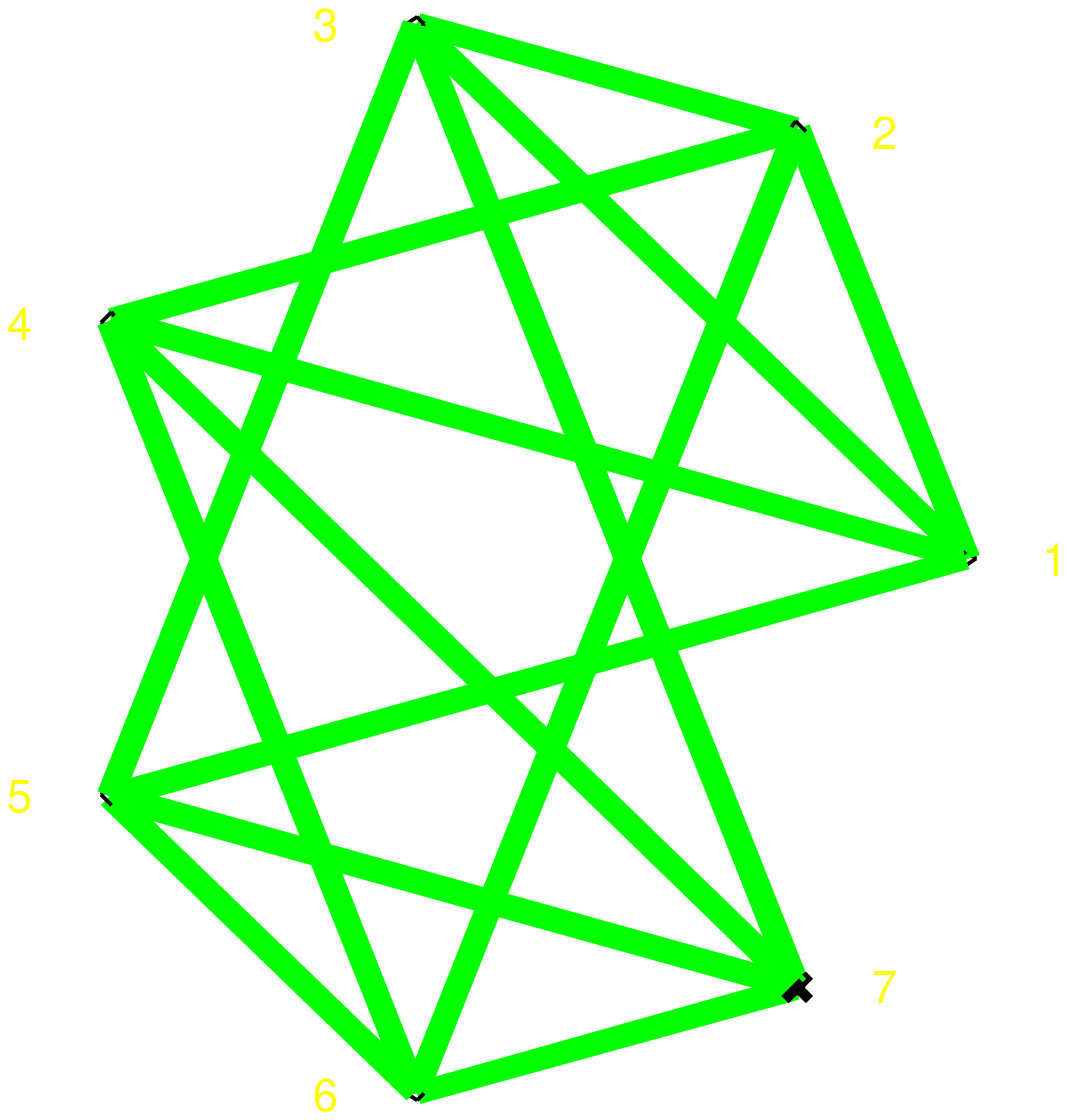}}&55.585~253~915&14&882&$P_3$&$C^7_{1,2}$, $\overline{C_7}$, \cite{K2}\\[-6mm]
7&&\multicolumn{5}{l}{$\frac{441}{8}Q_7$}\\[1ex]\hline\hline
$P_{6,1}$&\hspace*{-2mm}\raisebox{-9mm}{\includegraphics[width=12mm]{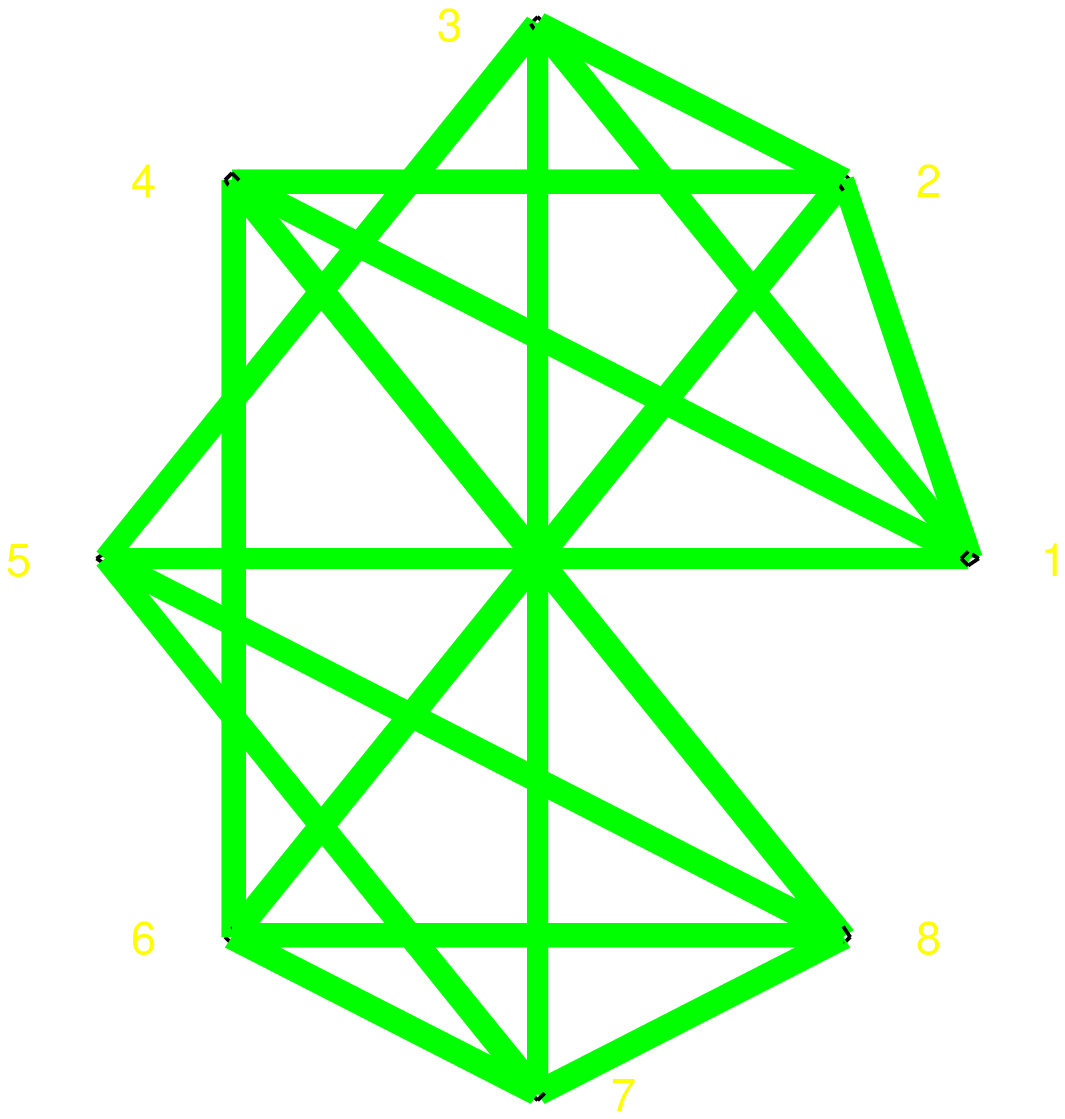}}&168.337~409~994&16&24192&$P_3$&$C^8_{1,2}$ \cite{U1}\\[-6mm]
9&&\multicolumn{5}{l}{$168Q_9$}\\[1ex]\hline
$P_{6,2}$&\hspace*{-2mm}\raisebox{-9mm}{\includegraphics[width=12mm]{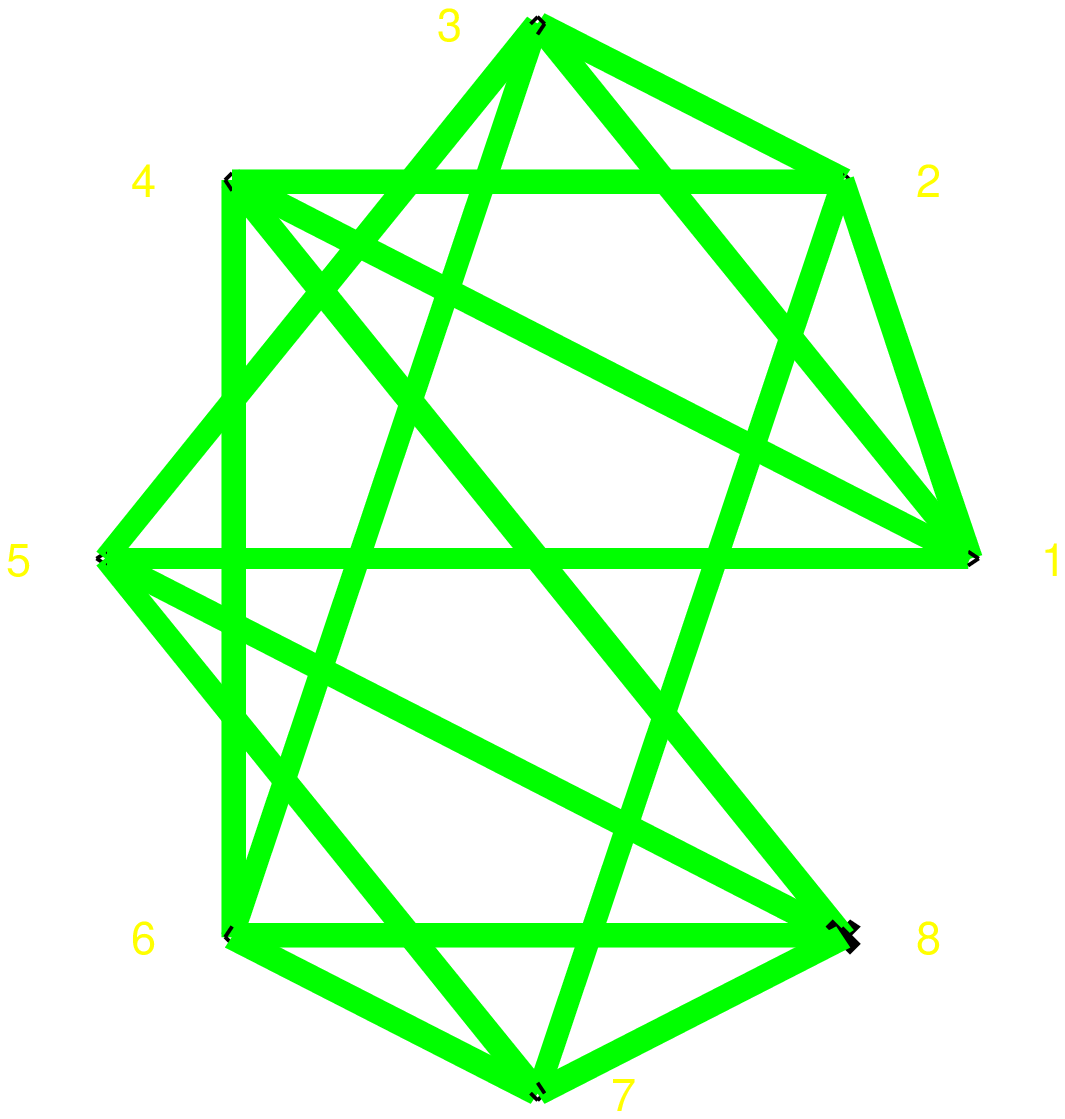}}&132.243~533~110&4&16&$P_3$&\cite{BK}\\[-6mm]
9&&\multicolumn{5}{l}{$\frac{1063}{9}Q_9+8Q_3^3$}\\[1ex]\hline
$P_{6,3}$&\hspace*{-2mm}\raisebox{-9mm}{\includegraphics[width=12mm]{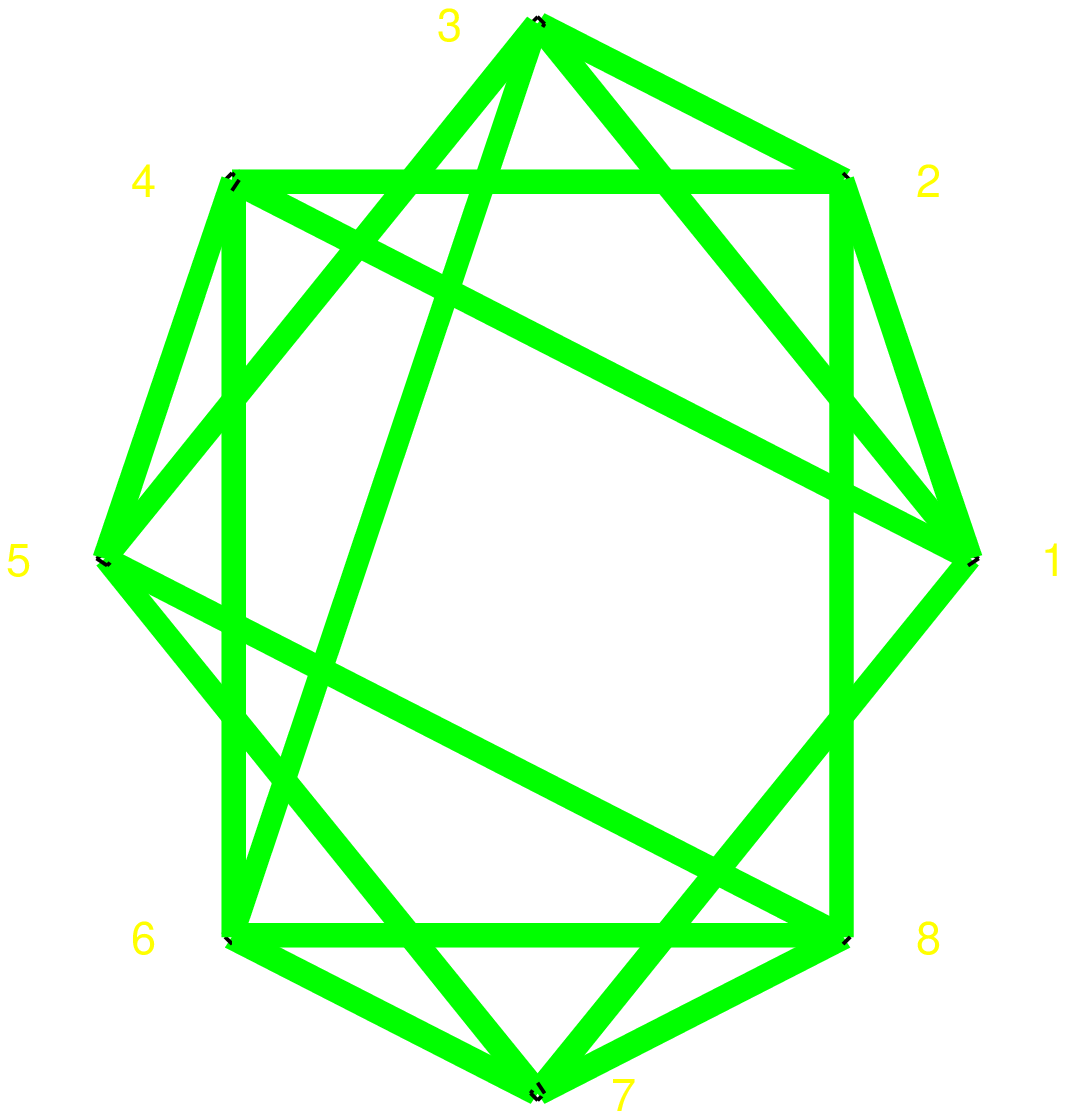}}&107.711~024~841&16&72&$P_3^2$&\cite{BK}\\[-6mm]
8&&\multicolumn{5}{l}{$256Q_8+72Q_3Q_5$}\\[1ex]\hline
$P_{6,4}$&\hspace*{-2mm}\raisebox{-9mm}{\includegraphics[width=12mm]{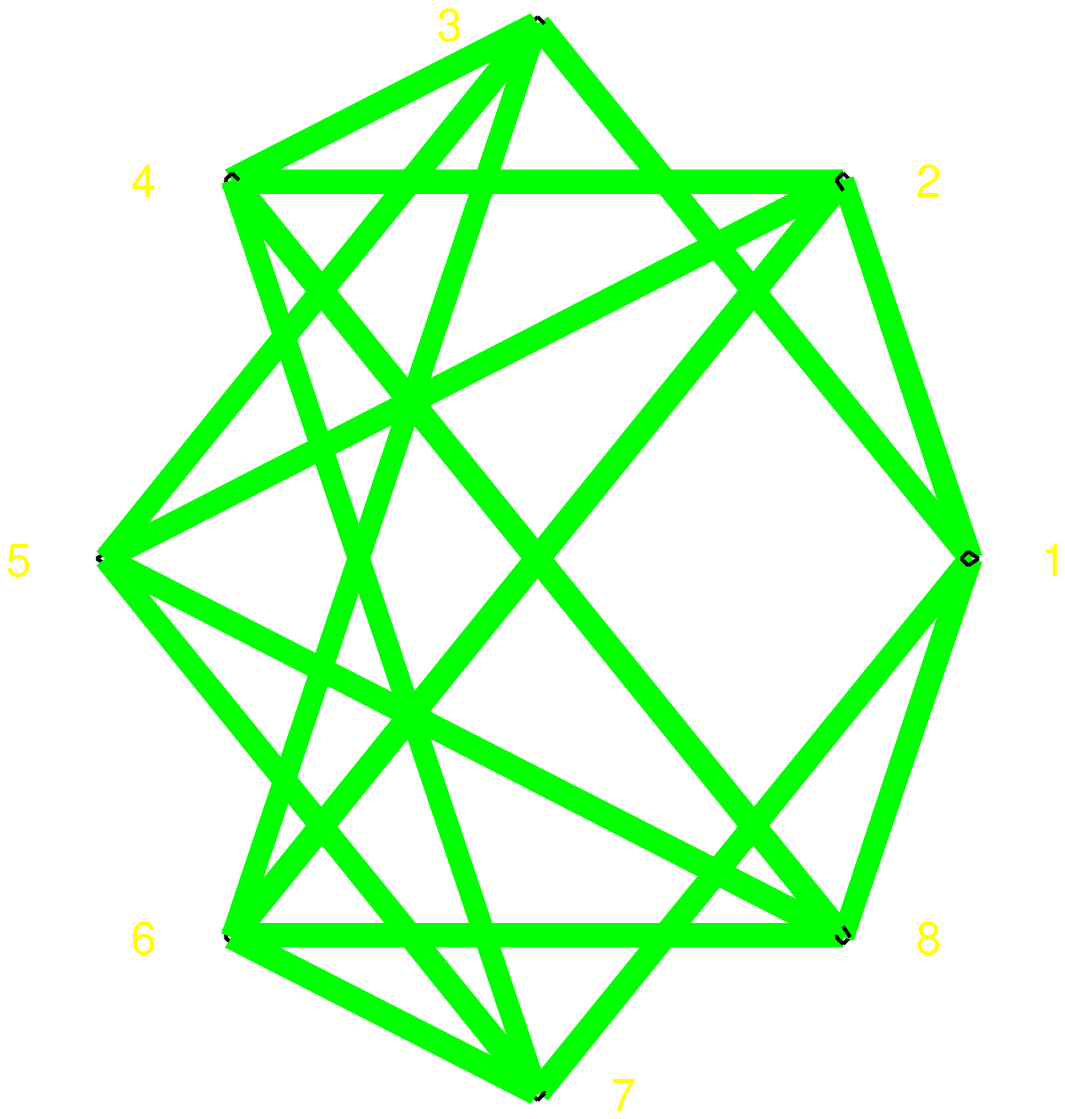}}&71.506~081~796&1152&1728&$P_{6,4}$&$C^8_{1,3}$ \cite{BK}, \cite{S3}\\[-6mm]
8&&\multicolumn{5}{l}{$-4096Q_8+288Q_3Q_5$}\\[1ex]\hline\hline
$P_{7,1}$&\hspace*{-2mm}\raisebox{-9mm}{\includegraphics[width=12mm]{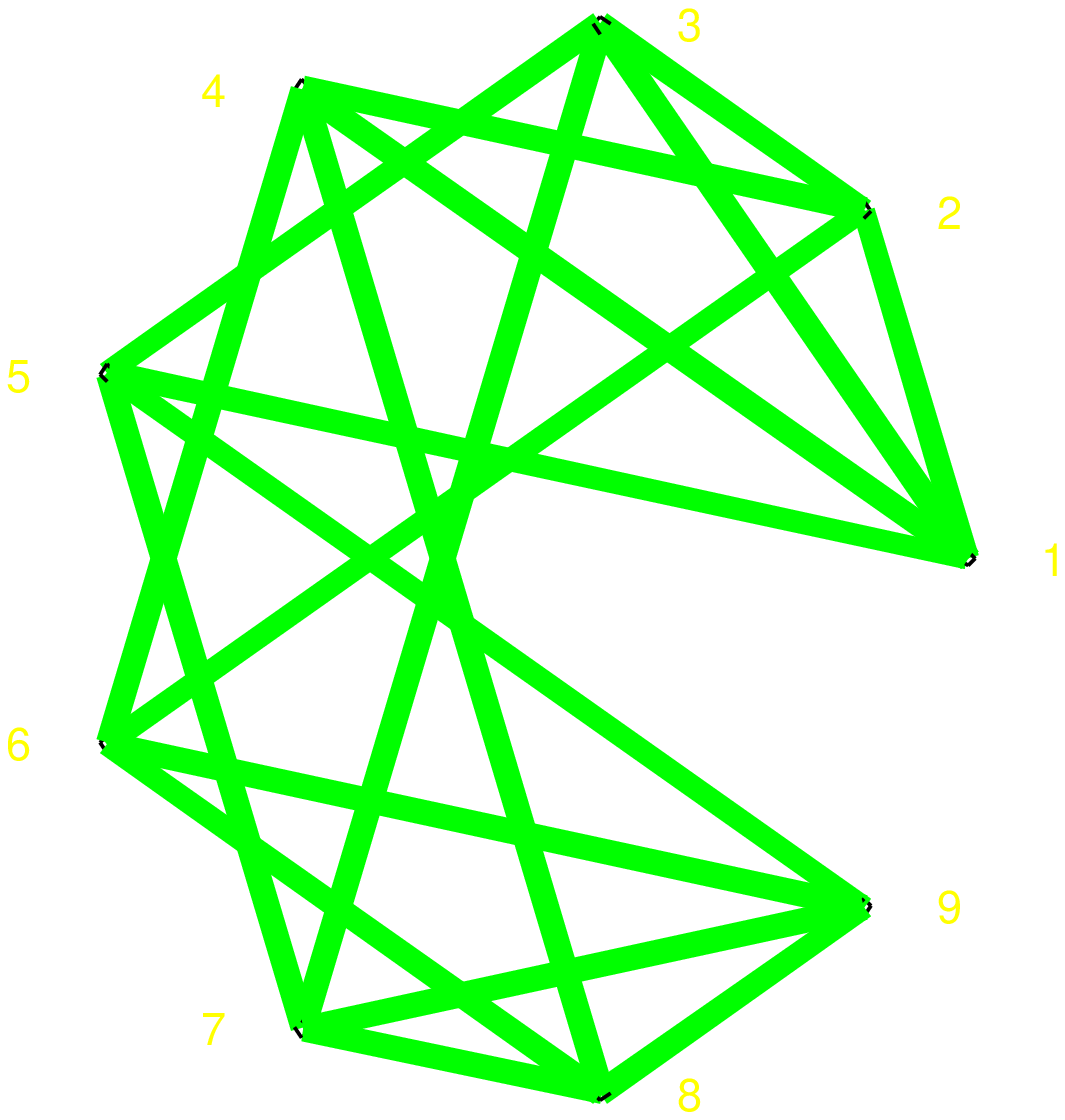}}&527.745~051~766&18&405108&$P_3$&$C^9_{1,2}$ \cite{BK}\\[-6mm]
11&&\multicolumn{5}{l}{$\frac{33759}{64}Q_{11,1}$}\\[1ex]\hline
$P_{7,2}$&\hspace*{-2mm}\raisebox{-9mm}{\includegraphics[width=12mm]{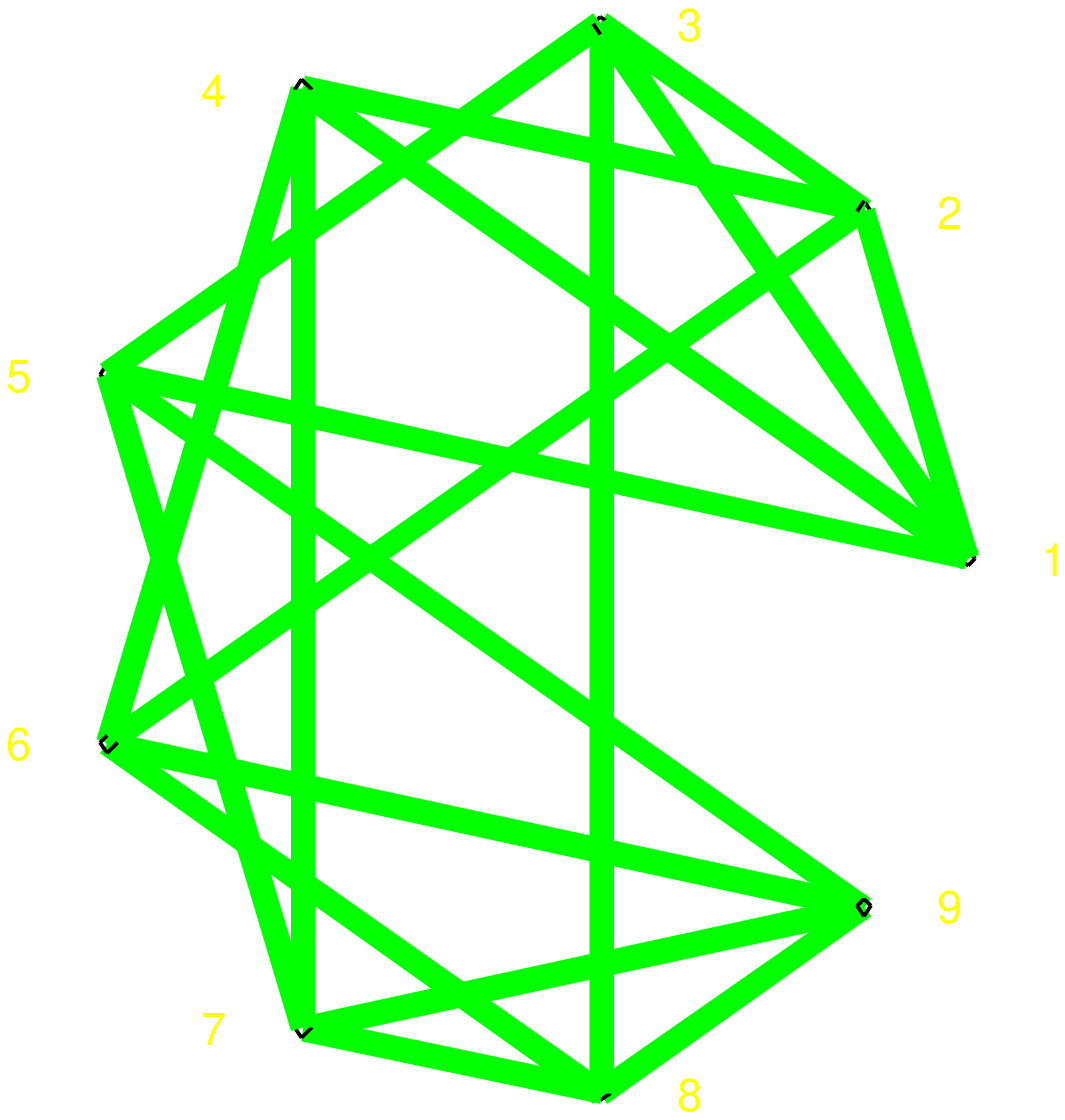}}&380.887~829~534&2&20&$P_3$&(56), \cite{BK}\\[-6mm]
11&&\multicolumn{5}{l}{$\frac{62957}{192}Q_{11,1}+9Q_{11,2}+35Q_3^2Q_5$}\\[1ex]\hline
$P_{7,3}$&\hspace*{-2mm}\raisebox{-9mm}{\includegraphics[width=12mm]{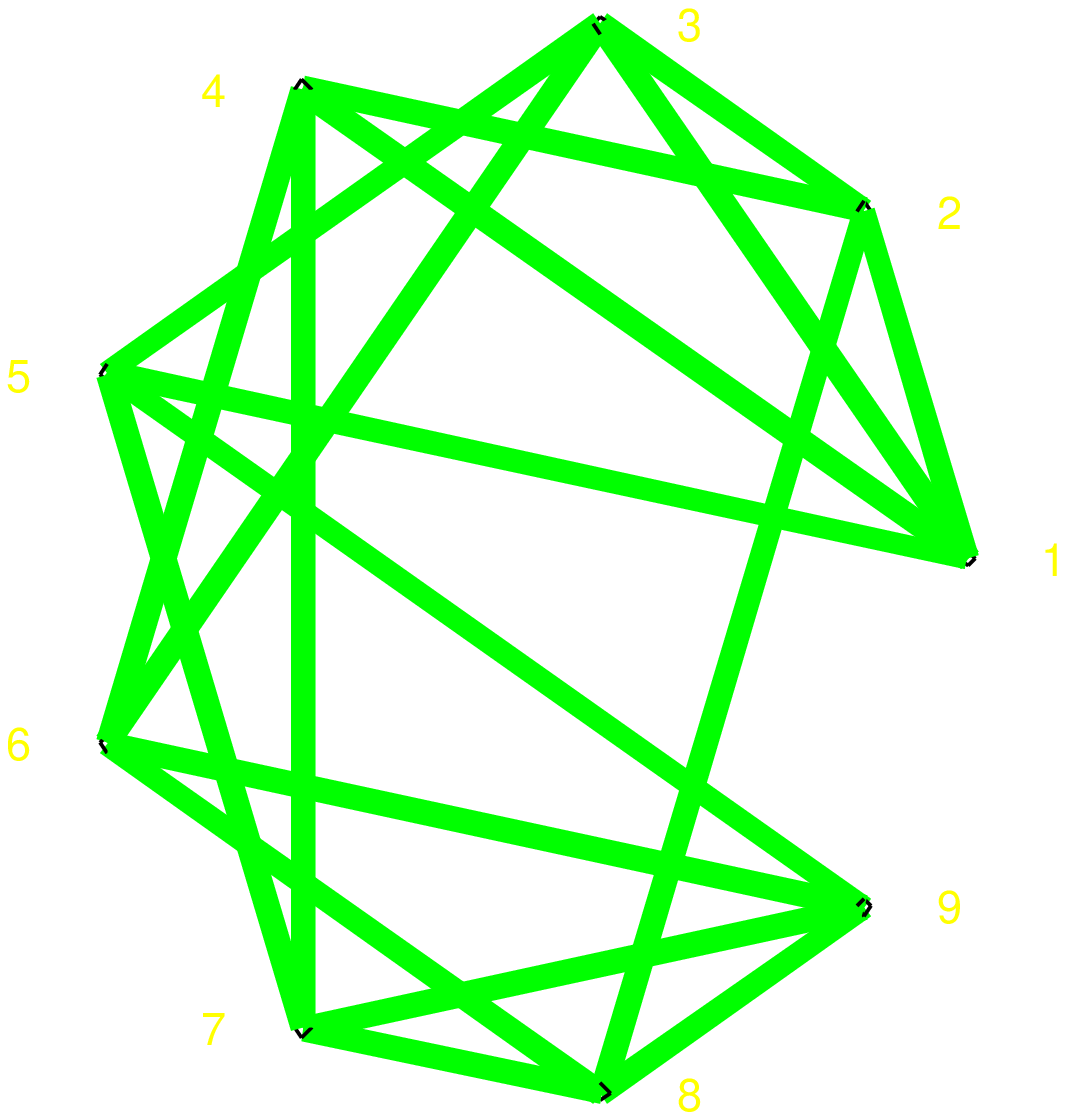}}&336.067~072~110&2&16&$P_3$&(56), \cite{BK}\\[-6mm]
11&&\multicolumn{5}{l}{$\frac{73133}{240}Q_{11,1}+\frac{144}{5}Q_{11,2}+20Q_3^2Q_5$}\\[1ex]\hline
$P_{7,4}$&\hspace*{-2mm}\raisebox{-9mm}{\includegraphics[width=12mm]{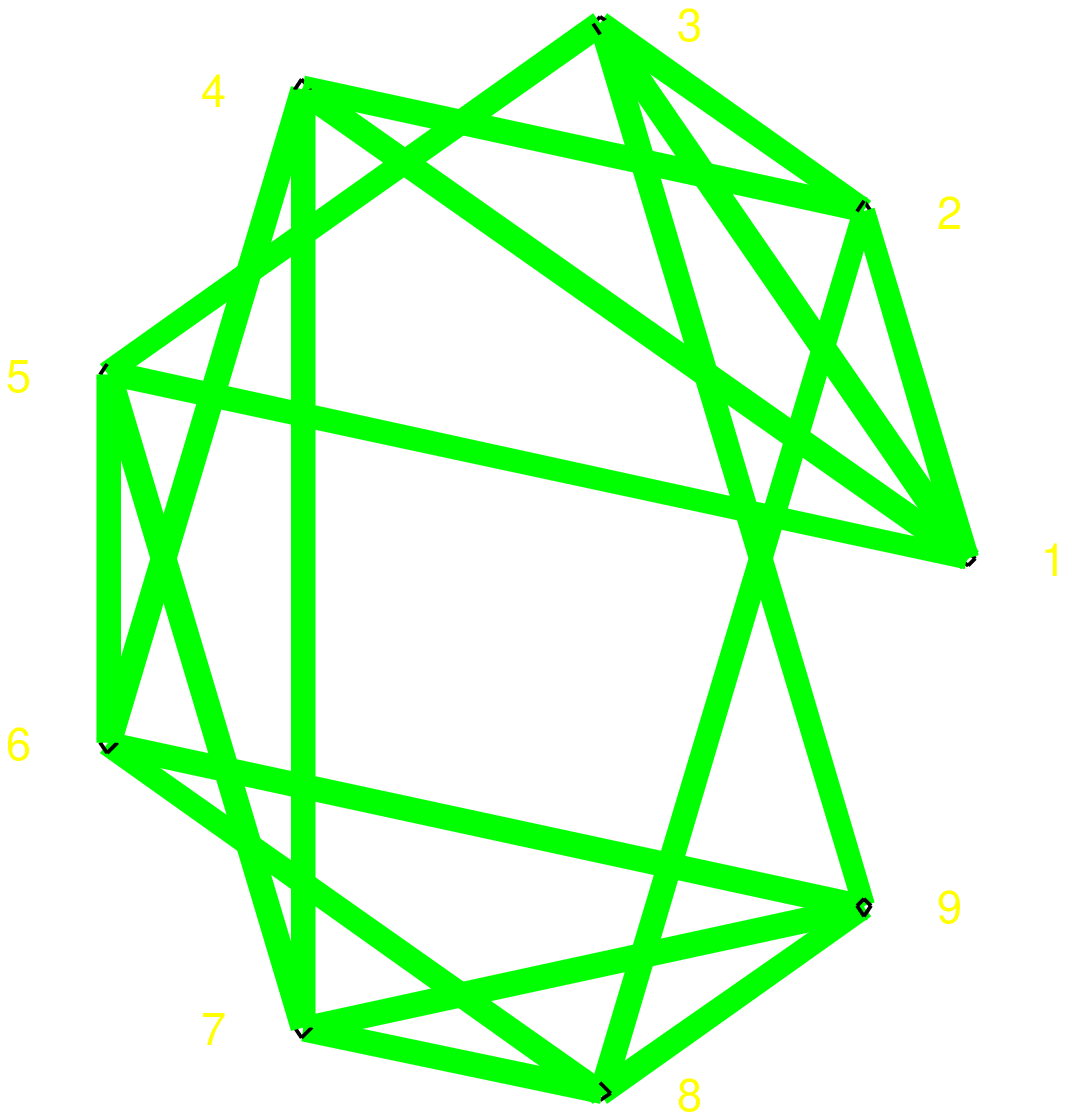}}&294.035~314~185&4&320&$P_3^2$&(23), \cite{BK}\\[-6mm]
10&&\multicolumn{5}{l}{$420Q_3Q_7-200Q_5^2$}\\[1ex]\hline
$P_{7,5}$&\hspace*{-2mm}\raisebox{-9mm}{\includegraphics[width=12mm]{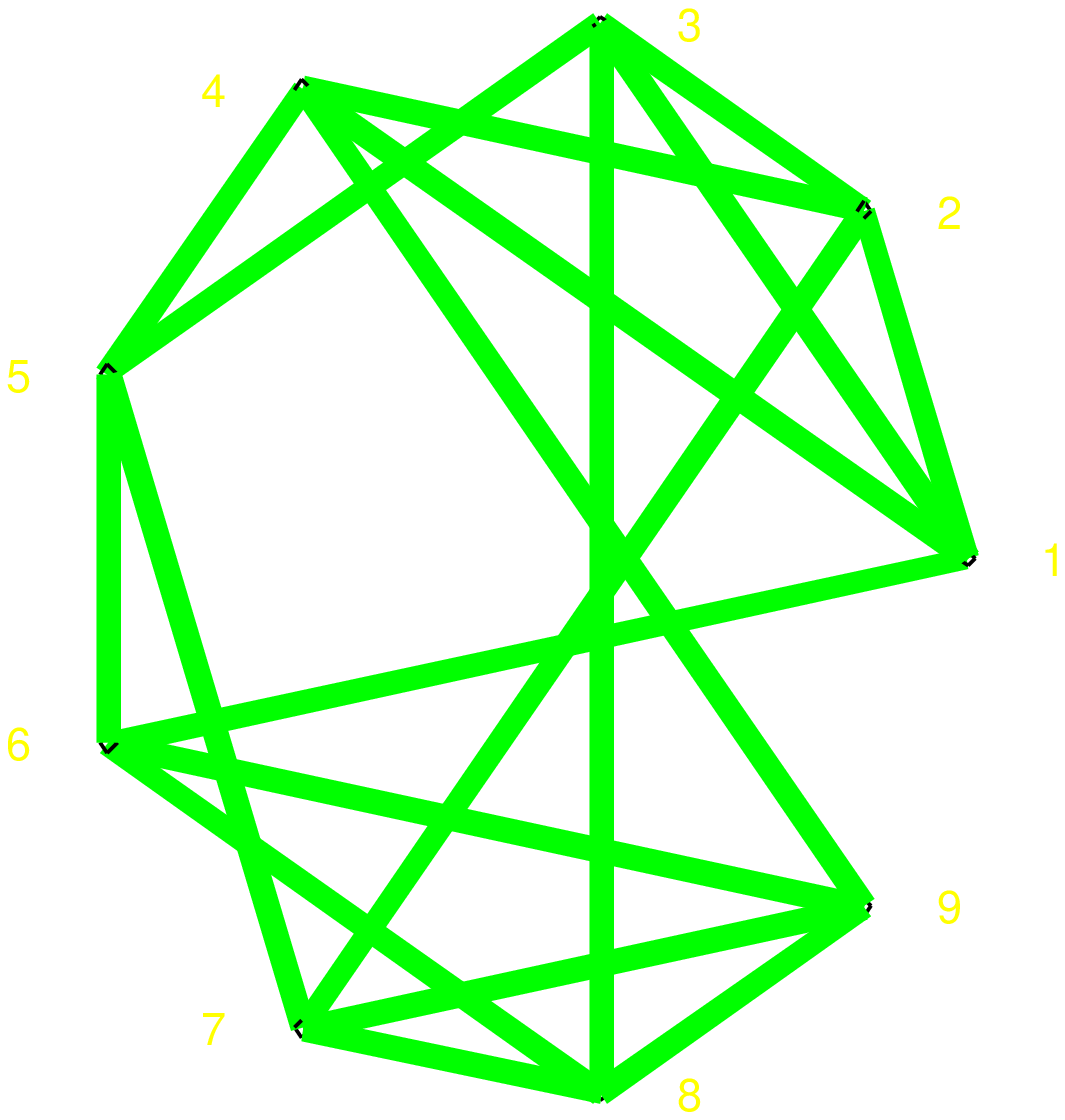}}&254.763~009~595&8&144&$P_3^2$&(23), \cite{BK}\\[-6mm]
10&&\multicolumn{5}{l}{$-189Q_3Q_7+450Q_5^2$}\\[1ex]\hline
$P_{7,6}$&\hspace*{-2mm}\raisebox{-9mm}{\includegraphics[width=12mm]{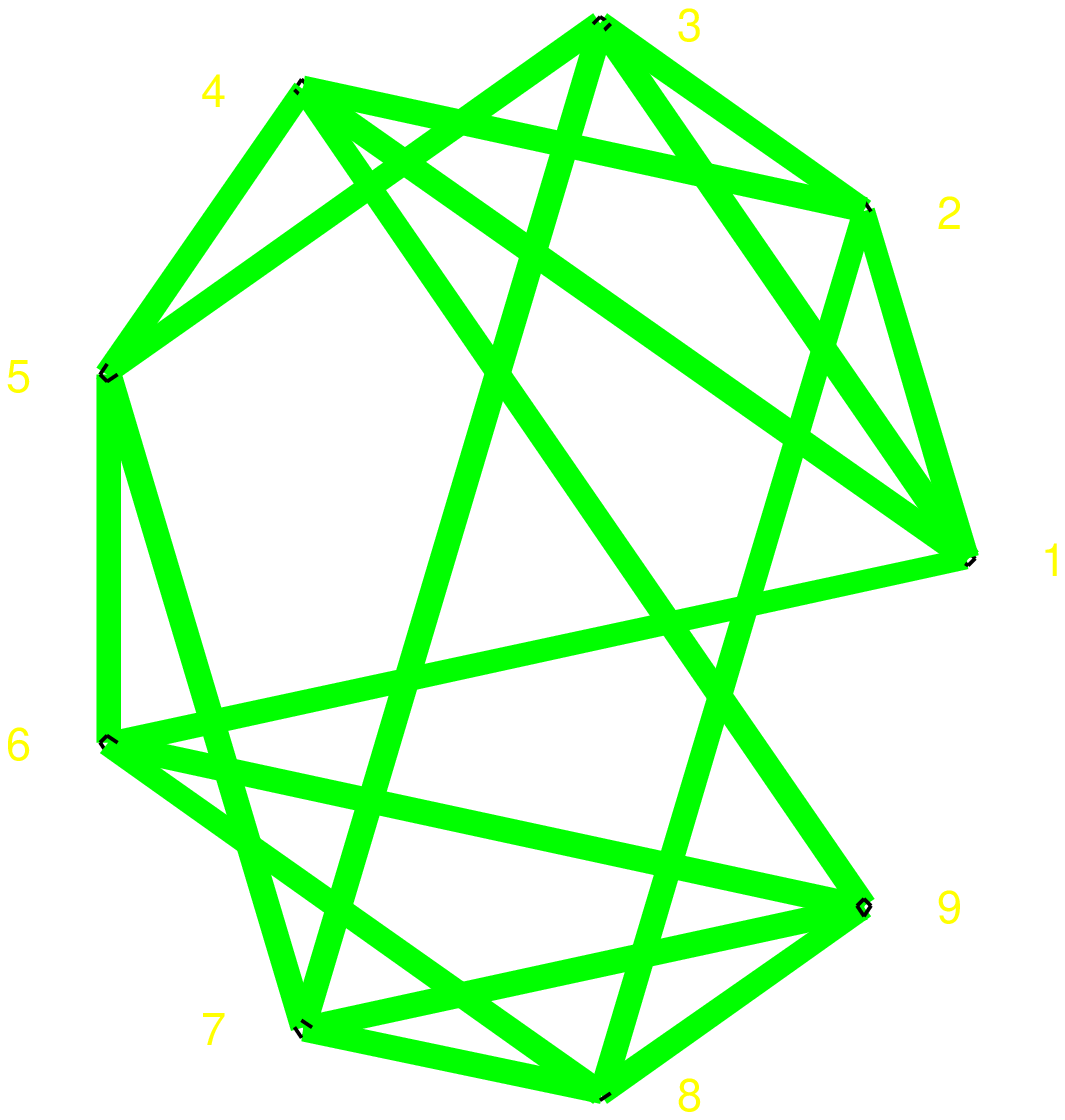}}&273.482~574~258&2&20&$P_3$&(34), \cite{BK}\\[-6mm]
11&&\multicolumn{5}{l}{$\frac{14279}{64}Q_{11,1}-51Q_{11,2}+35Q_3^2Q_5$}\\[1ex]\hline
$P_{7,7}$&\hspace*{-2mm}\raisebox{-9mm}{\includegraphics[width=12mm]{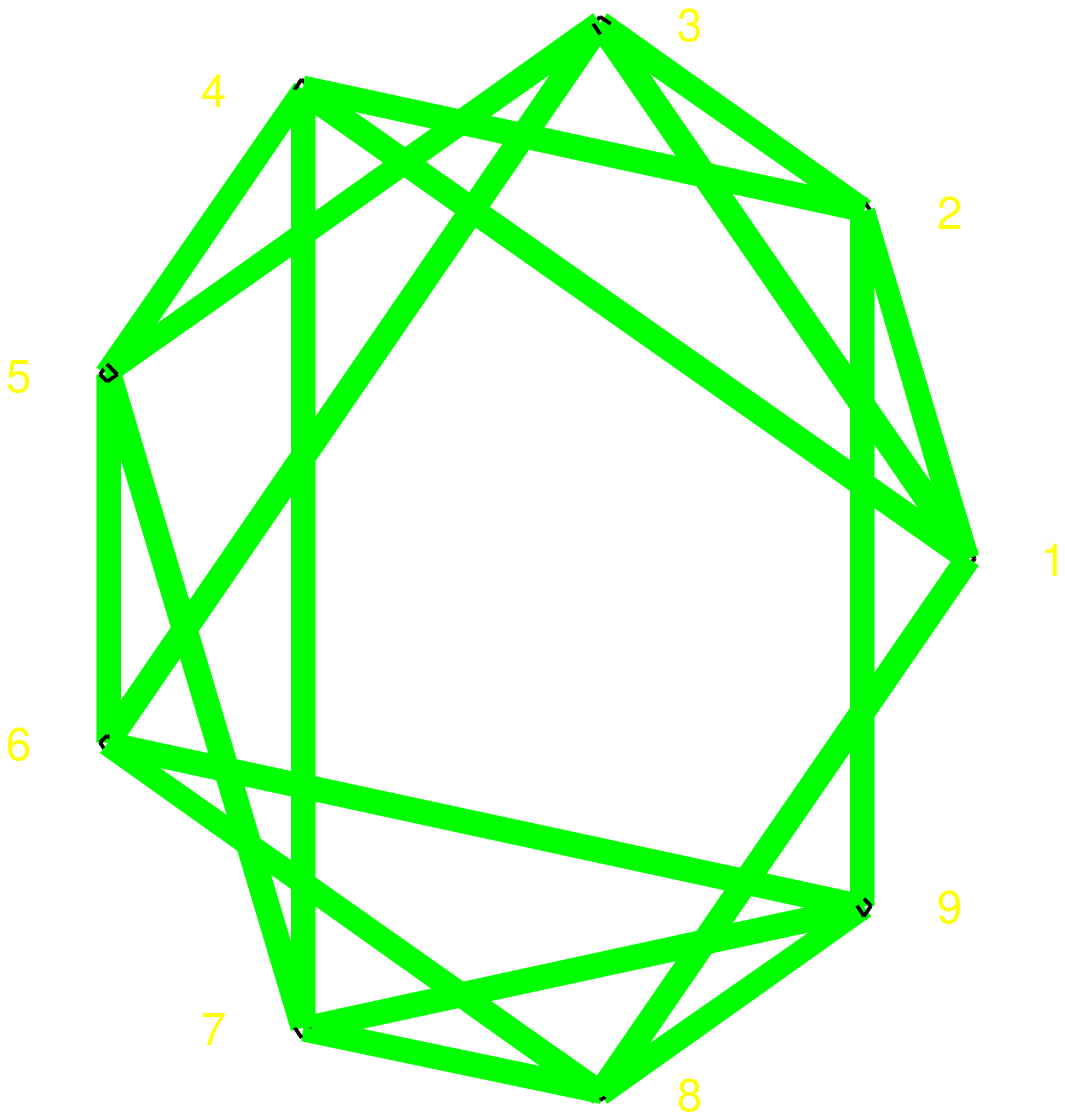}}&294.035~314~185&8&320&$P_3^2$&Fourier, twist\\[-6mm]
10&&\multicolumn{5}{l}{$P_{7,4}$\hspace*{91mm}}
\end{tabular}

\begin{tabular}{lllllll}
name&graph&numerical value&$|$Aut$|$&index&ancestor&rem, (sf), [Lit]\\[1ex]
\multicolumn{2}{l}{weight}&\multicolumn{5}{l}{exact value}\\[1ex]\hline\hline
$P_{7,8}$&\hspace*{-2mm}\raisebox{-9mm}{\includegraphics[width=12mm]{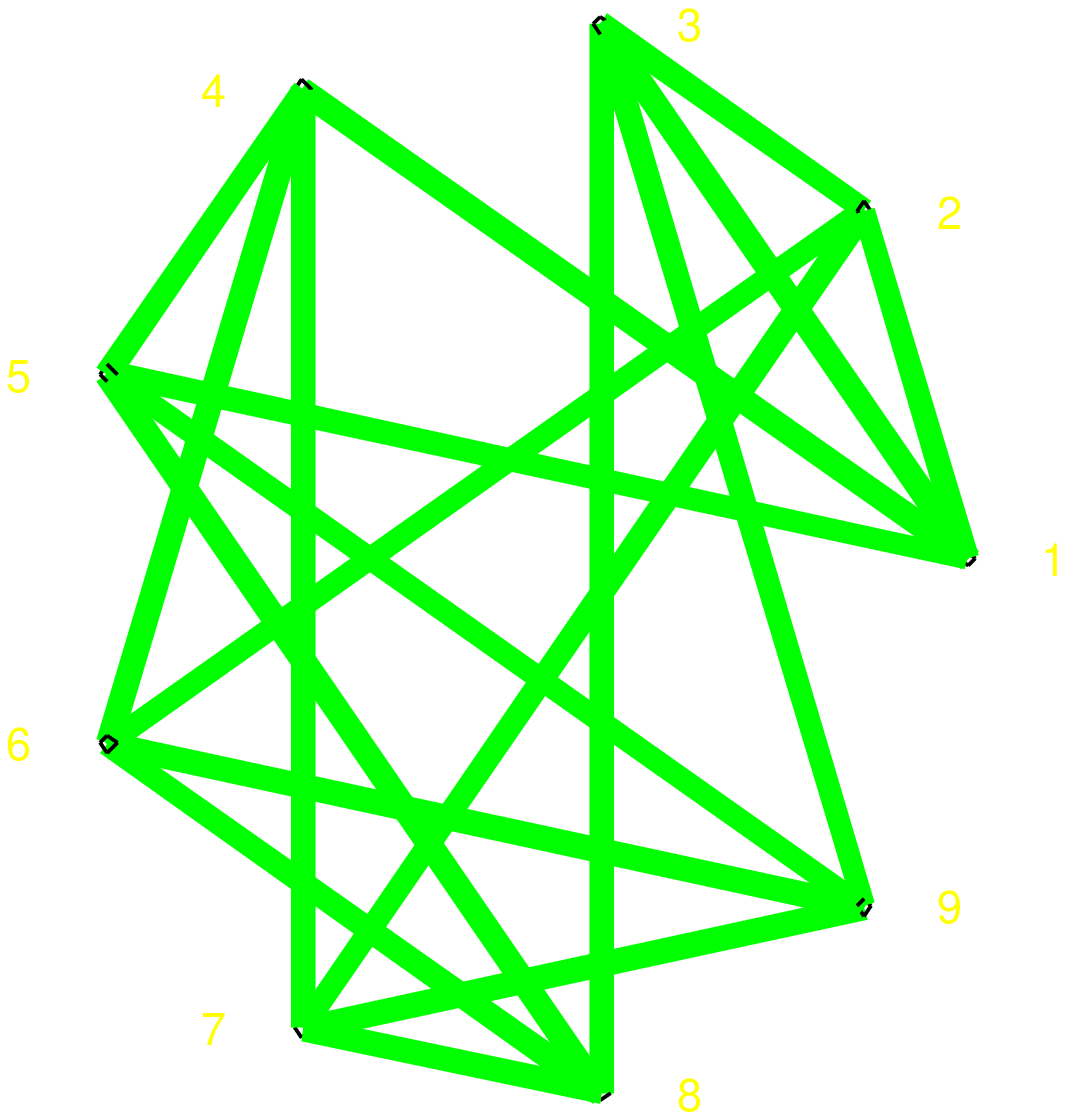}}&183.032~420~030&16&16&$P_{7,8}$&(37)\\[-6mm]
11&&\multicolumn{5}{l}{$\frac{22383}{20}Q_{11,1}-\frac{4572}{5}Q_{11,2}+1792Q_3Q_8-700Q_3^2Q_5$}\\[1ex]\hline
$P_{7,9}$&\hspace*{-2mm}\raisebox{-9mm}{\includegraphics[width=12mm]{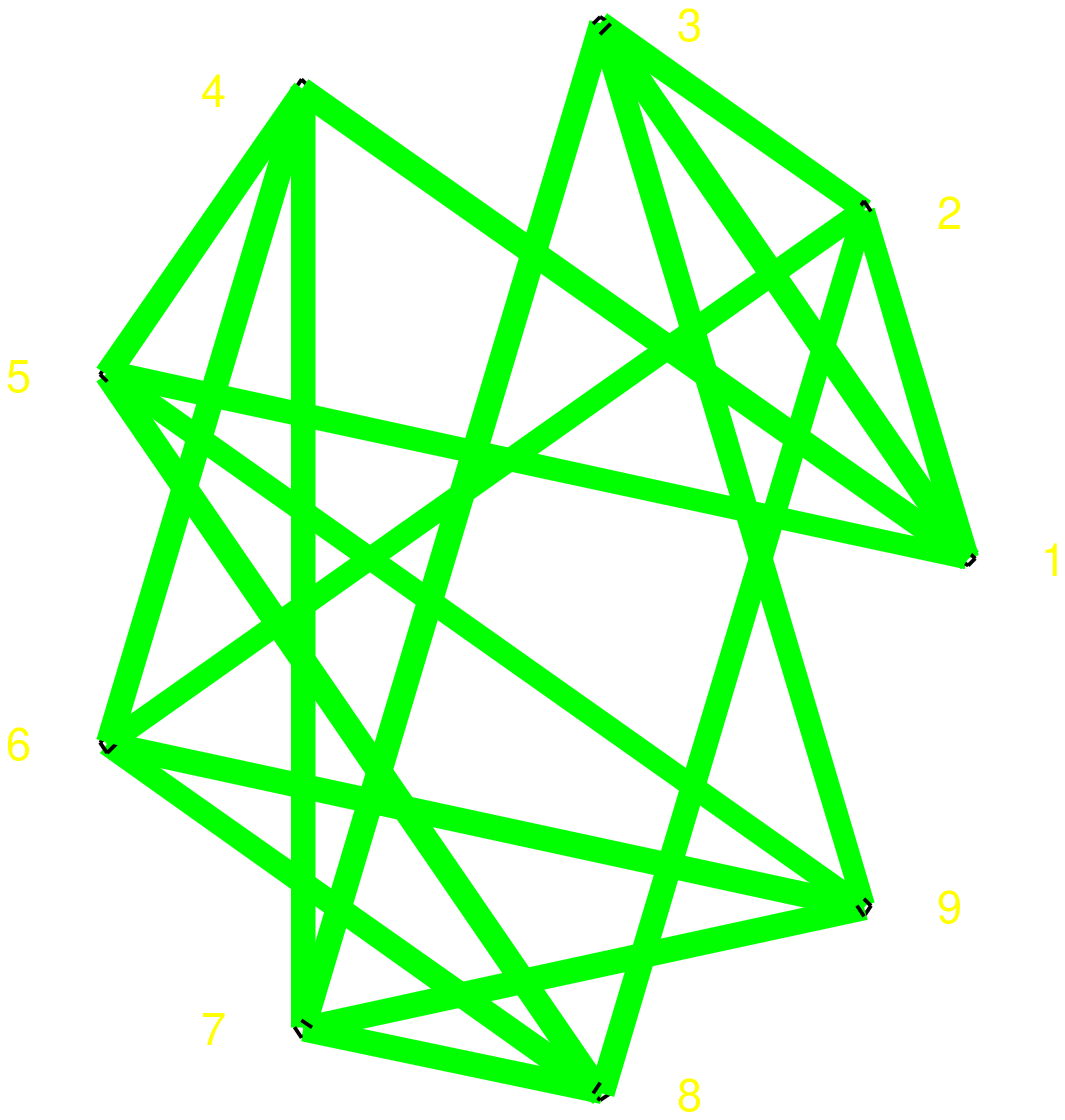}}&216.919~375~55(6)&12&?&$P_{7,9}$&\\[-6mm]
11?&&\multicolumn{5}{l}{?}\\[1ex]\hline
$P_{7,10}$&\hspace*{-2mm}\raisebox{-9mm}{\includegraphics[width=12mm]{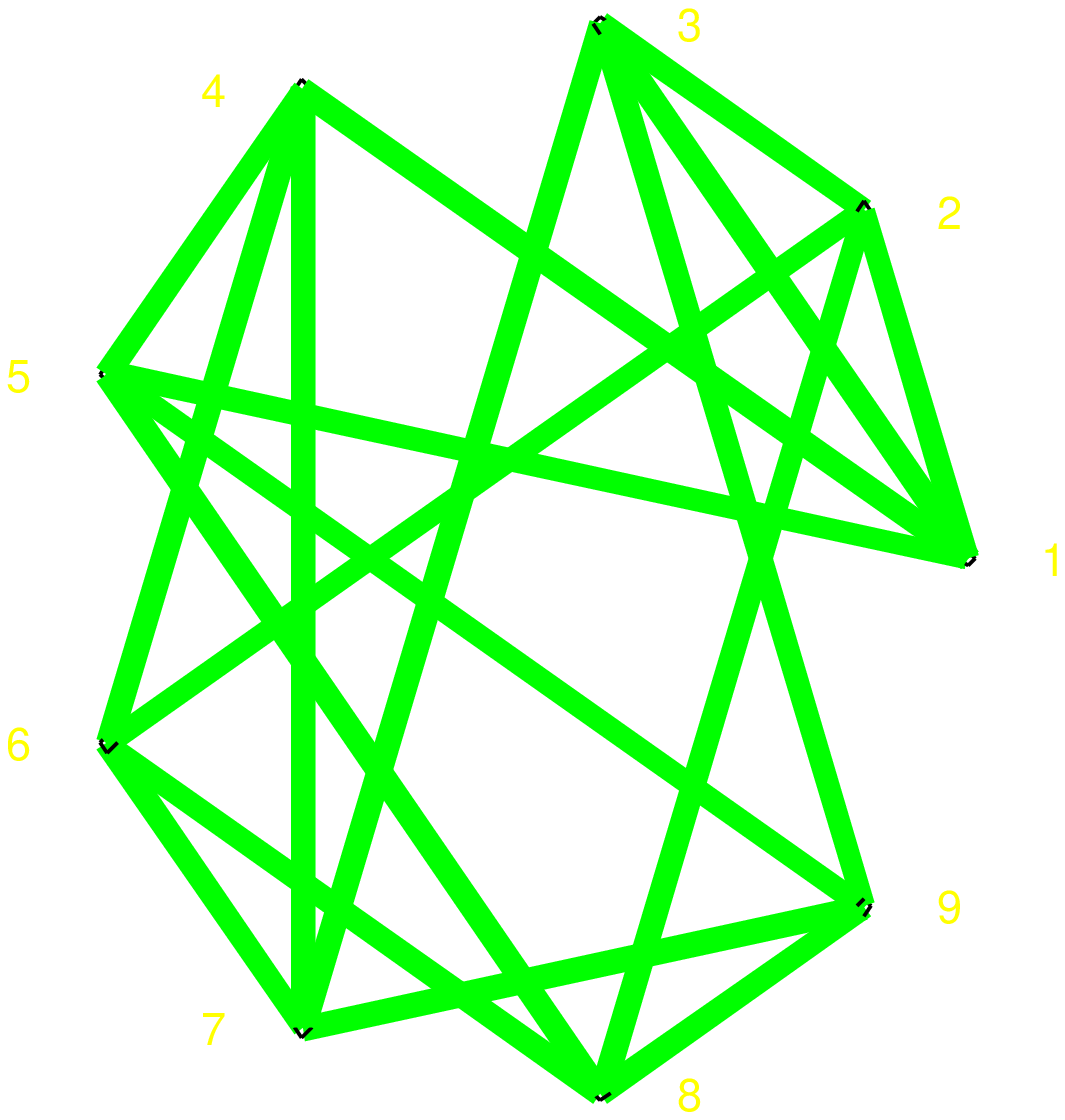}}&254.763~009~595&72&144&$P_{7,10}$&$K_3$\raisebox{.7ex}{\fbox{}}$K_3$, Fourier\\[-6mm]
10&&\multicolumn{5}{l}{$P_{7,5}$}\\[1ex]\hline
$P_{7,11}$&\hspace*{-2mm}\raisebox{-9mm}{\includegraphics[width=12mm]{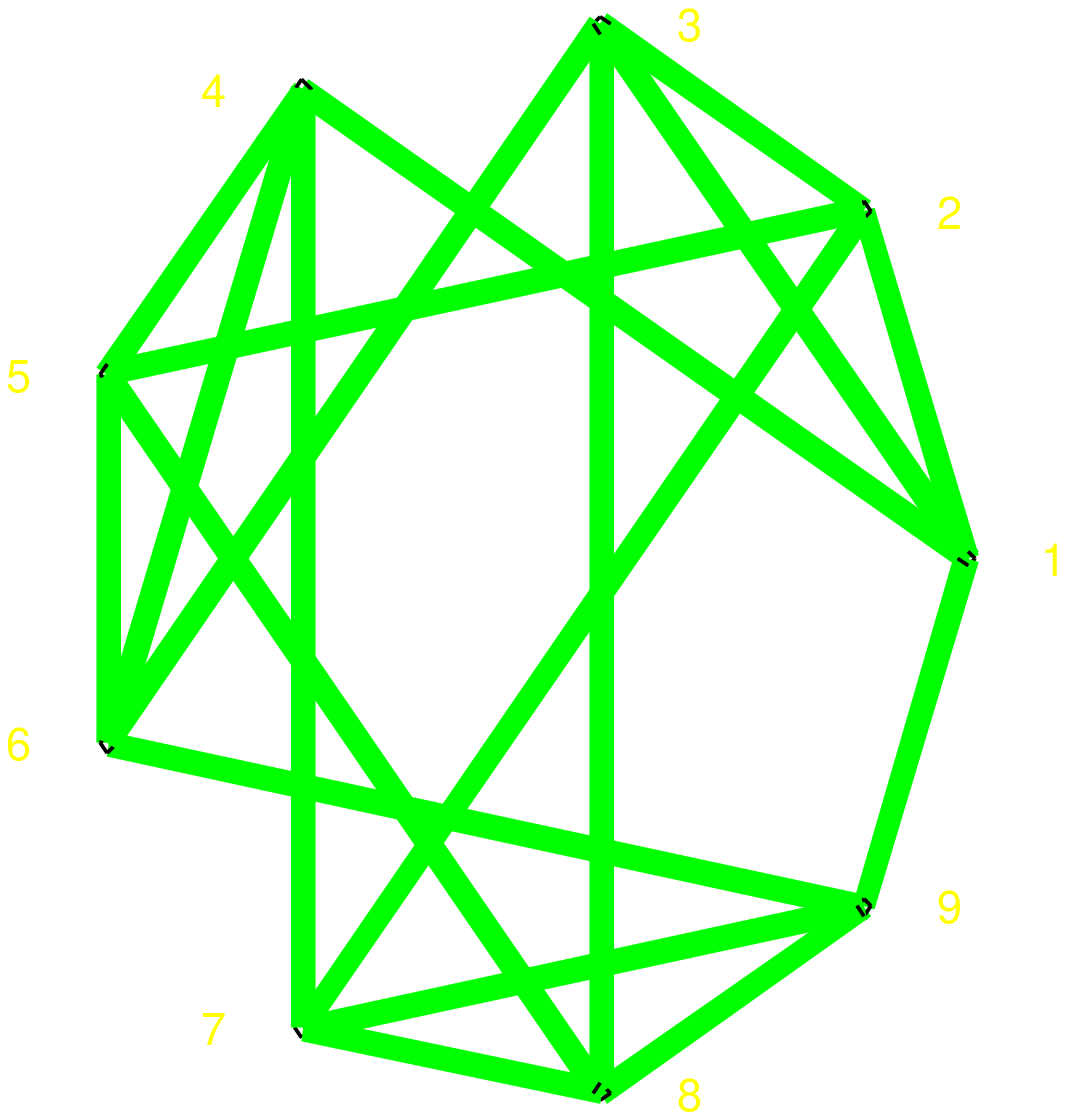}}&200.357~566~429&18&?&$P_{7,11}$&$C^9_{1,3}$\\[-6mm]
11?&&\multicolumn{5}{l}{?}\\[1ex]\hline\hline
$P_{8,1}$&\hspace*{-2mm}\raisebox{-9mm}{\includegraphics[width=12mm]{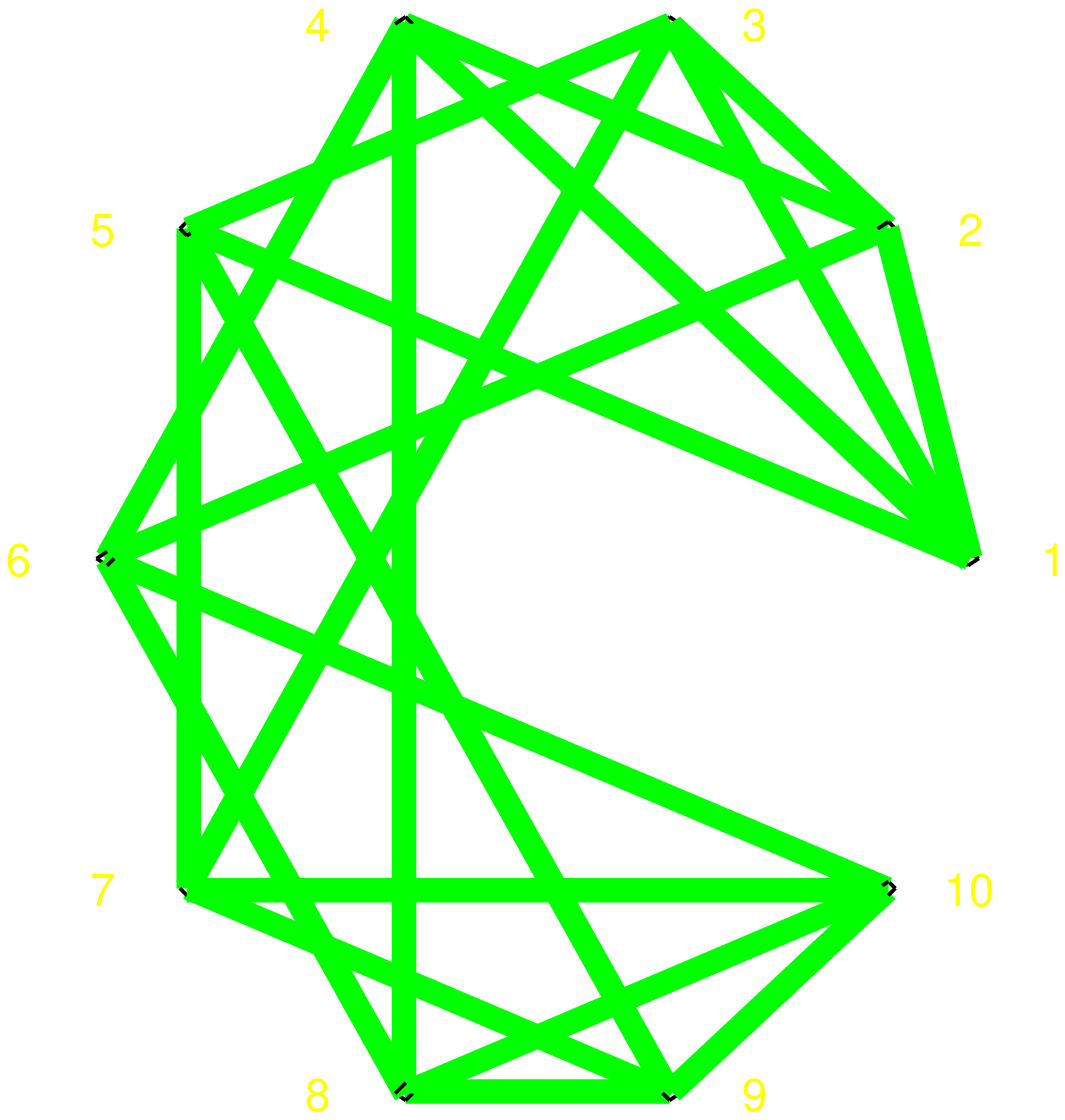}}&1~716.210~576~104&20&2635776&$P_3$&$C^{10}_{1,2}$\\[-6mm]
13&&\multicolumn{5}{l}{$1716Q_{13,1}$}\\[1ex]\hline
$P_{8,2}$&\hspace*{-2mm}\raisebox{-9mm}{\includegraphics[width=12mm]{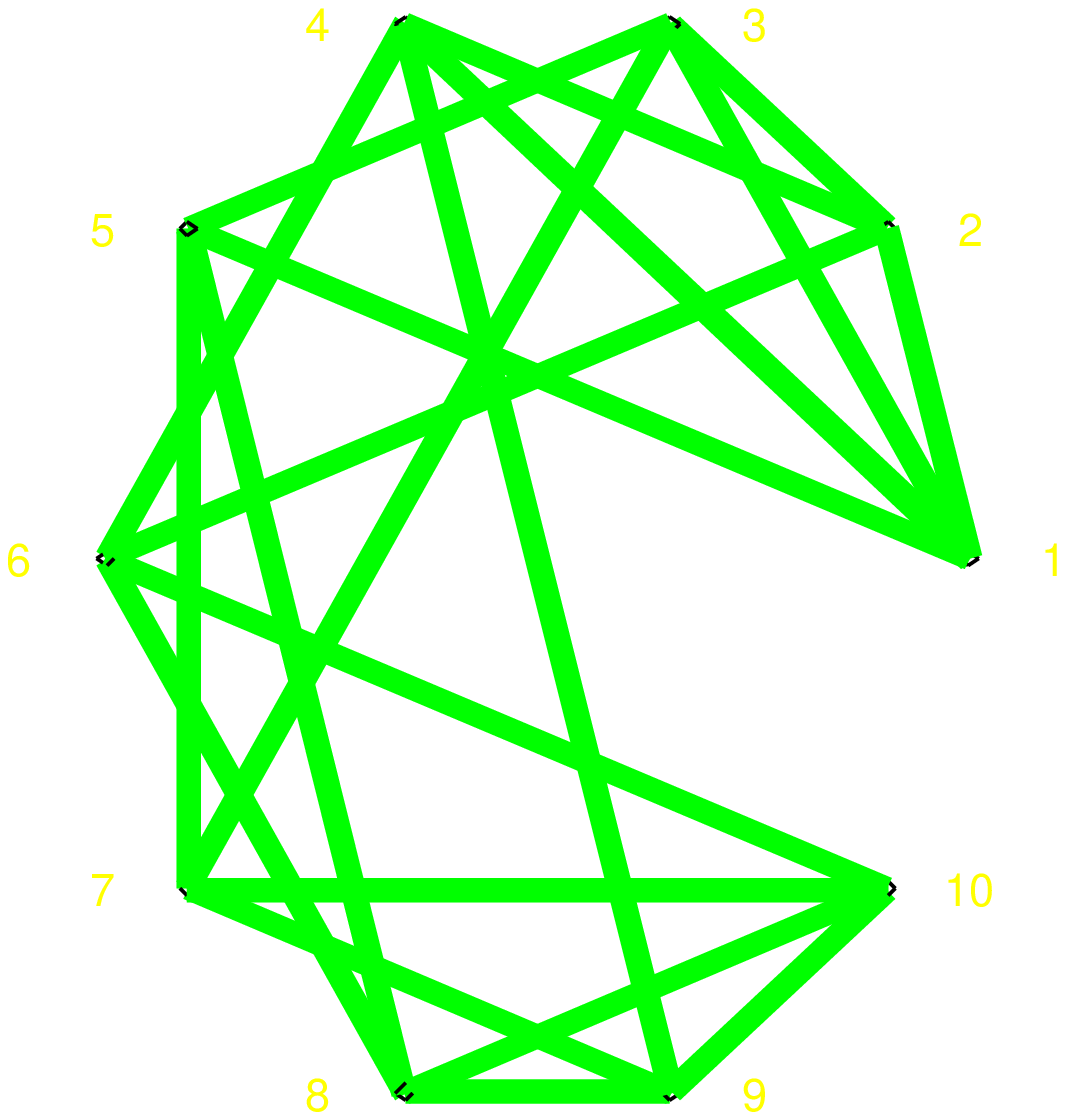}}&1~145.592~929~599&2&12&$P_3$&(35)\\[-6mm]
13&&\multicolumn{5}{l}{$\frac{25147347}{22400}Q_{13,1}-\frac{16881}{1400}Q_{13,2}+\frac{459}{112}Q_{13,3}+\frac{1305}{8}Q_3^2Q_7-135Q_3Q_5^2$}\\[1ex]\hline
$P_{8,3}$&\hspace*{-2mm}\raisebox{-9mm}{\includegraphics[width=12mm]{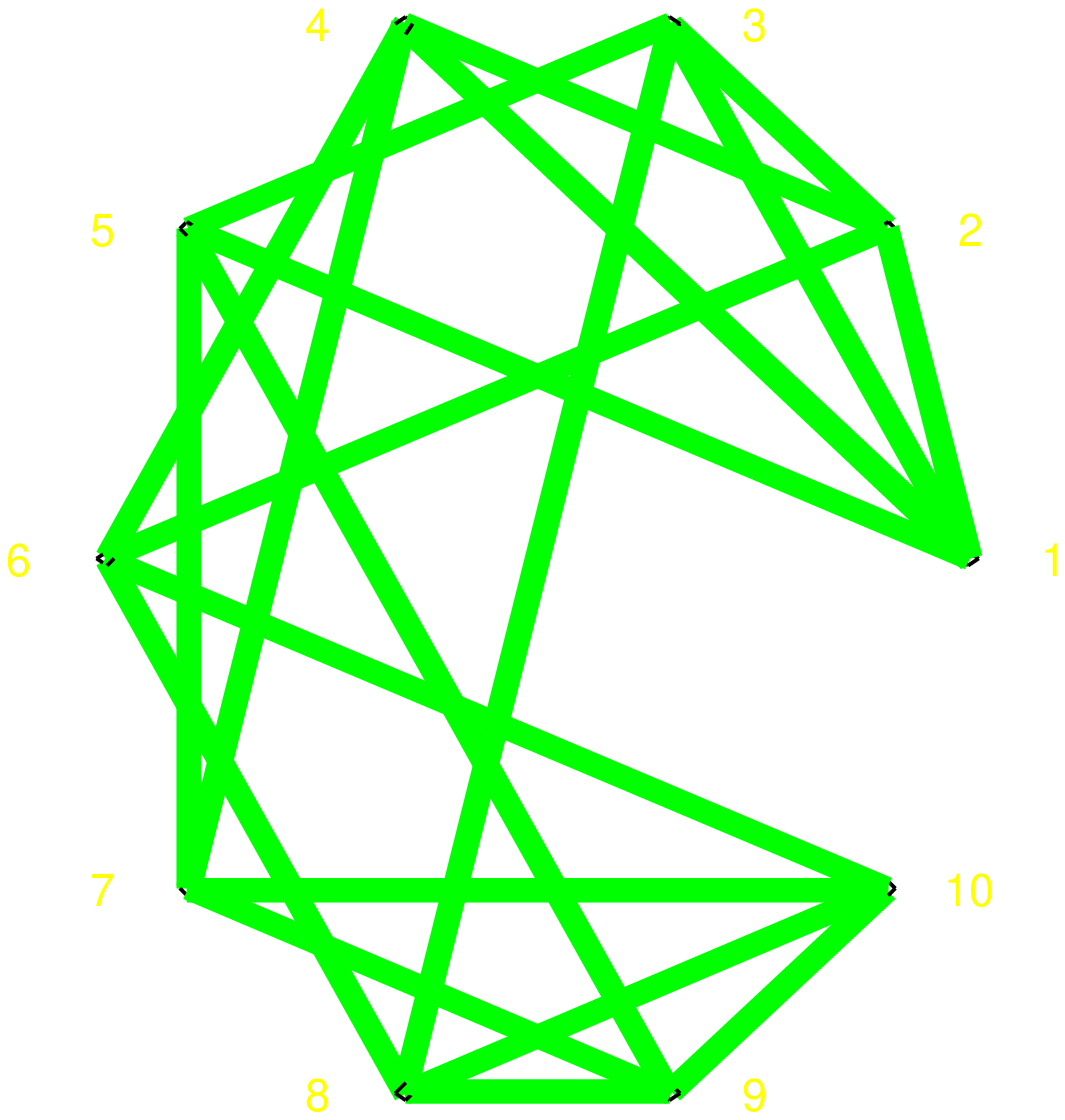}}&1~105.107~697~390&4&1280&$P_3$&$Z_{4,5}$, (120)\\[-6mm]
13&&\multicolumn{5}{l}{$298Q_{13,1}+56Q_{13,2}-20Q_{13,3}-280Q_3^2Q_7+800Q_3Q_5^2$}\\[1ex]\hline
$P_{8,4}$&\hspace*{-2mm}\raisebox{-9mm}{\includegraphics[width=12mm]{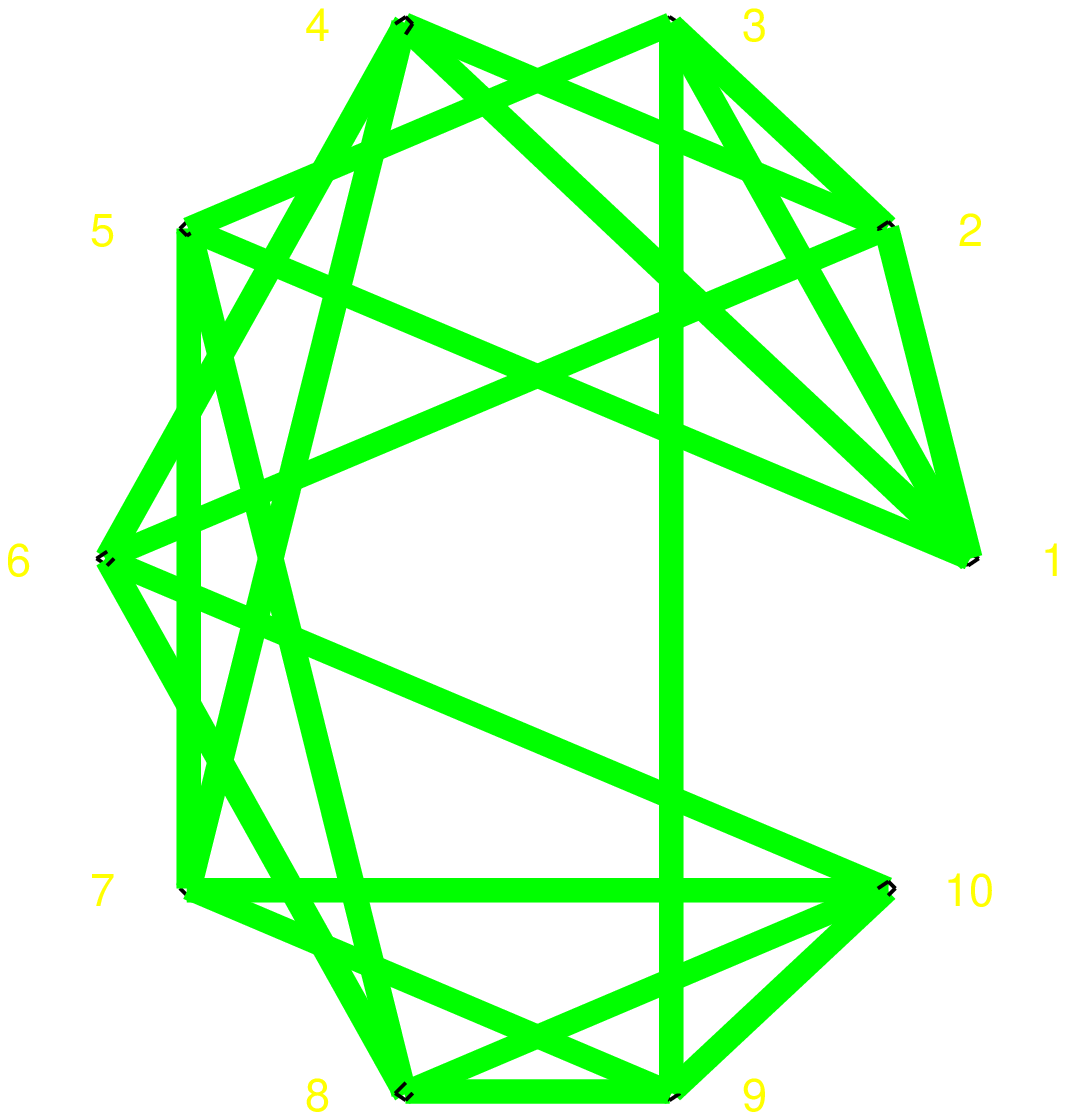}}&966.830~801~986&1&12&$P_3$&(76)\\[-6mm]
13&&\multicolumn{5}{l}{$\frac{17124243}{22400}Q_{13,1}-\frac{19689}{1400}Q_{13,2}+\frac{1755}{112}Q_{13,3}+\frac{9}{8}Q_3^2Q_7+135Q_3Q_5^2$}\\[1ex]\hline
$P_{8,5}$&\hspace*{-2mm}\raisebox{-9mm}{\includegraphics[width=12mm]{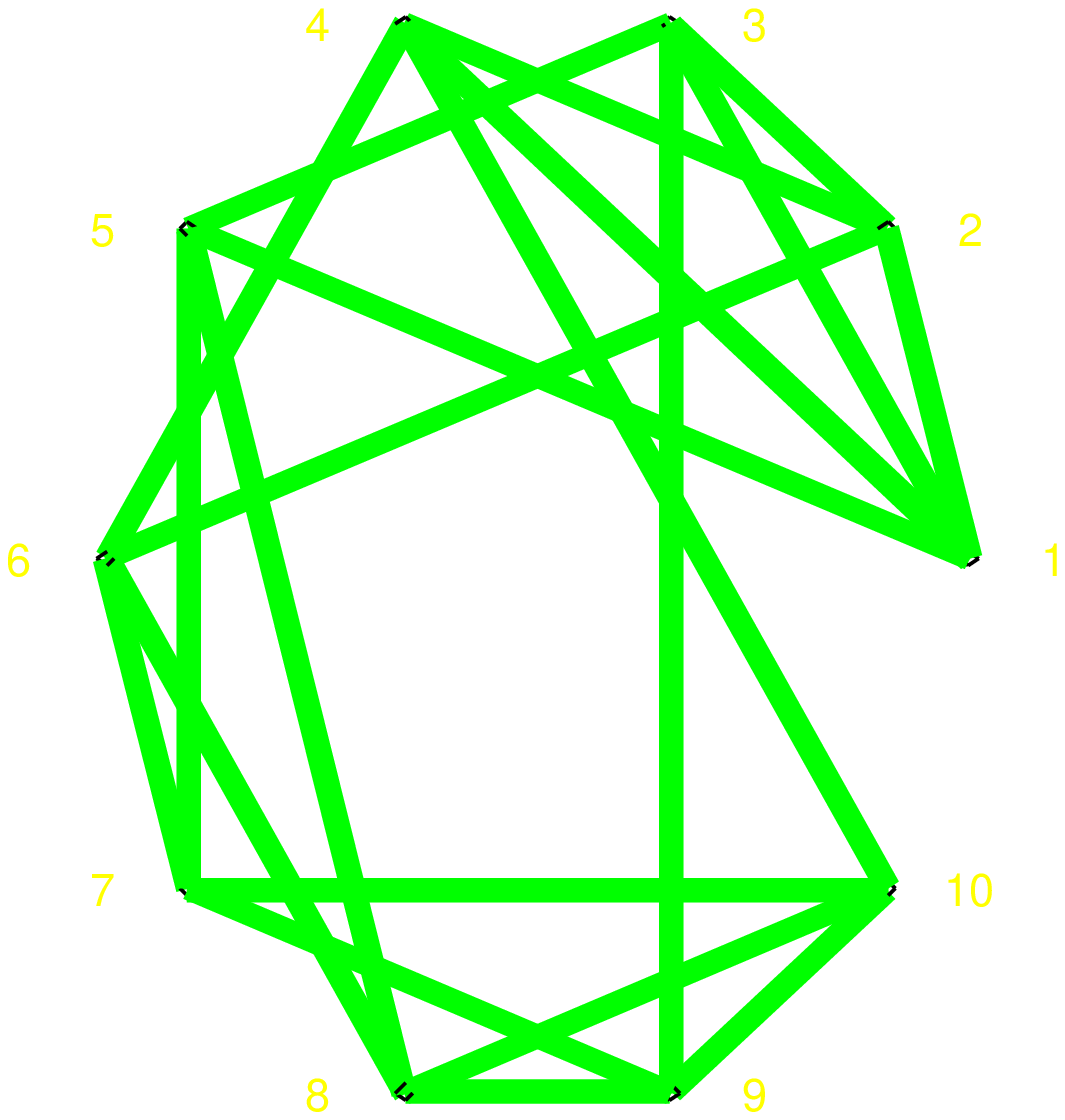}}&844.512~518~603&4&24&$P_3^2$&(54)\\[-6mm]
12&&\multicolumn{5}{l}{$1536Q_{12,1}-1280Q_{12,2}+36Q_3Q_9+\frac{1299}{2}Q_5Q_7$}\\[1ex]\hline
$P_{8,6}$&\hspace*{-2mm}\raisebox{-9mm}{\includegraphics[width=12mm]{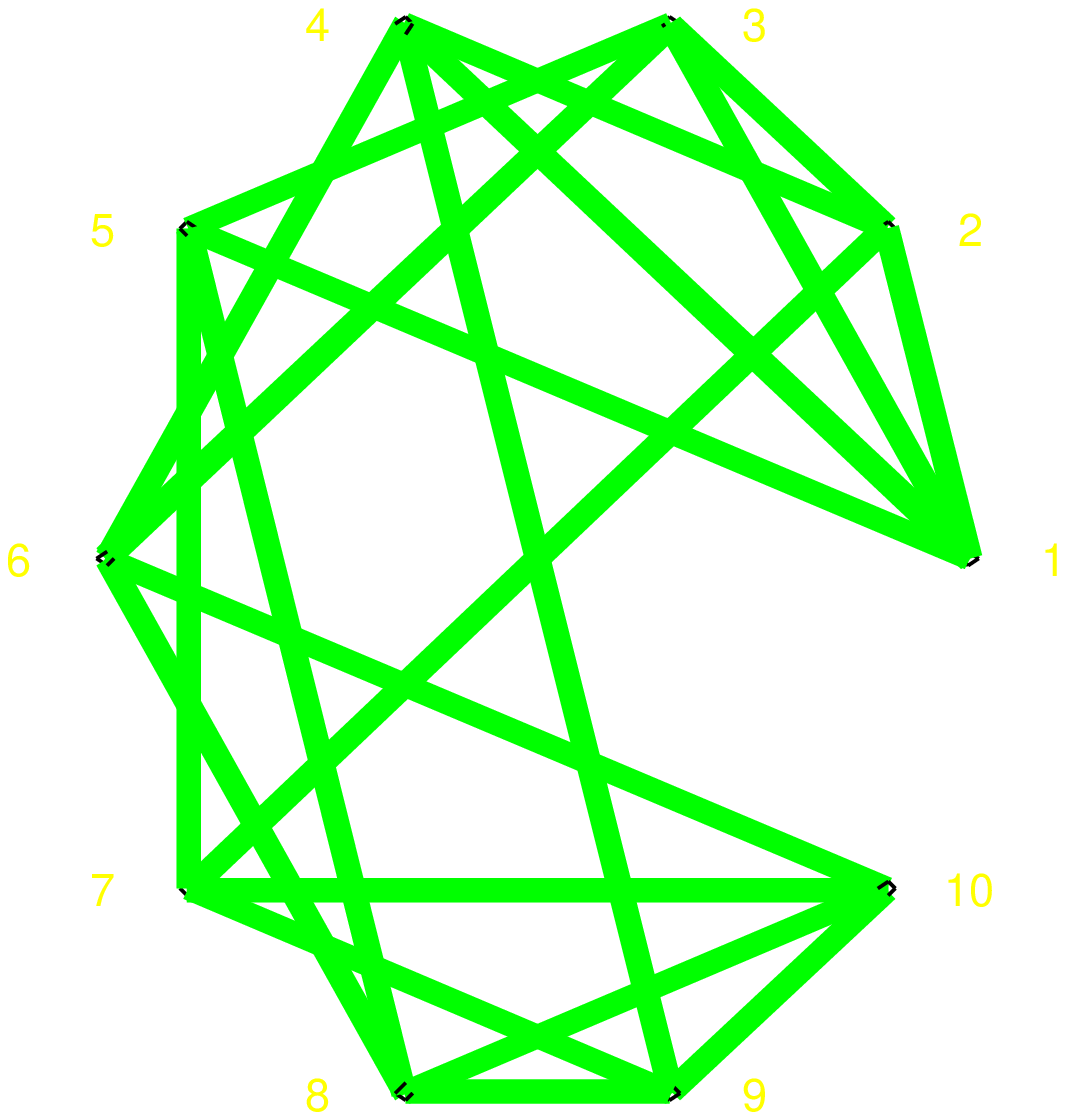}}&904.280~824~357&4&32&$P_3$&(54)\\[-6mm]
13&&\multicolumn{5}{l}{$\frac{214841}{336}Q_{13,1}-\frac{423}{7}Q_{13,2}+\frac{705}{14}Q_{13,3}+183Q_3^2Q_7$}\\[1ex]\hline
$P_{8,7}$&\hspace*{-2mm}\raisebox{-9mm}{\includegraphics[width=12mm]{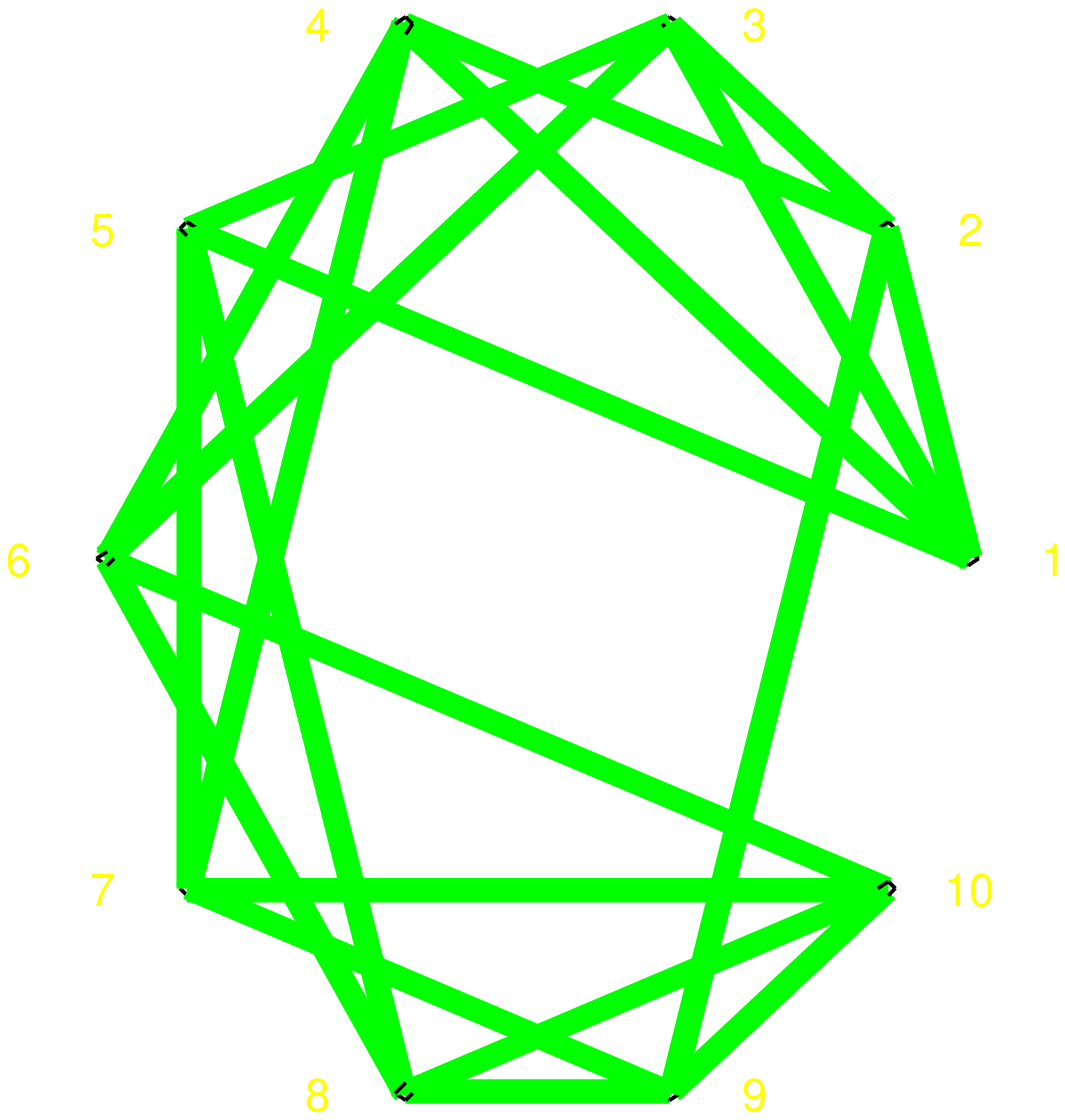}}&847.646~115~639&2&144&$P_3$&(74)\\[-6mm]
13&&\multicolumn{5}{l}{$\frac{2061501}{2800}Q_{13,1}+\frac{13527}{175}Q_{13,2}-\frac{675}{14}Q_{13,3}$}\\[1ex]\hline
$P_{8,8}$&\hspace*{-2mm}\raisebox{-9mm}{\includegraphics[width=12mm]{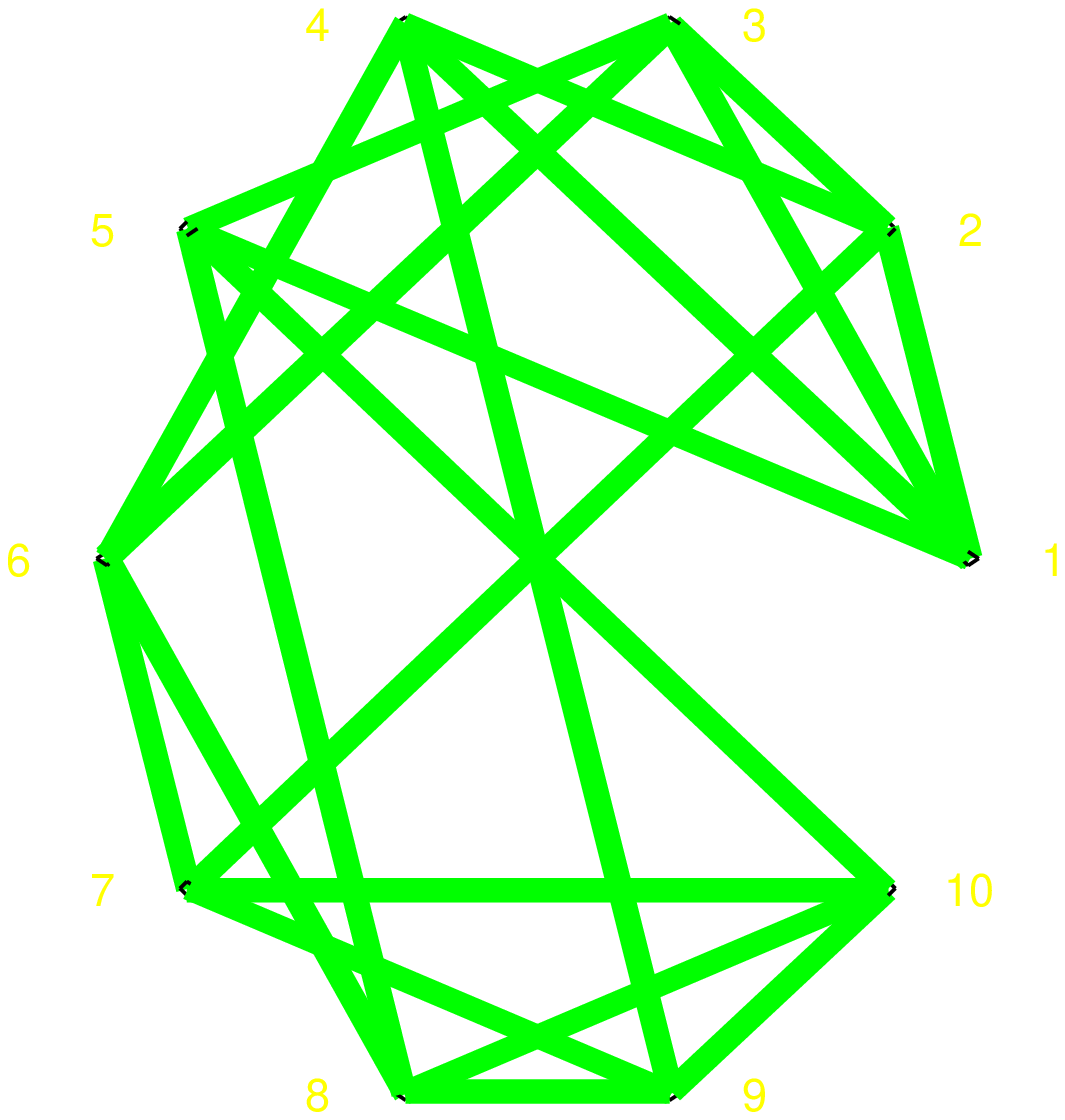}}&847.646~115~639&2&144&$P_3$&twist\\[-6mm]
13&&\multicolumn{5}{l}{$P_{8,7}$}\\[1ex]\hline
$P_{8,9}$&\hspace*{-2mm}\raisebox{-9mm}{\includegraphics[width=12mm]{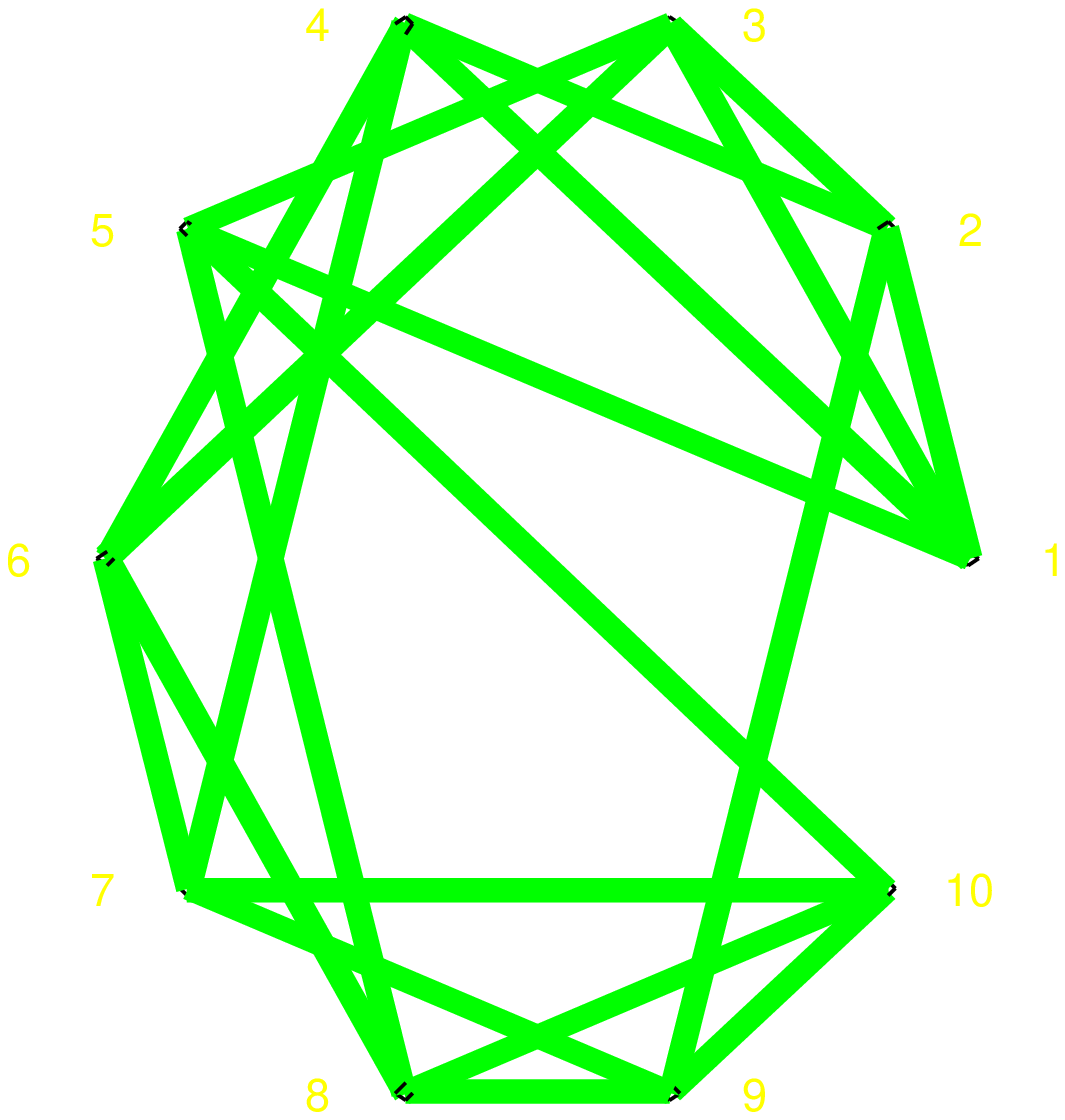}}&904.280~824~357&2&32&$P_3$&twist\\[-6mm]
13&&\multicolumn{5}{l}{$P_{8,6}$}\\[1ex]\hline
$P_{8,10}$&\hspace*{-2mm}\raisebox{-9mm}{\includegraphics[width=12mm]{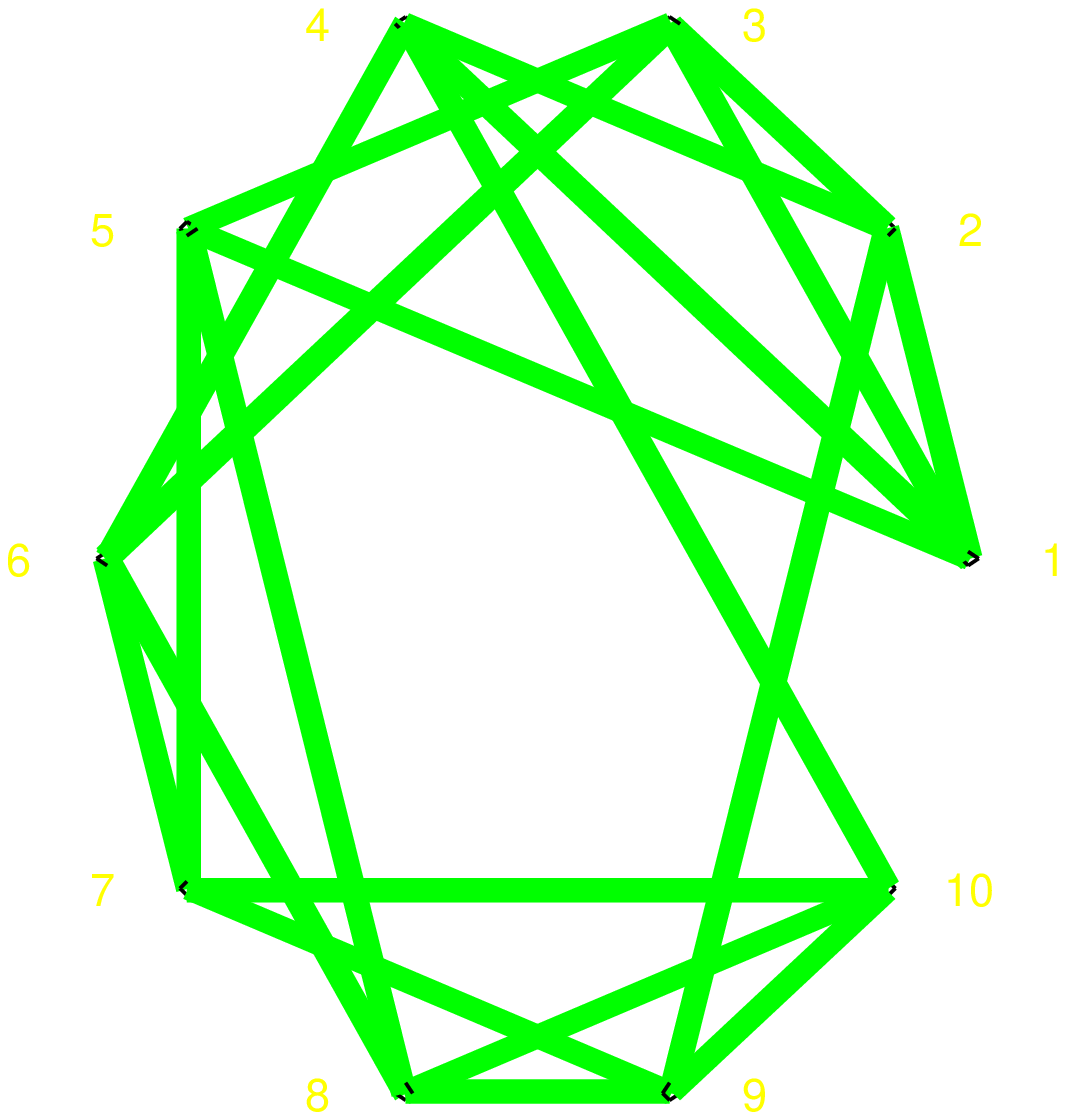}}&735.764~103~468&2&72&$P_3^2$&(37)\\[-6mm]
12&&\multicolumn{5}{l}{$1536Q_{12,1}-1280Q_{12,2}-\frac{63}{2}Q_3Q_9+\frac{2493}{4}Q_5Q_7$}\\[1ex]\hline
$P_{8,11}$&\hspace*{-2mm}\raisebox{-9mm}{\includegraphics[width=12mm]{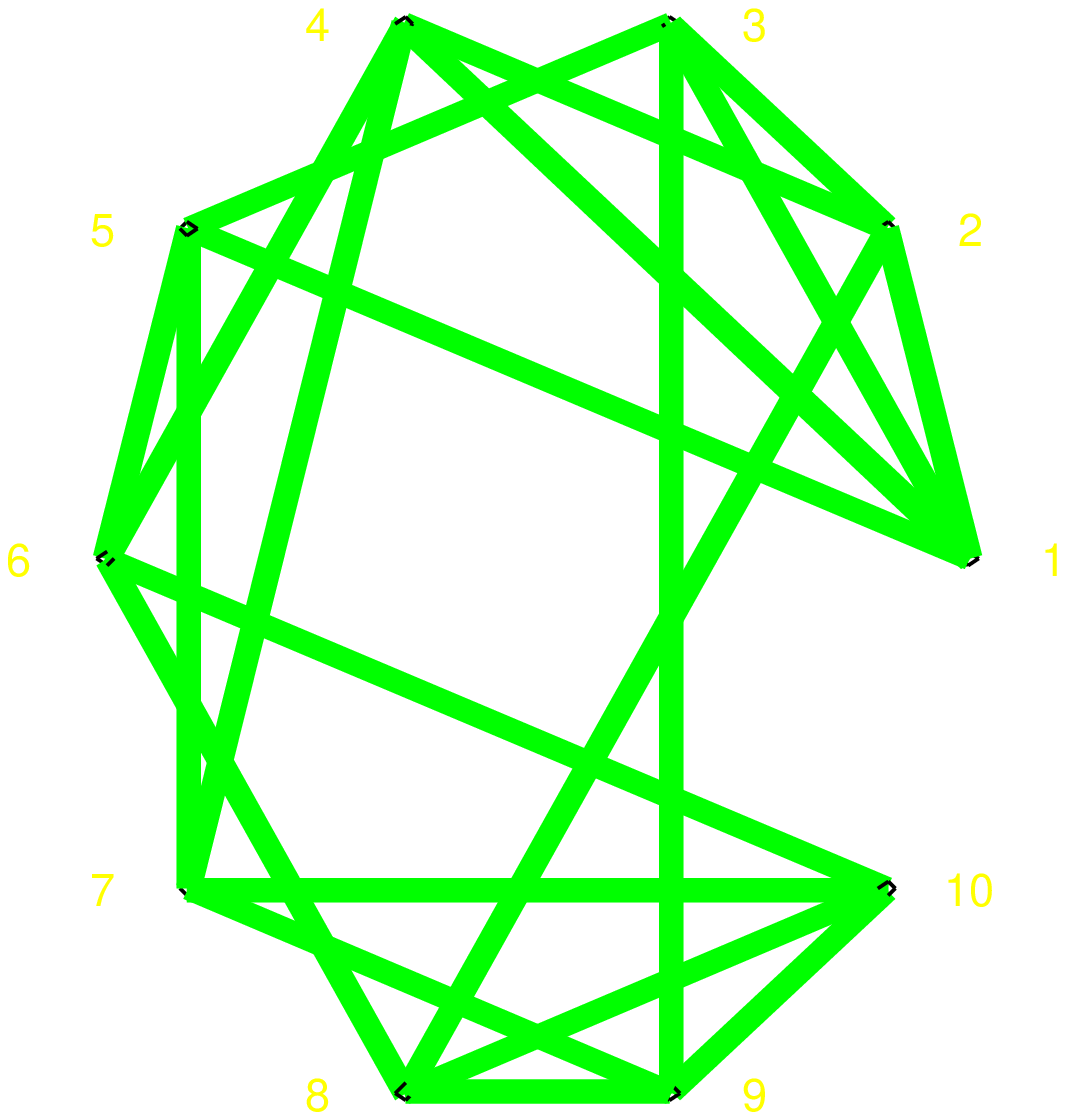}}&805.347~388~507&4&16&$P_3^2$&(33)\\[-6mm]
12&&\multicolumn{5}{l}{$\frac{10240}{69}Q_{12,1}+\frac{81920}{69}Q_{12,2}-\frac{2560}{69}Q_{12,3}+\frac{45503}{69}Q_3Q_9+\frac{305}{46}Q_5Q_7-12Q_3^4$}
\end{tabular}

\begin{tabular}{lllllll}
name&graph&numerical value&$|$Aut$|$&index&ancestor&rem, (sf), [Lit]\\[1ex]
\multicolumn{2}{l}{weight}&\multicolumn{5}{l}{exact value}\\[1ex]\hline\hline
$P_{8,12}$&\hspace*{-2mm}\raisebox{-9mm}{\includegraphics[width=12mm]{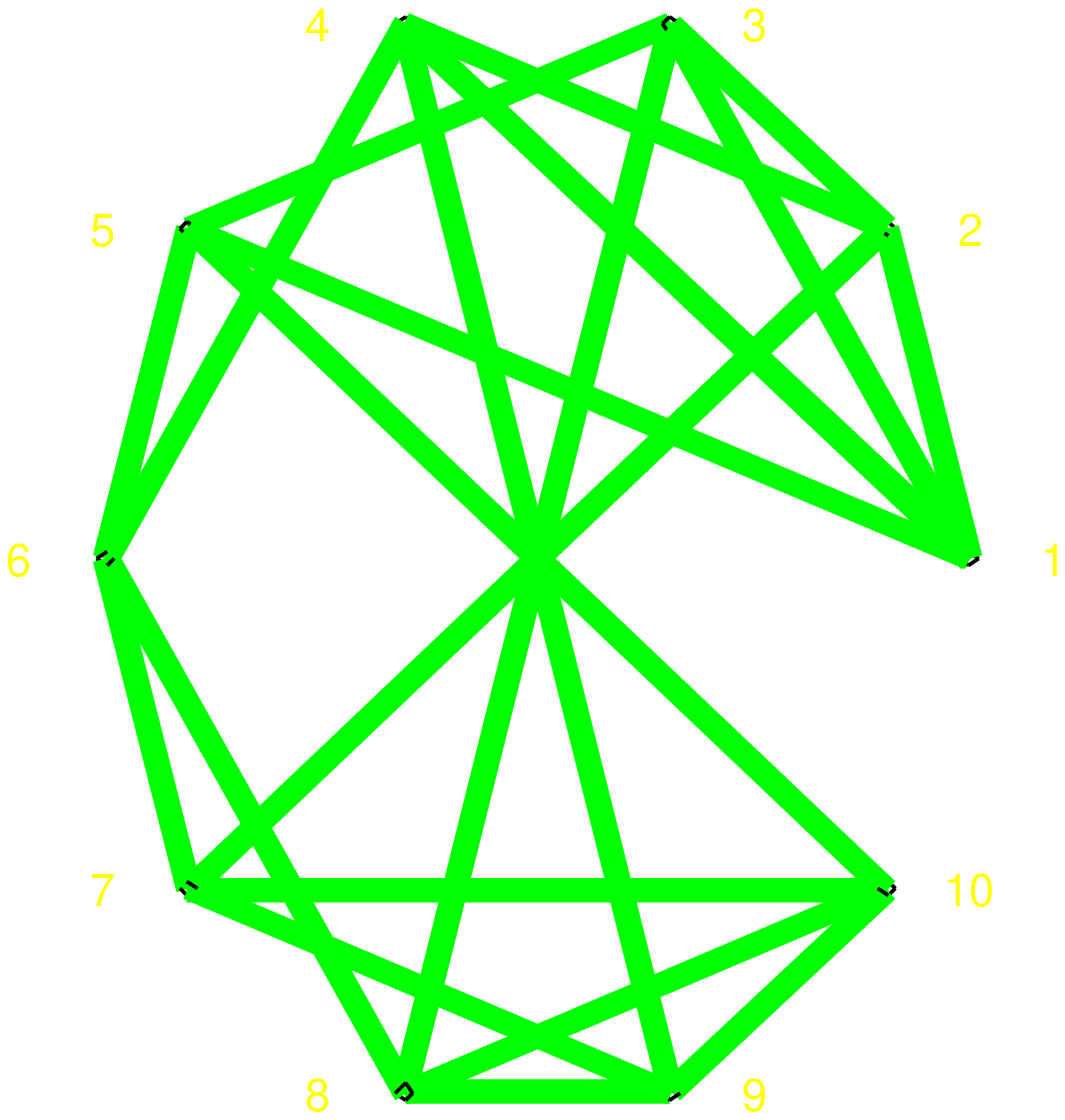}}&688.898~361~296&2&288&$P_3^2$&(40)\\[-6mm]
12&&\multicolumn{5}{l}{$1024Q_{12,2}-1008Q_3Q_9+1800Q_5Q_7$}\\[1ex]\hline
$P_{8,13}$&\hspace*{-2mm}\raisebox{-9mm}{\includegraphics[width=12mm]{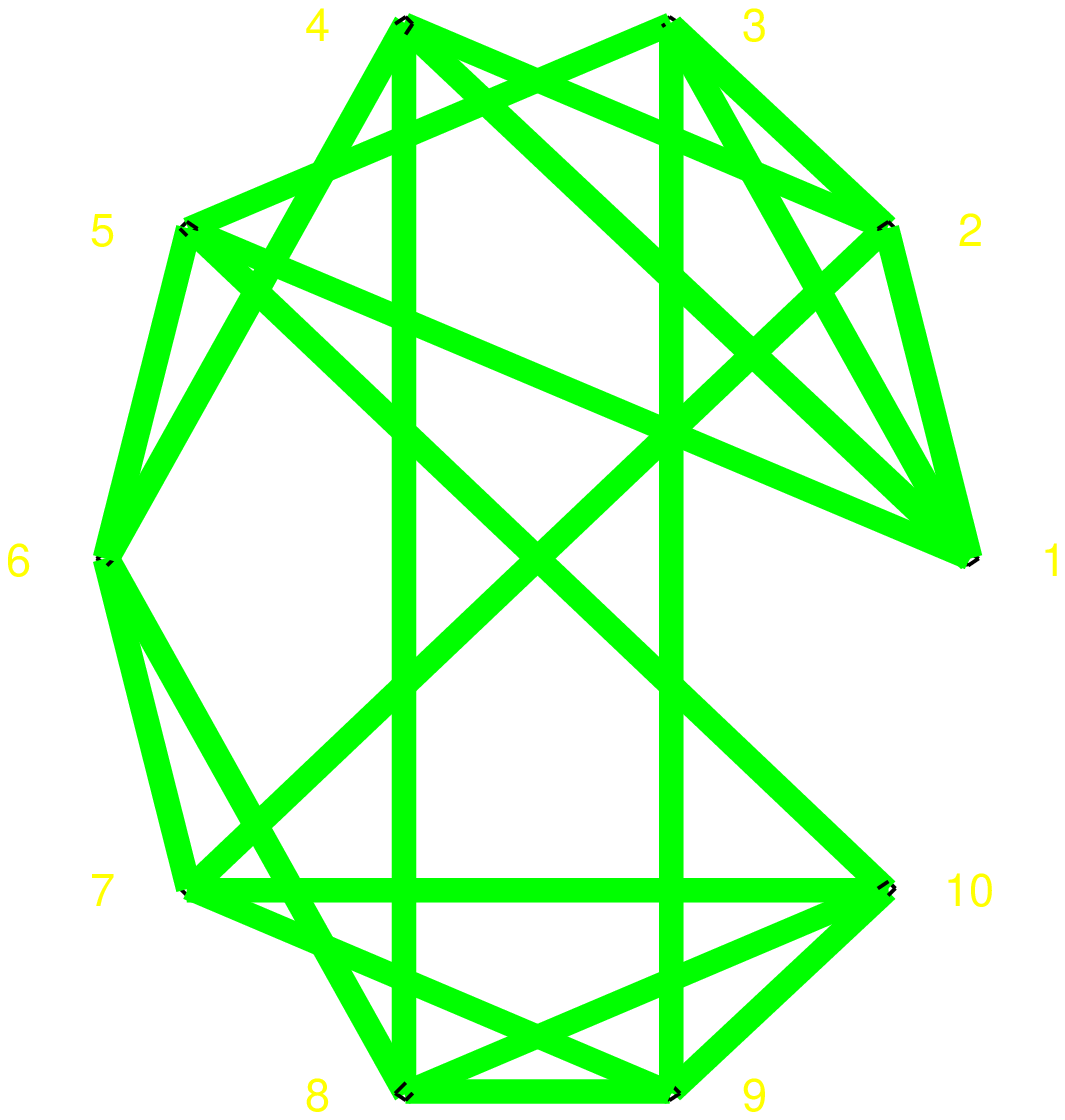}}&742.977~090~366&1&4&$P_3$&(105)\\[-6mm]
13&&\multicolumn{5}{l}{$\frac{10087273}{9600}Q_{13,1}+\frac{8007}{200}Q_{13,2}-\frac{813}{16}Q_{13,3}+\frac{2247}{8}Q_3^2Q_7-465Q_3Q_5^2$}\\[1ex]\hline
$P_{8,14}$&\hspace*{-2mm}\raisebox{-9mm}{\includegraphics[width=12mm]{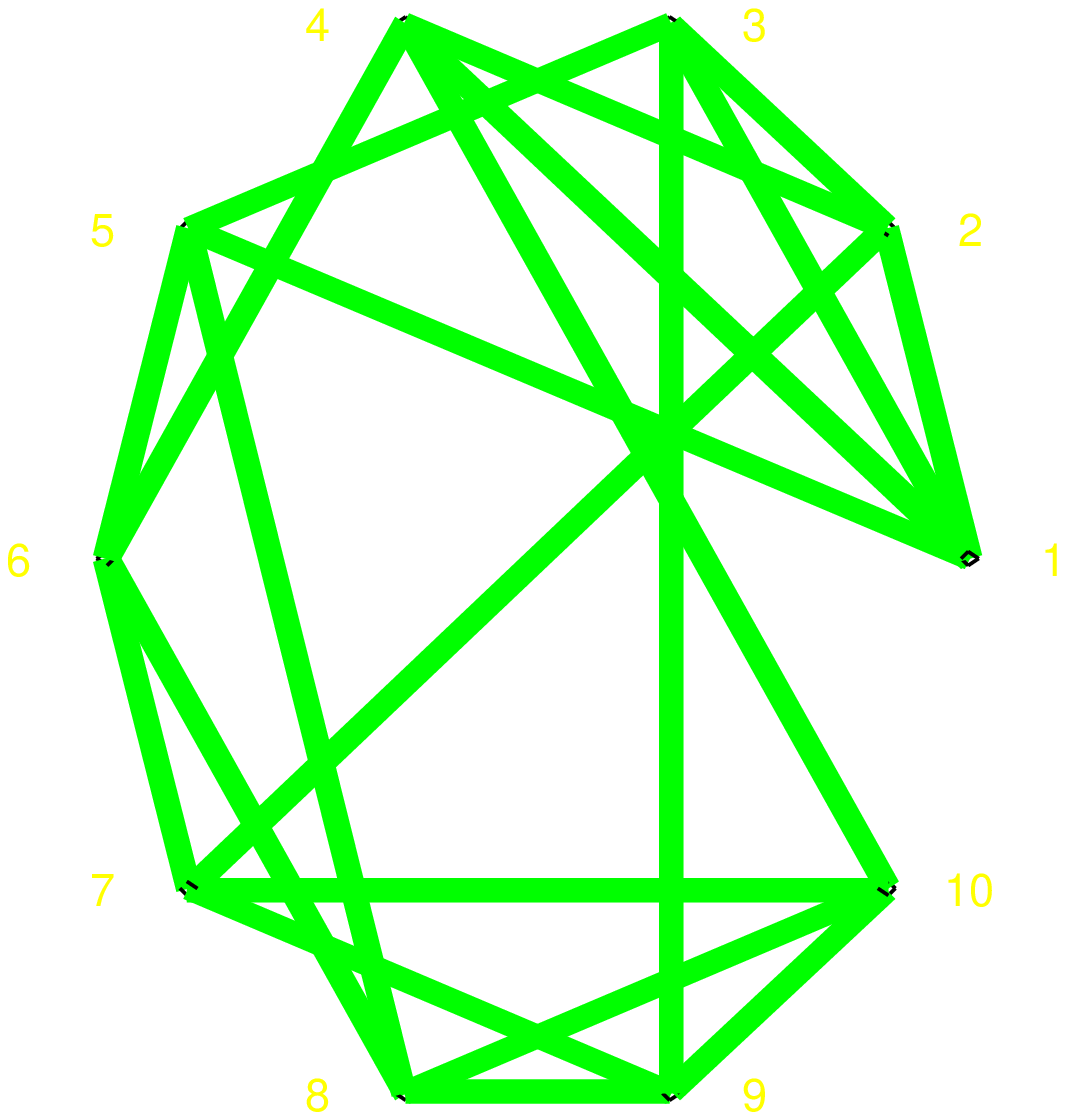}}&749.818~622~995&1&4&$P_3$&(35)\\[-6mm]
13&&\multicolumn{5}{l}{$\frac{41038969}{67200}Q_{13,1}-\frac{30129}{1400}Q_{13,2}+\frac{1611}{112}Q_{13,3}+\frac{153}{8}Q_3^2Q_7+105Q_3Q_5^2$}\\[1ex]\hline
$P_{8,15}$&\hspace*{-2mm}\raisebox{-9mm}{\includegraphics[width=12mm]{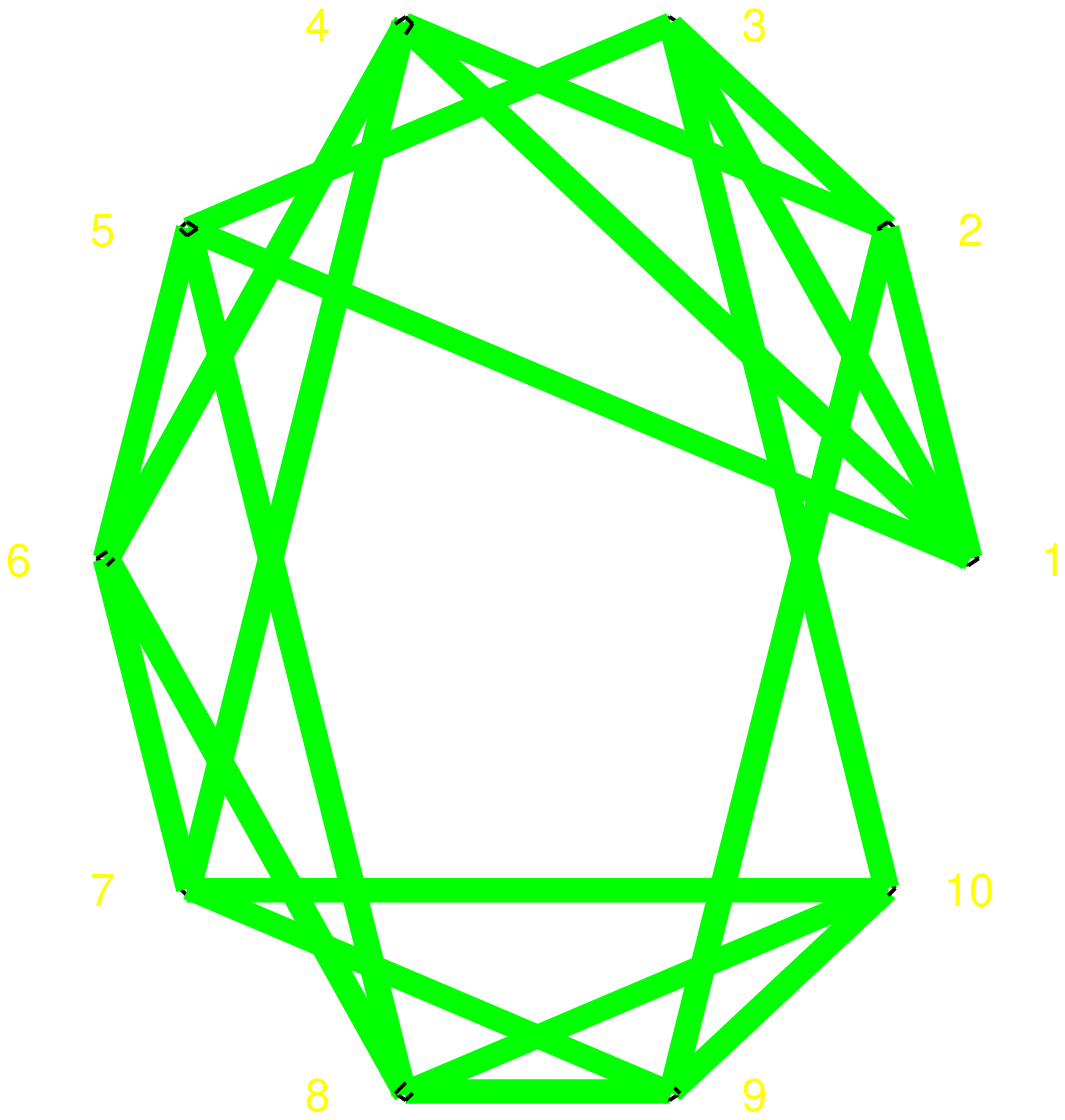}}&805.347~388~507&2&16&$P_3^2$&twist\\[-6mm]
12&&\multicolumn{5}{l}{$P_{8,11}$}\\[1ex]\hline
$P_{8,16}$&\hspace*{-2mm}\raisebox{-9mm}{\includegraphics[width=12mm]{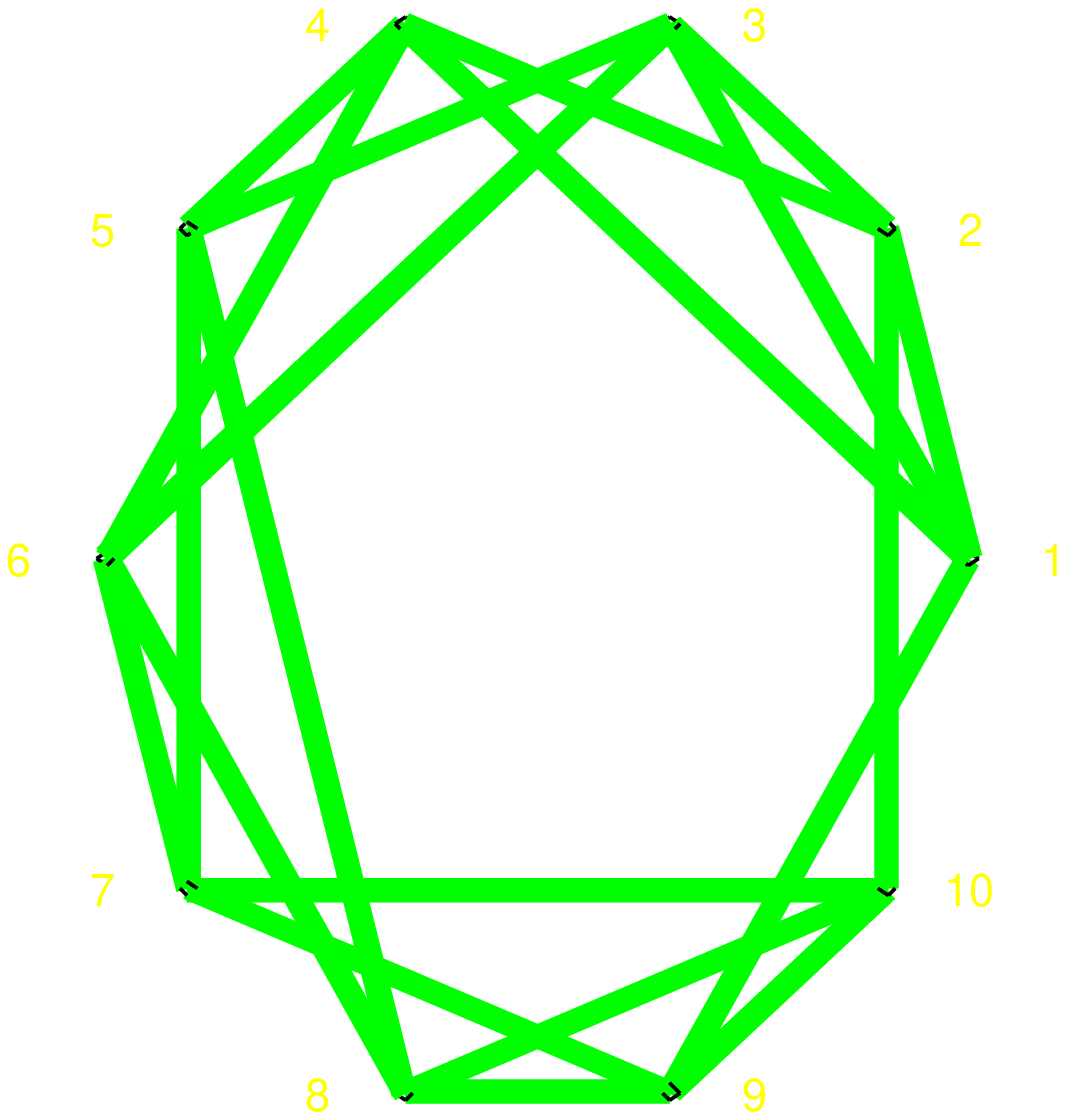}}&633.438~914~549&32&576&$P_3^3$&(92)\hfill$[-10080Q_5^2$\\[-6mm]
11,10&&\multicolumn{5}{l}{$-\frac{31851}{5}Q_{11,1}\!+\!\frac{24336}{5}Q_{11,2}\!-\!10240Q_3Q_8\!+\!5040Q_3^2Q_5\!-\!8192Q_{10}\!+\!9648Q_3Q_7$}\\[1ex]\hline
$P_{8,17}$&\hspace*{-2mm}\raisebox{-9mm}{\includegraphics[width=12mm]{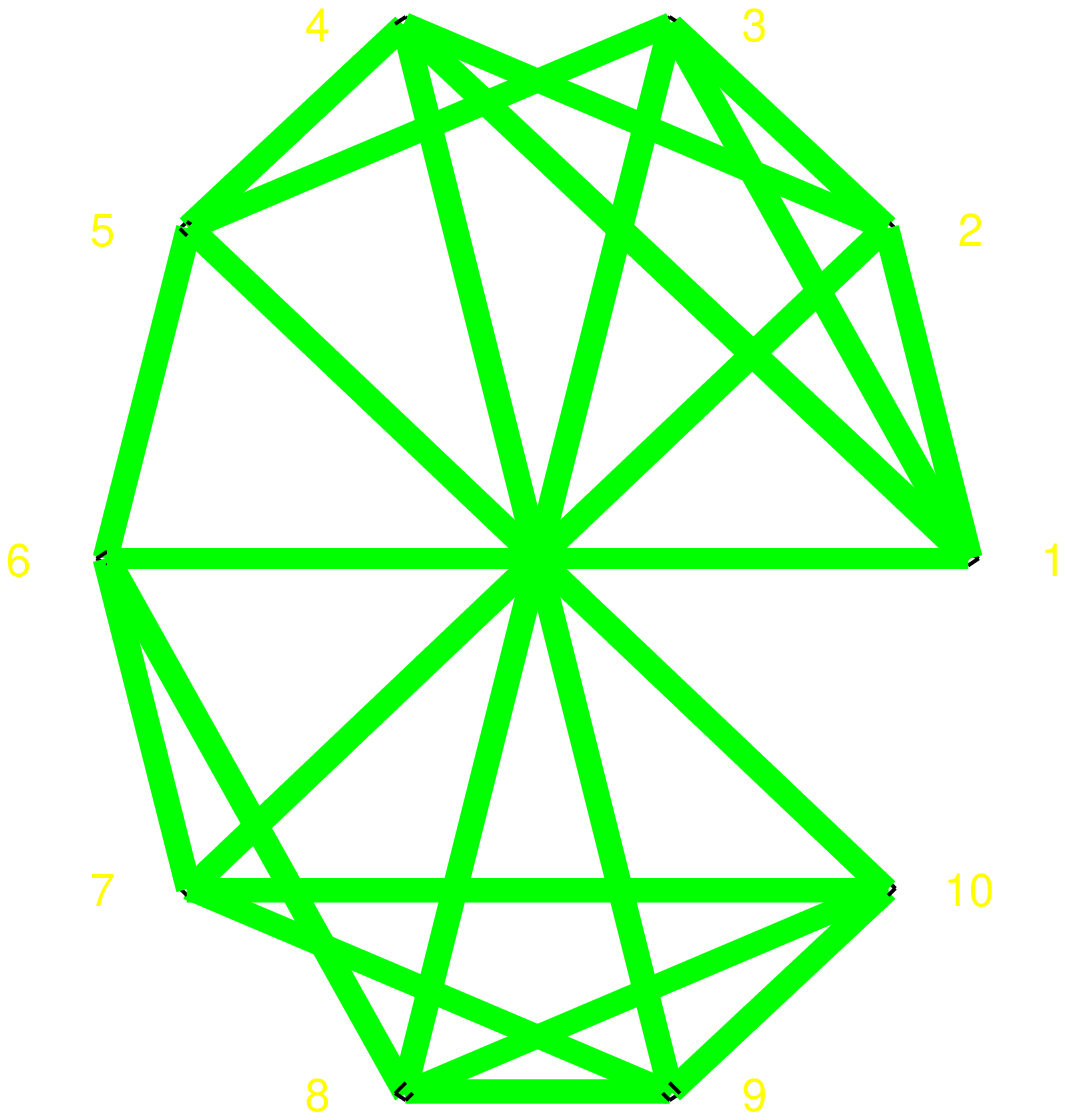}}&?&2&?&$P_3$&\\[-6mm]
13?&&\multicolumn{5}{l}{?}\\[1ex]\hline
$P_{8,18}$&\hspace*{-2mm}\raisebox{-9mm}{\includegraphics[width=12mm]{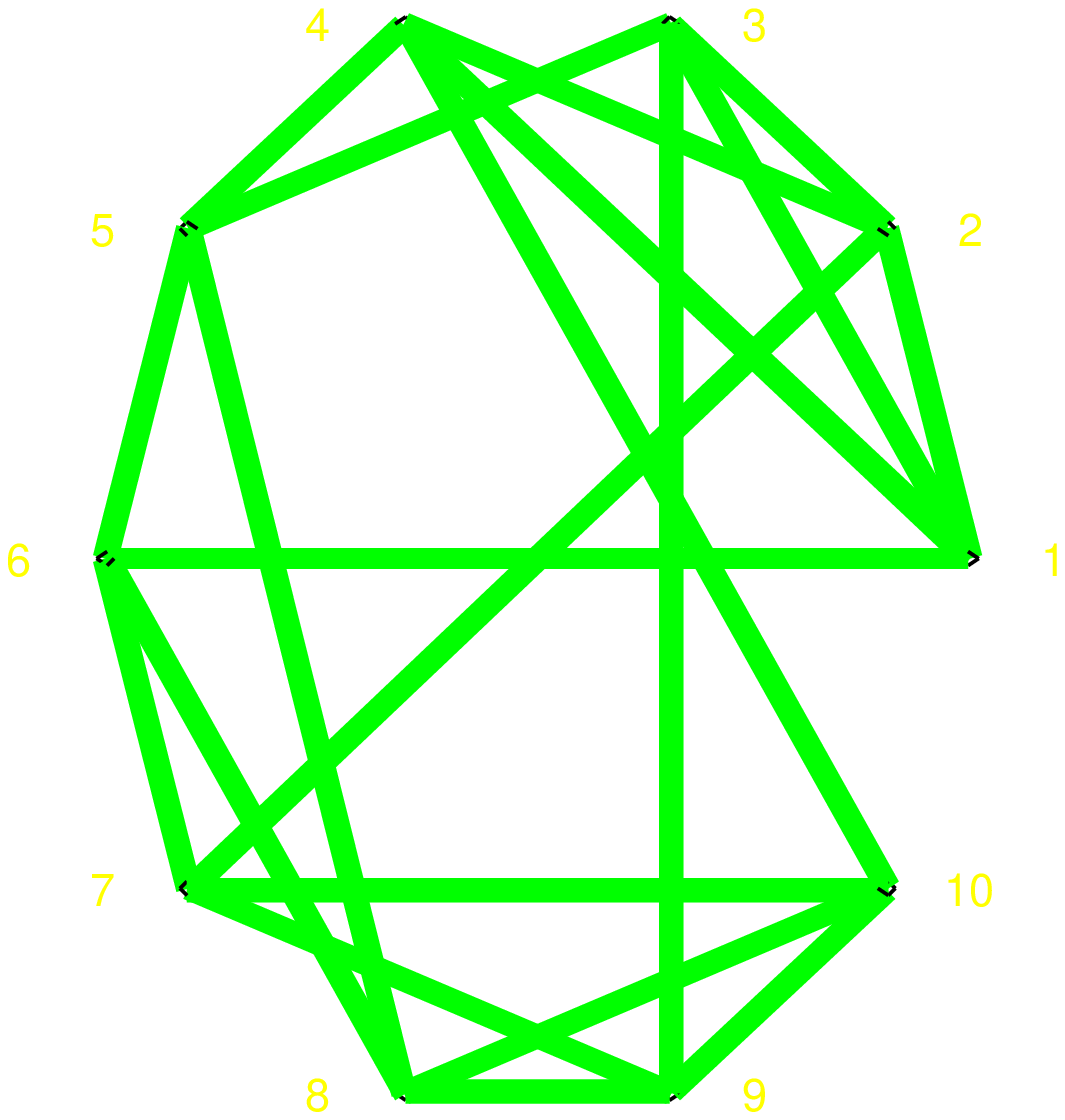}}&641.723~358~297&2&48&$P_3^2$&(79)\\[-6mm]
12&&\multicolumn{5}{l}{$727Q_3Q_9-\frac{735}{2}Q_5Q_7+72Q_3^4$}\\[1ex]\hline
$P_{8,19}$&\hspace*{-2mm}\raisebox{-9mm}{\includegraphics[width=12mm]{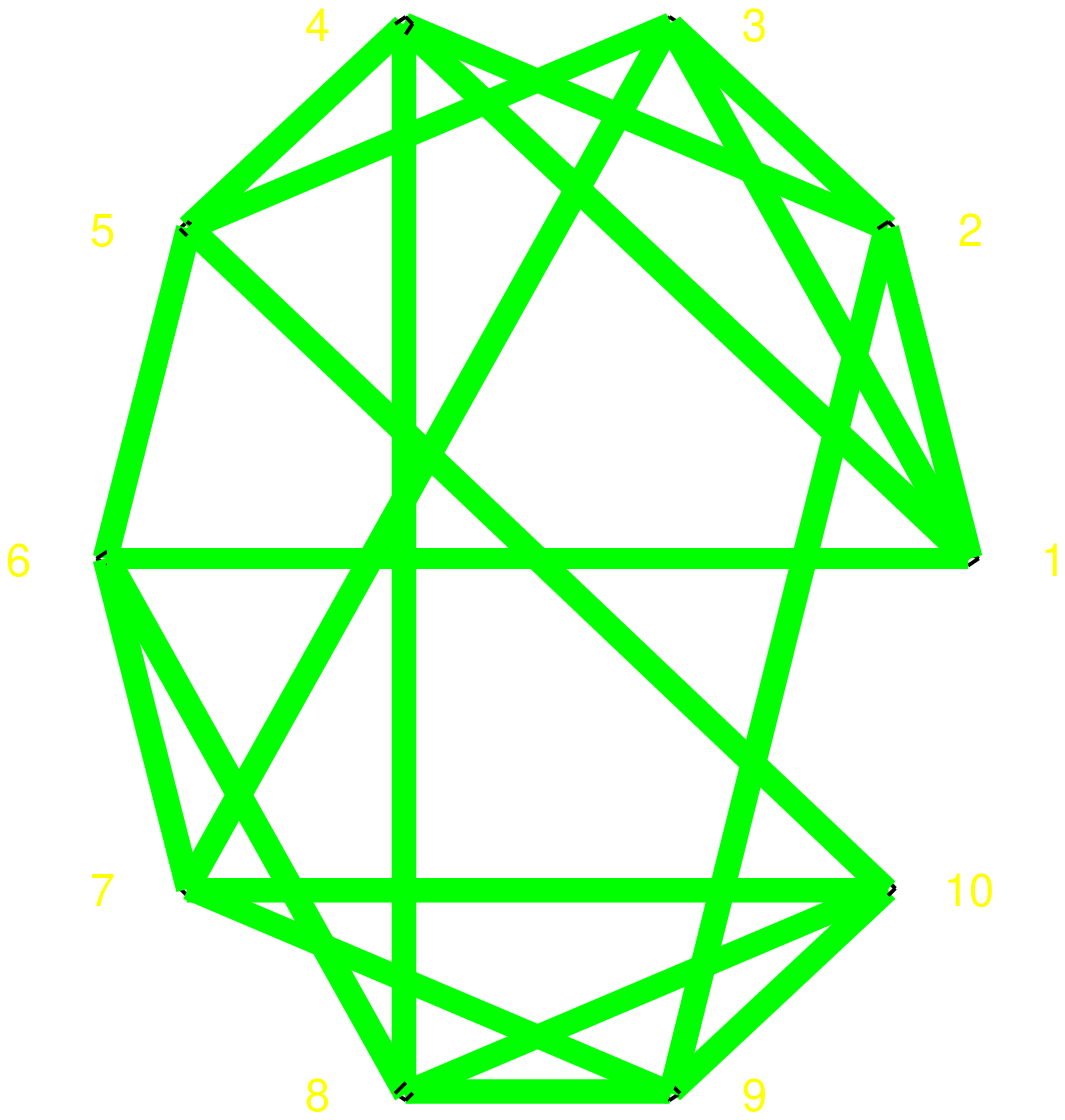}}&?&4&?&$P_3^2$&\\[-6mm]
12?&&\multicolumn{5}{l}{?}\\[1ex]\hline
$P_{8,20}$&\hspace*{-2mm}\raisebox{-9mm}{\includegraphics[width=12mm]{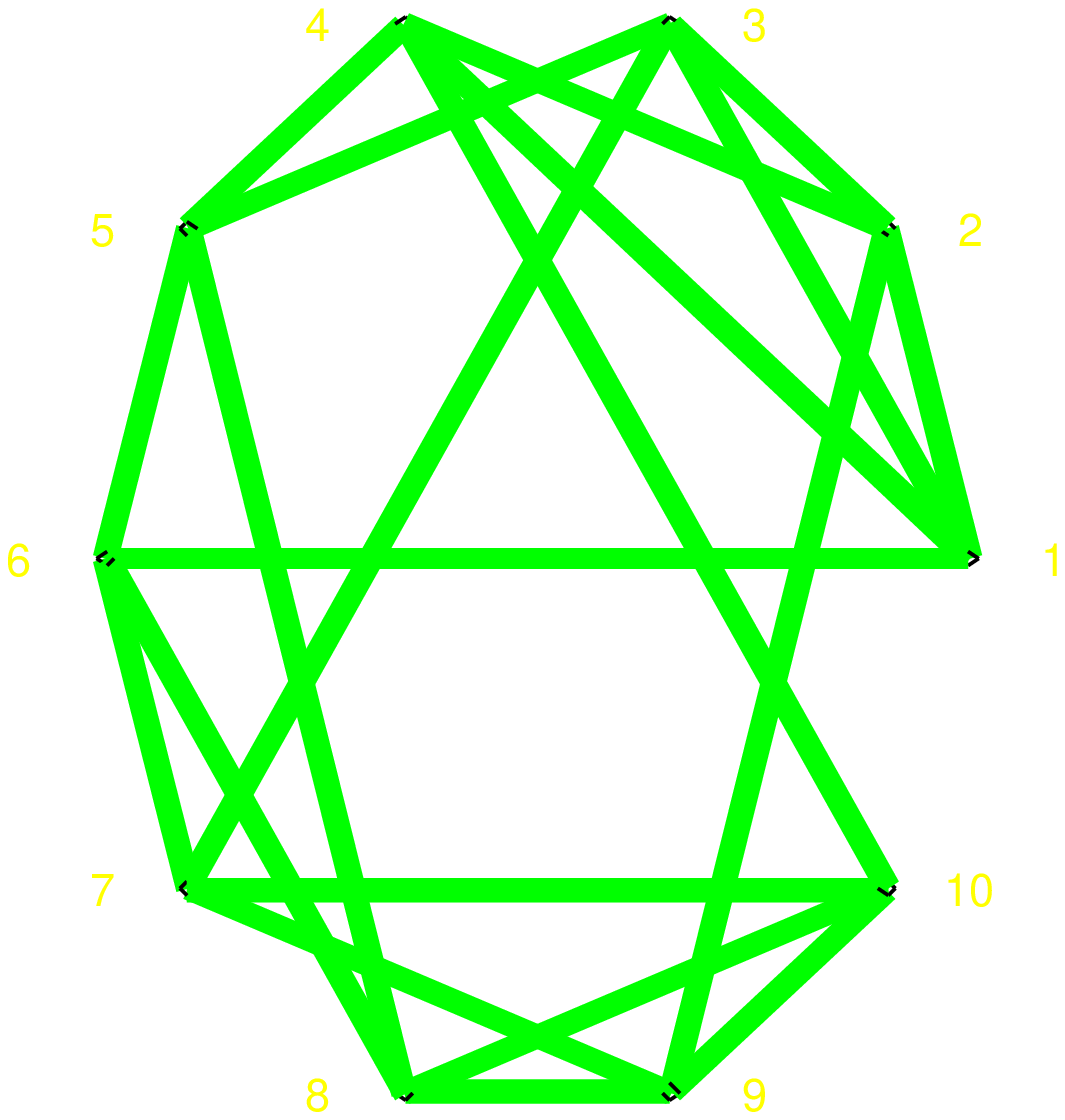}}&?&1&?&$P_3$&\\[-6mm]
13?&&\multicolumn{5}{l}{?}\\[1ex]\hline
$P_{8,21}$&\hspace*{-2mm}\raisebox{-9mm}{\includegraphics[width=12mm]{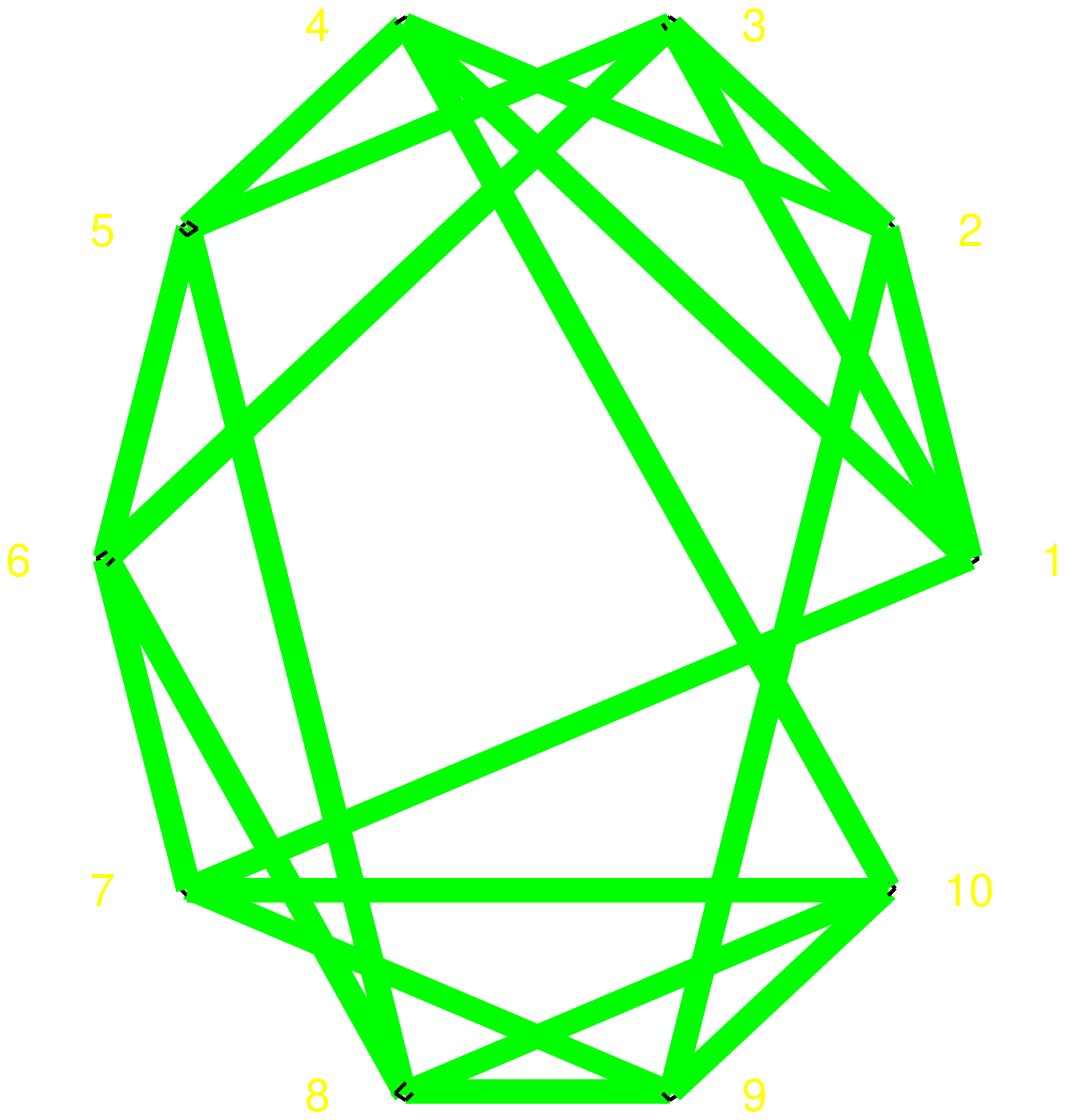}}&742.977~090~366&2&4&$P_3$&Fourier, twist\\[-6mm]
13&&\multicolumn{5}{l}{$P_{8,13}$}\\[1ex]\hline
$P_{8,22}$&\hspace*{-2mm}\raisebox{-9mm}{\includegraphics[width=12mm]{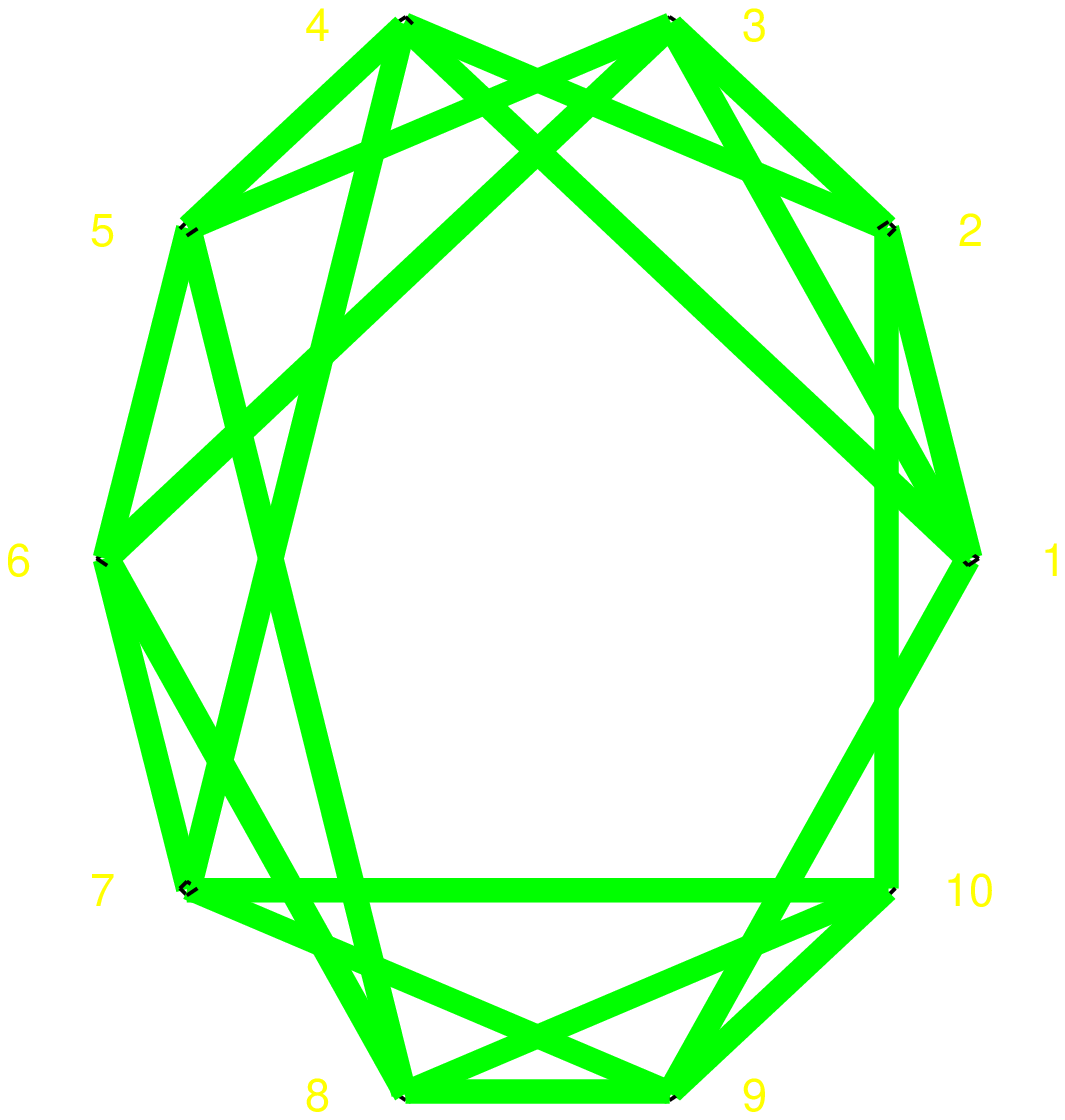}}&735.764~103~468&4&72&$P_3^2$&twist\\[-6mm]
12&&\multicolumn{5}{l}{$P_{8,10}$}\\[1ex]\hline
$P_{8,23}$&\hspace*{-2mm}\raisebox{-9mm}{\includegraphics[width=12mm]{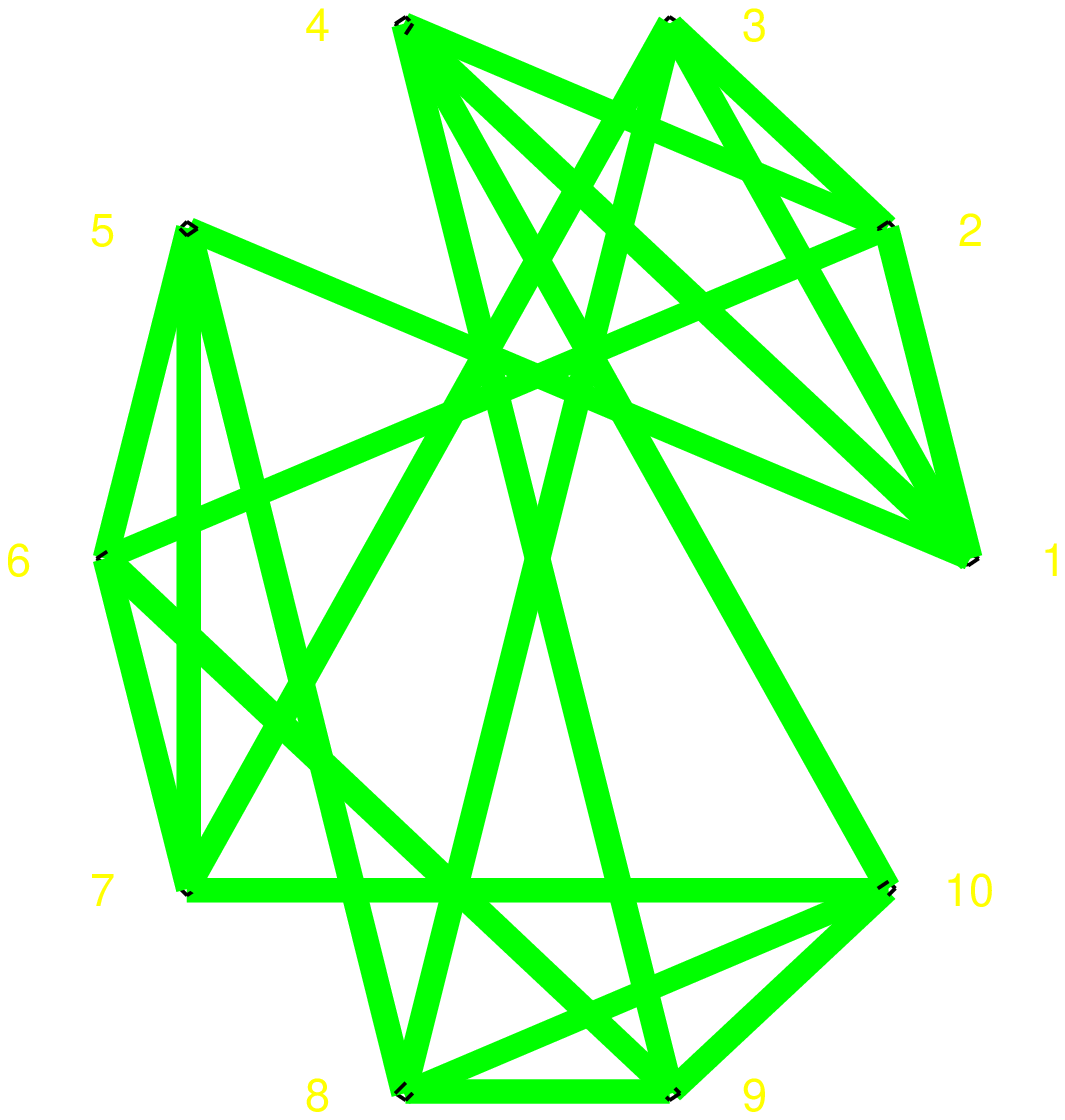}}&?&2&?&$P_3$&twist\\[-6mm]
13?&&\multicolumn{5}{l}{$P_{8,17}$}\\[1ex]\hline
$P_{8,24}$&\hspace*{-2mm}\raisebox{-9mm}{\includegraphics[width=12mm]{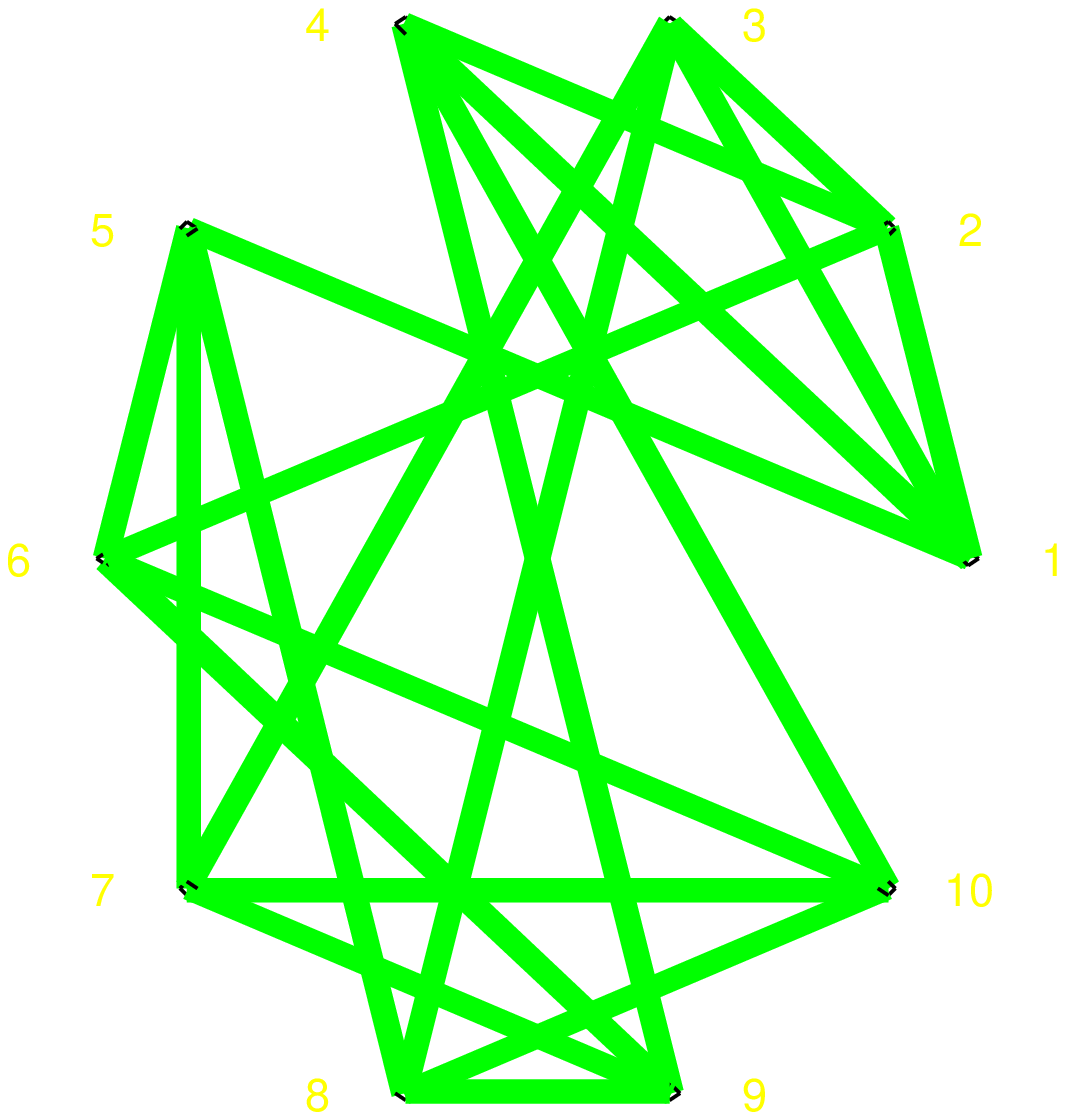}}&?&8&?&$P_{7,8}$&\\[-6mm]
13?&&\multicolumn{5}{l}{?}\\[1ex]\hline
$P_{8,25}$&\hspace*{-2mm}\raisebox{-9mm}{\includegraphics[width=12mm]{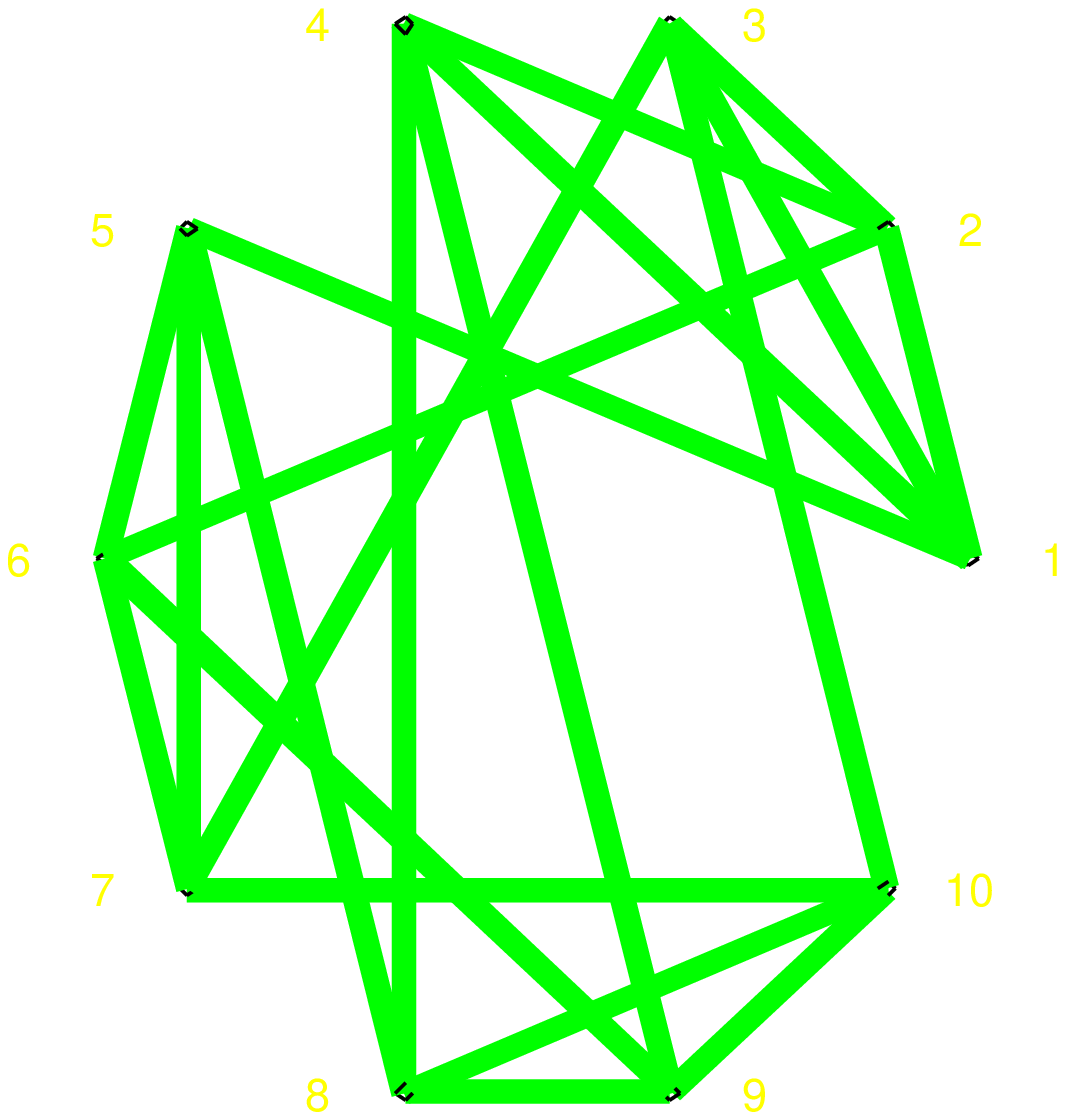}}&641.723~358~297&4&48&$P_3^2$&Fourier, twist\\[-6mm]
12&&\multicolumn{5}{l}{$P_{8,18}$}\\[1ex]\hline
$P_{8,26}$&\hspace*{-2mm}\raisebox{-9mm}{\includegraphics[width=12mm]{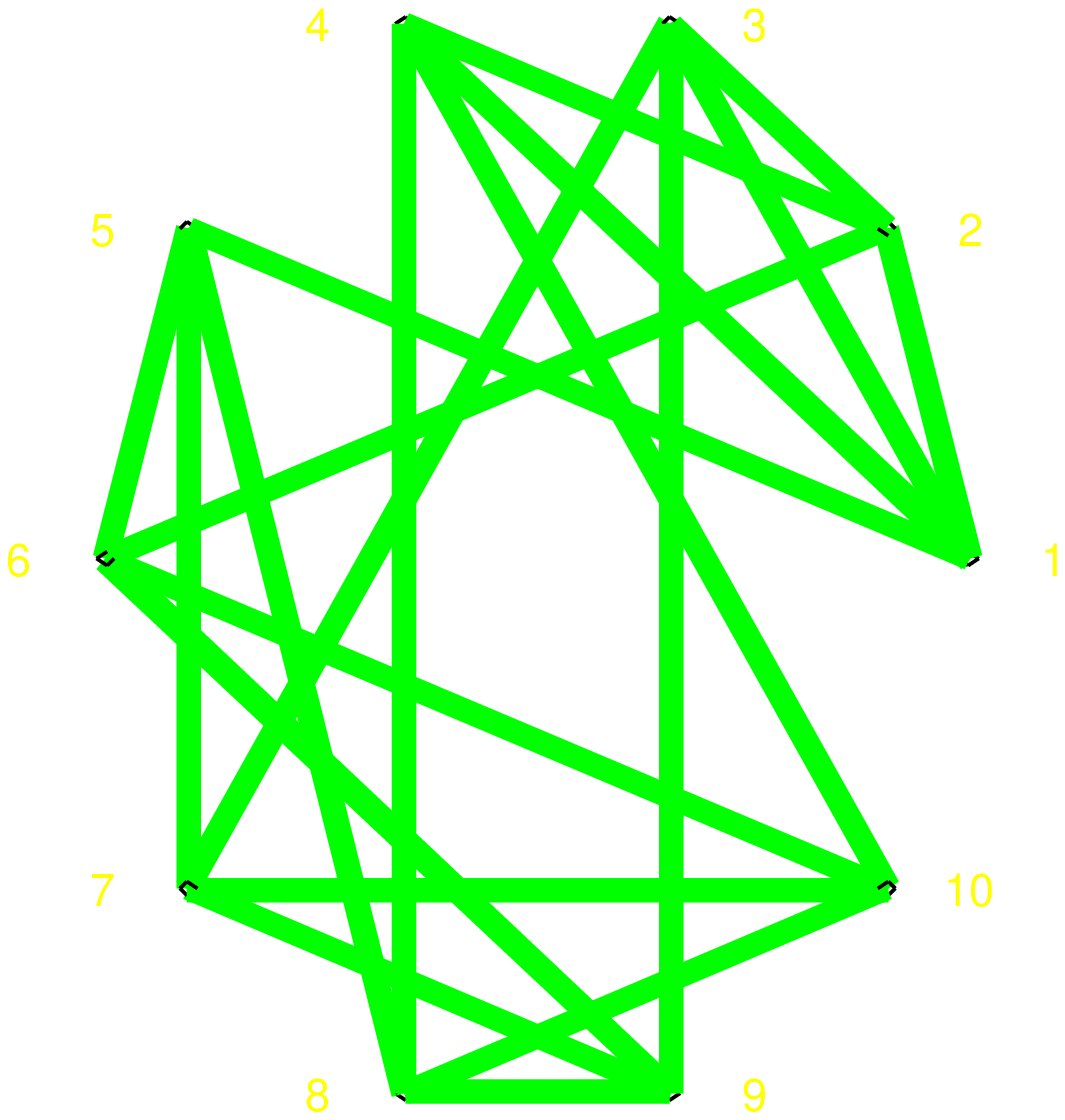}}&?&4&?&$P_{7,9}$&\\[-6mm]
13?&&\multicolumn{5}{l}{?}\\[1ex]\hline
$P_{8,27}$&\hspace*{-2mm}\raisebox{-9mm}{\includegraphics[width=12mm]{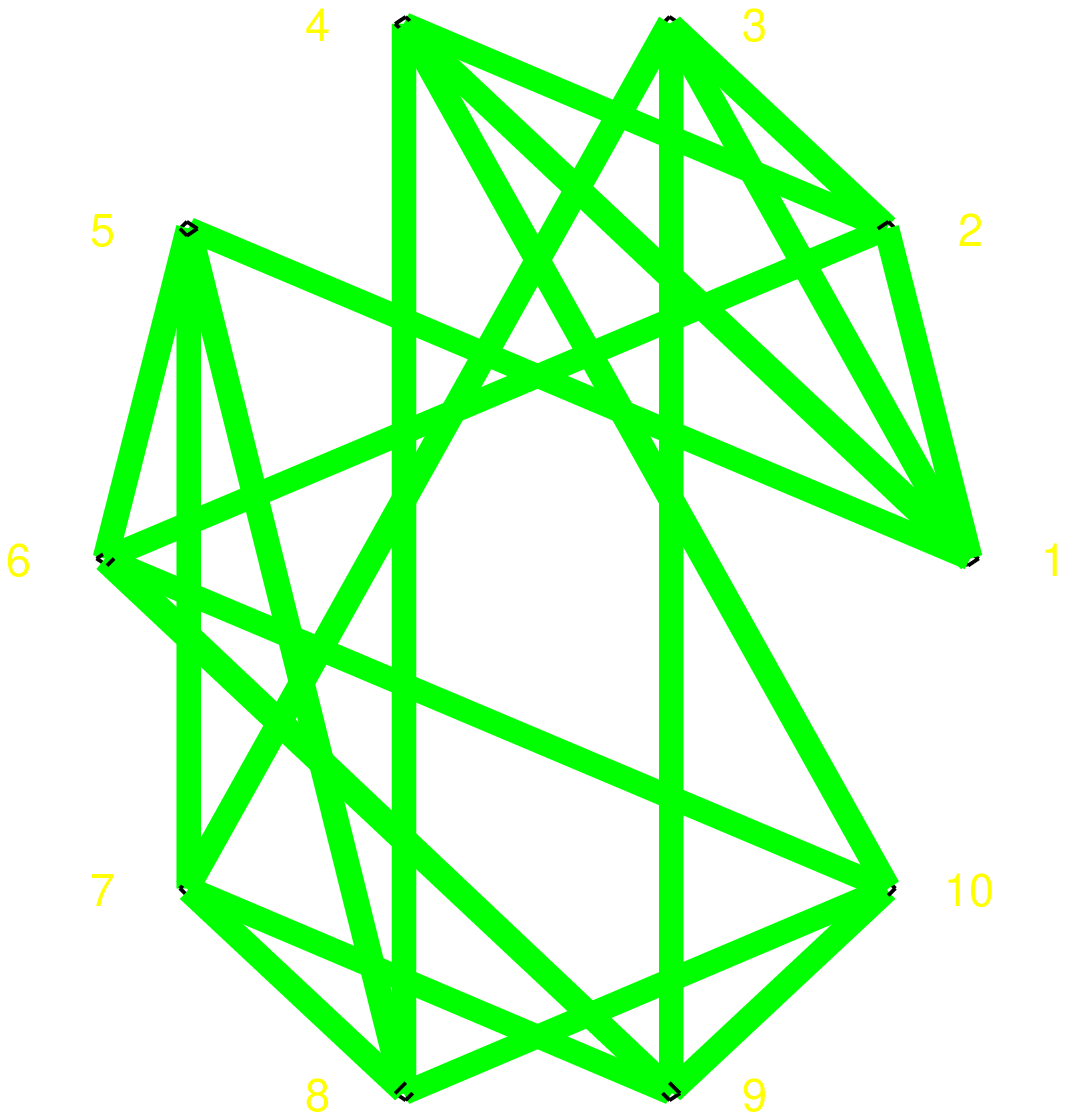}}&?&4&?&$P_{7,10}$&Fourier\\[-6mm]
12?&&\multicolumn{5}{l}{$P_{8,19}$}
\end{tabular}

\begin{tabular}{lllllll}
name&graph&numerical value&$|$Aut$|$&index&ancestor&rem, (sf), [Lit]\\[1ex]
\multicolumn{2}{l}{weight}&\multicolumn{5}{l}{exact value}\\[1ex]\hline\hline
$P_{8,28}$&\hspace*{-2mm}\raisebox{-9mm}{\includegraphics[width=12mm]{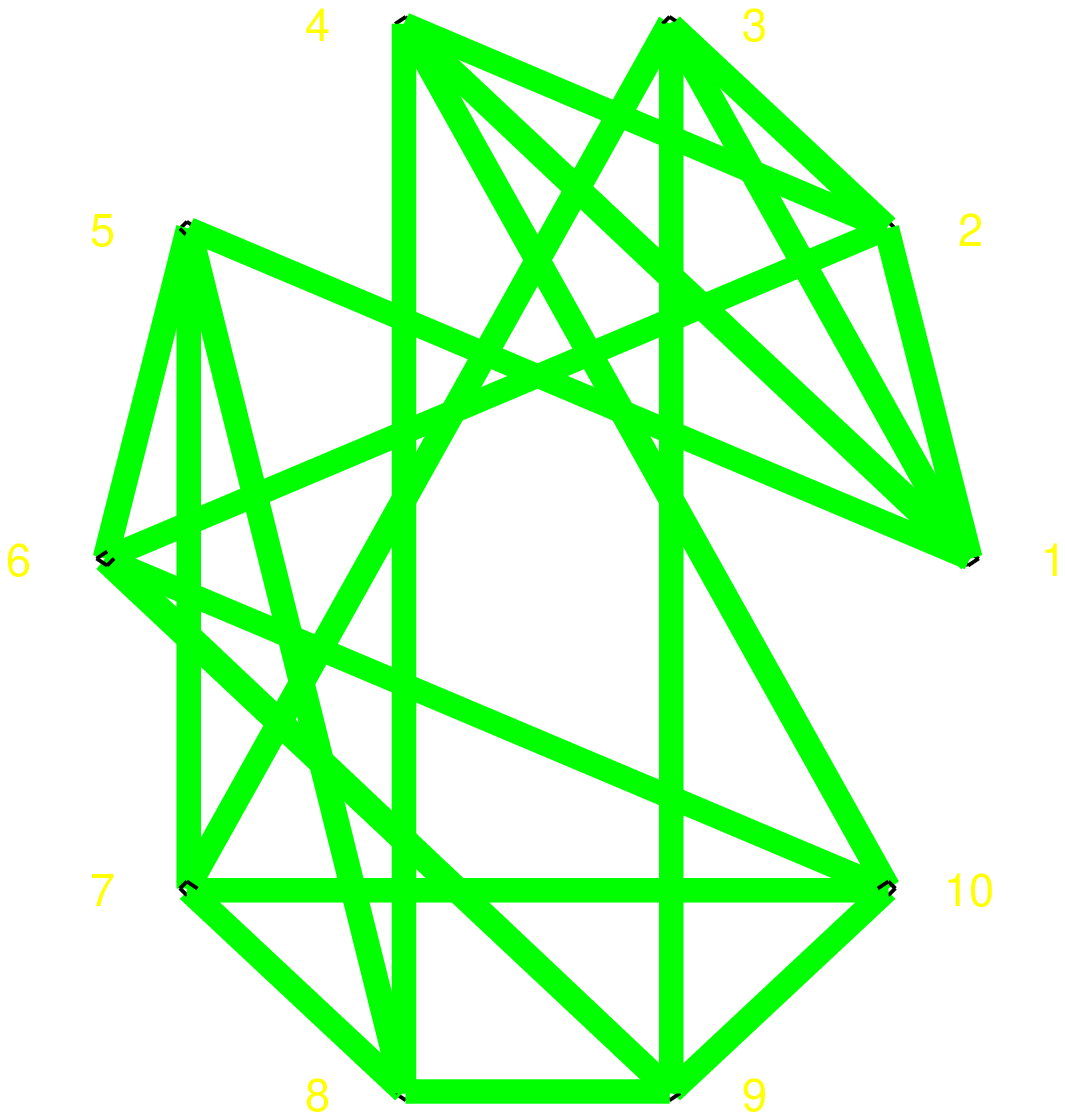}}&?&4&?&$P_{7,9}$&twist\\[-6mm]
13?&&\multicolumn{5}{l}{$P_{8,26}$}\\[1ex]\hline
$P_{8,29}$&\hspace*{-2mm}\raisebox{-9mm}{\includegraphics[width=12mm]{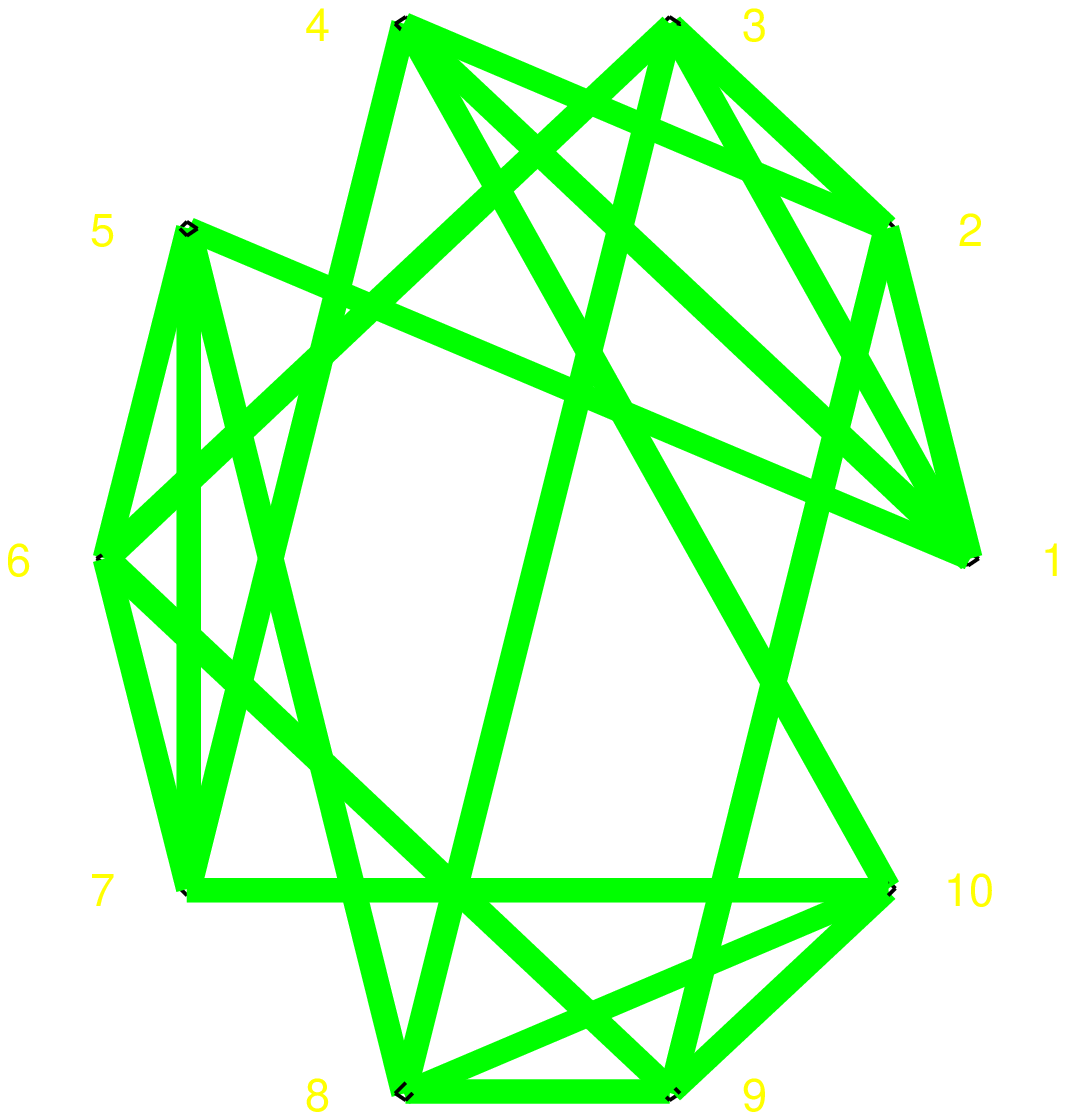}}&?&2&?&$P_{7,9}$&\\[-6mm]
13?&&\multicolumn{5}{l}{?}\\[1ex]\hline
$P_{8,30}$&\hspace*{-2mm}\raisebox{-9mm}{\includegraphics[width=12mm]{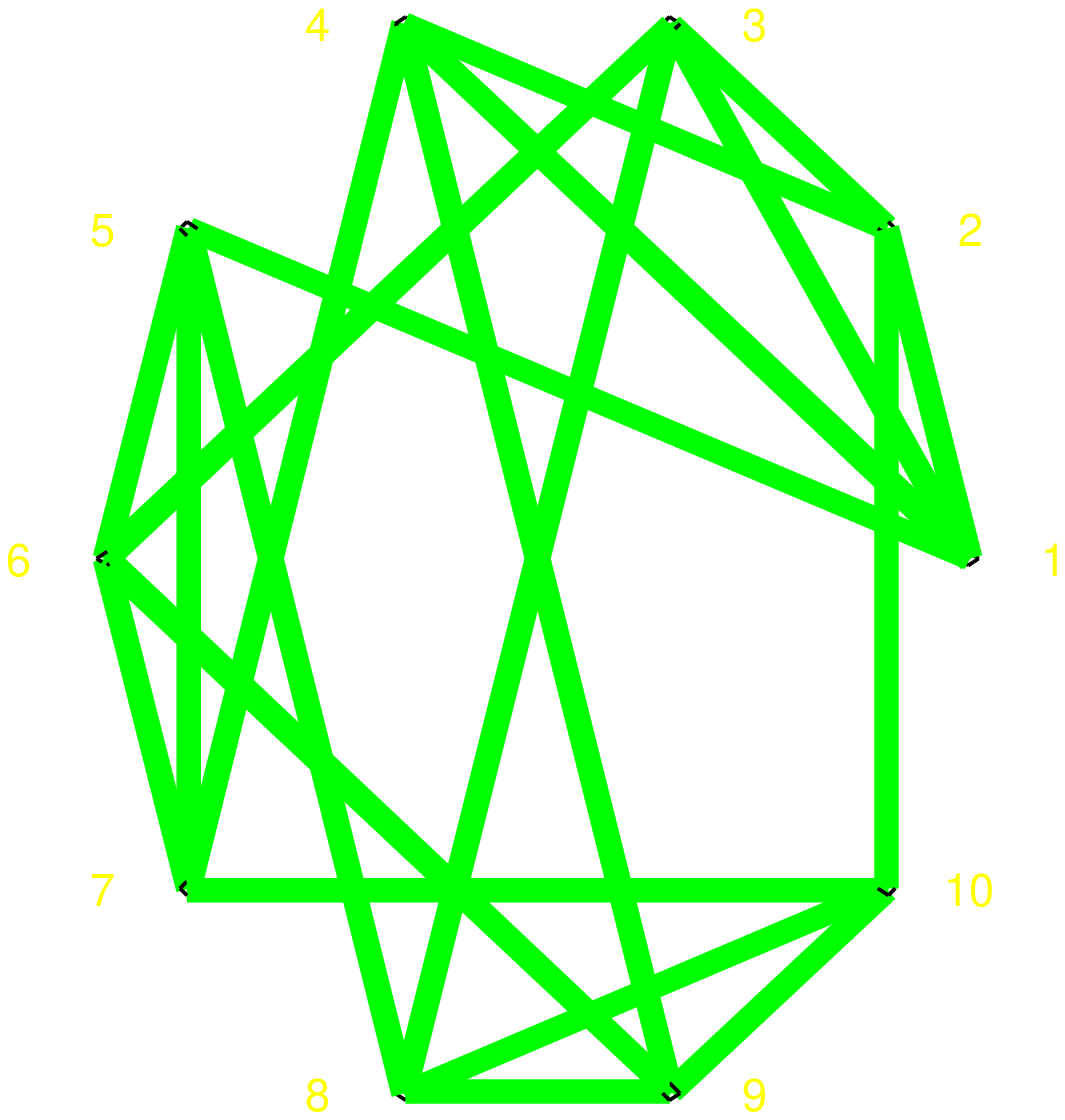}}&?&2&?&$P_{7,11}$&\\[-6mm]
13?&&\multicolumn{5}{l}{?}\\[1ex]\hline
$P_{8,31}$&\hspace*{-2mm}\raisebox{-9mm}{\includegraphics[width=12mm]{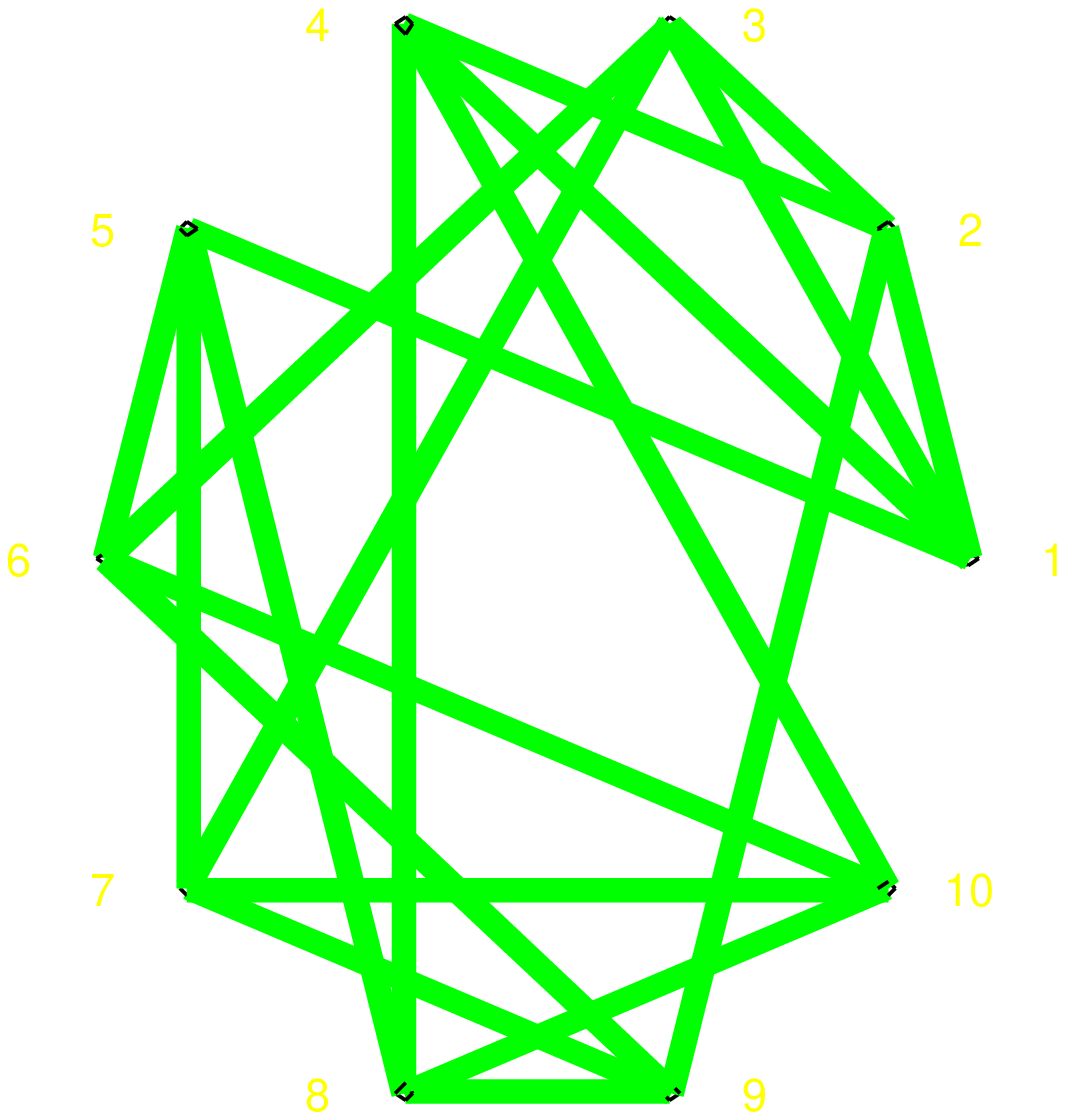}}&460.088~538~246&4&8&$P_{7,8}$&(64)\hfill$[-7330Q_3Q_5^2$\\[-6mm]
13&&\multicolumn{5}{l}{$\frac{67363763}{5600}Q_{13,1}\!-\!\frac{36487}{175}Q_{13,2}\!-\!\frac{1913}{7}Q_{13,3}\!+\!1792Q_3Q_{10}\!+\!7936Q_5Q_8\!+\!98Q_3^2Q_7$}\\[1ex]\hline
$P_{8,32}$&\hspace*{-2mm}\raisebox{-9mm}{\includegraphics[width=12mm]{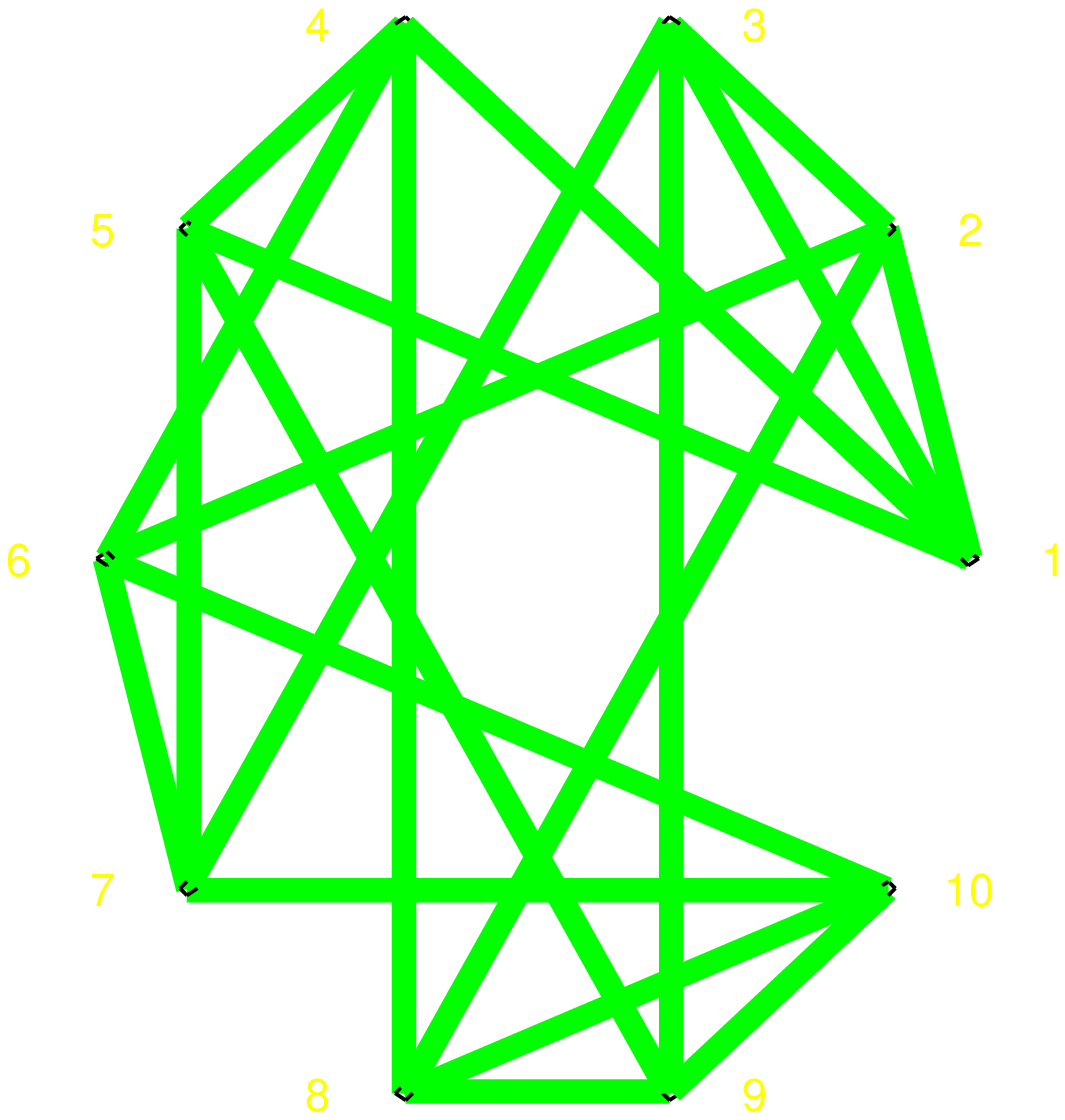}}&470.720~125~534&16&17280&$P_{8,32}$&(120)\\[-6mm]
12&&\multicolumn{5}{l}{$-\frac{81920}{23}Q_{12,1}-\frac{655360}{23}Q_{12,2}+\frac{20480}{23}Q_{12,3}+\frac{8760}{23}Q_3Q_9+\frac{15660}{23}Q_5Q_7$}\\[1ex]\hline
$P_{8,33}$&\hspace*{-2mm}\raisebox{-9mm}{\includegraphics[width=12mm]{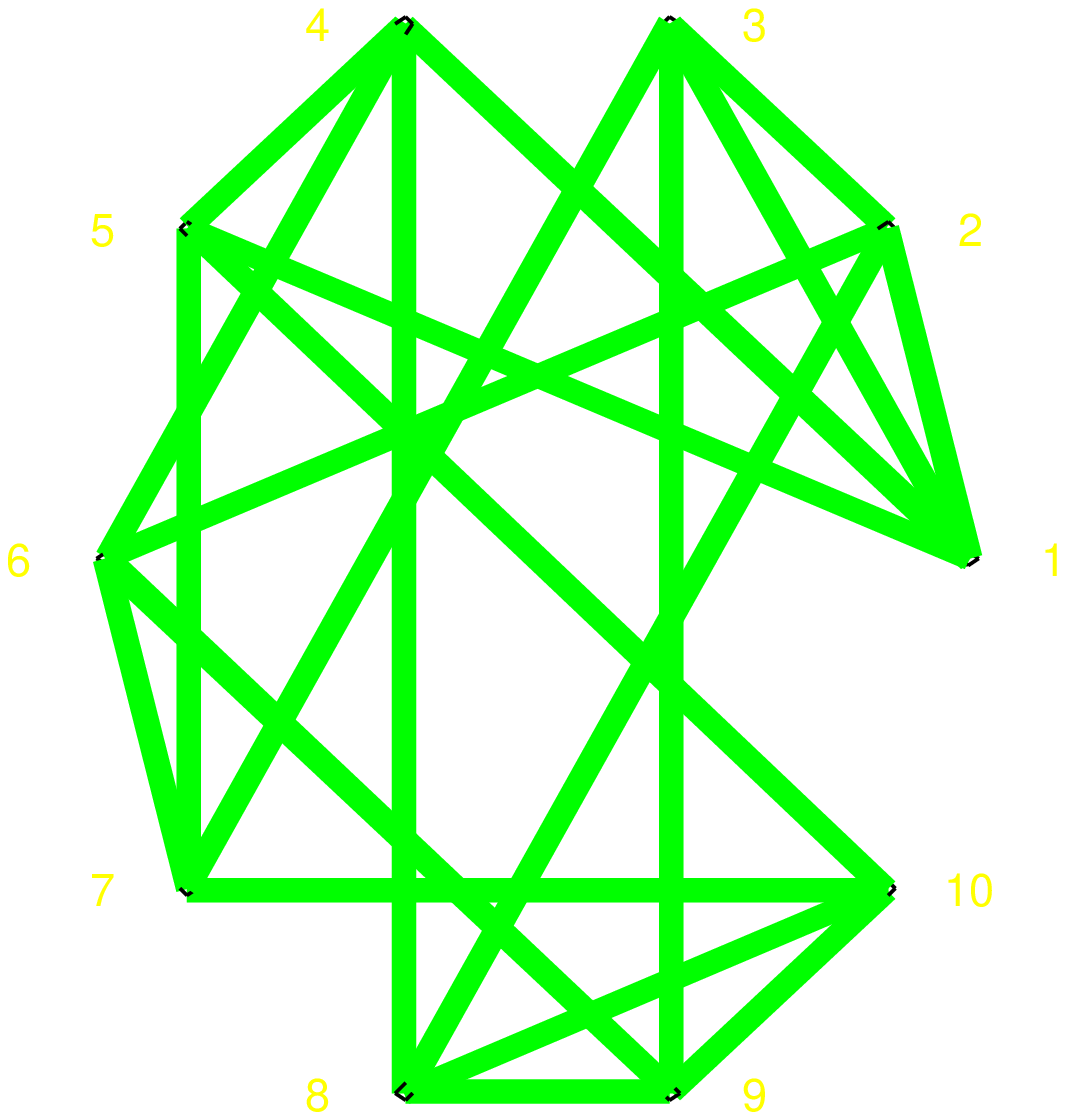}}&?&2&?&$P_{8,33}$&\\[-6mm]
13?&&\multicolumn{5}{l}{?}\\[1ex]\hline
$P_{8,34}$&\hspace*{-2mm}\raisebox{-9mm}{\includegraphics[width=12mm]{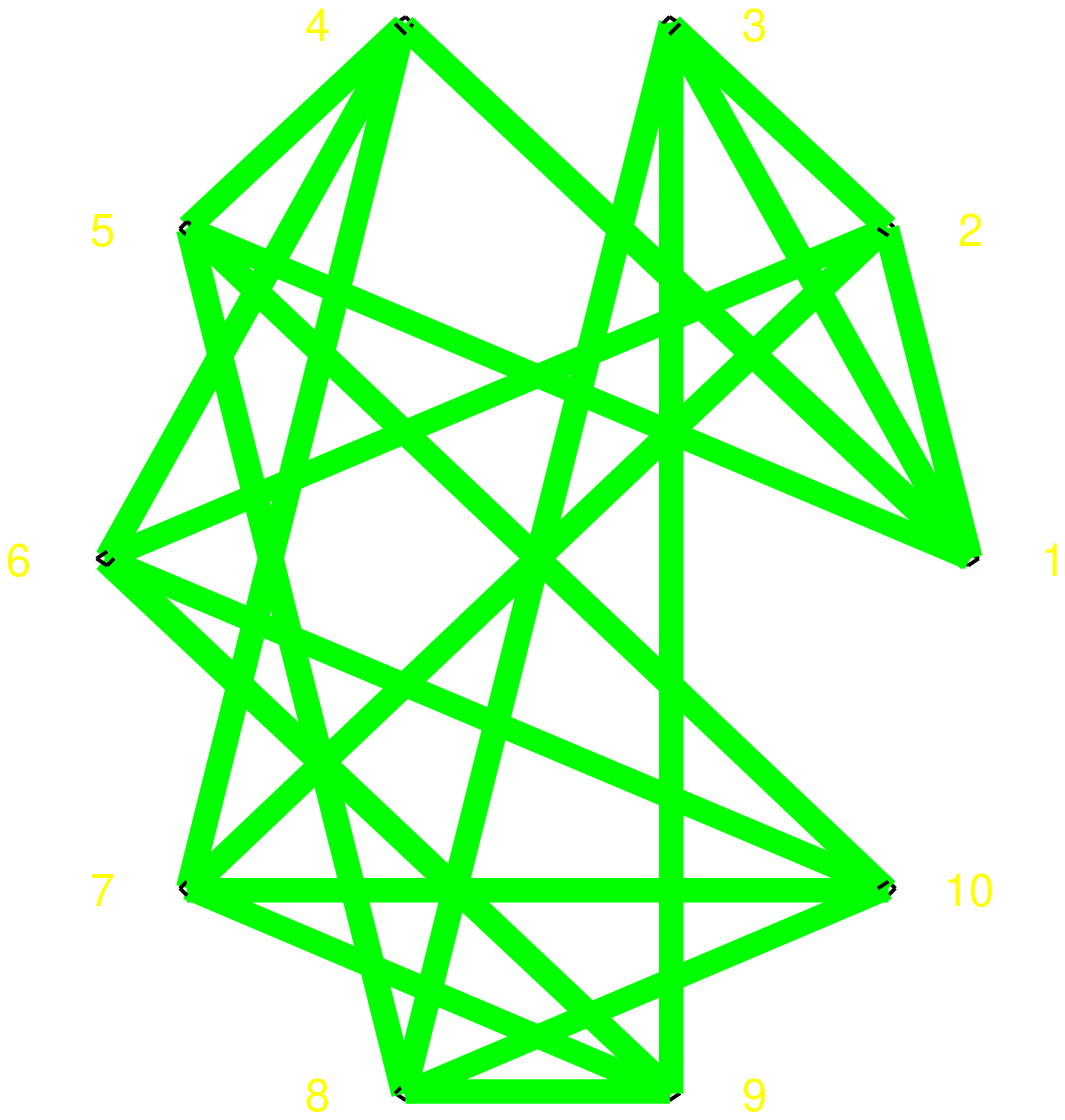}}&470.720~125~534&16&17280&$P_{8,34}$&twist\\[-6mm]
12&&\multicolumn{5}{l}{$P_{8,32}$}\\[1ex]\hline
$P_{8,35}$&\hspace*{-2mm}\raisebox{-9mm}{\includegraphics[width=12mm]{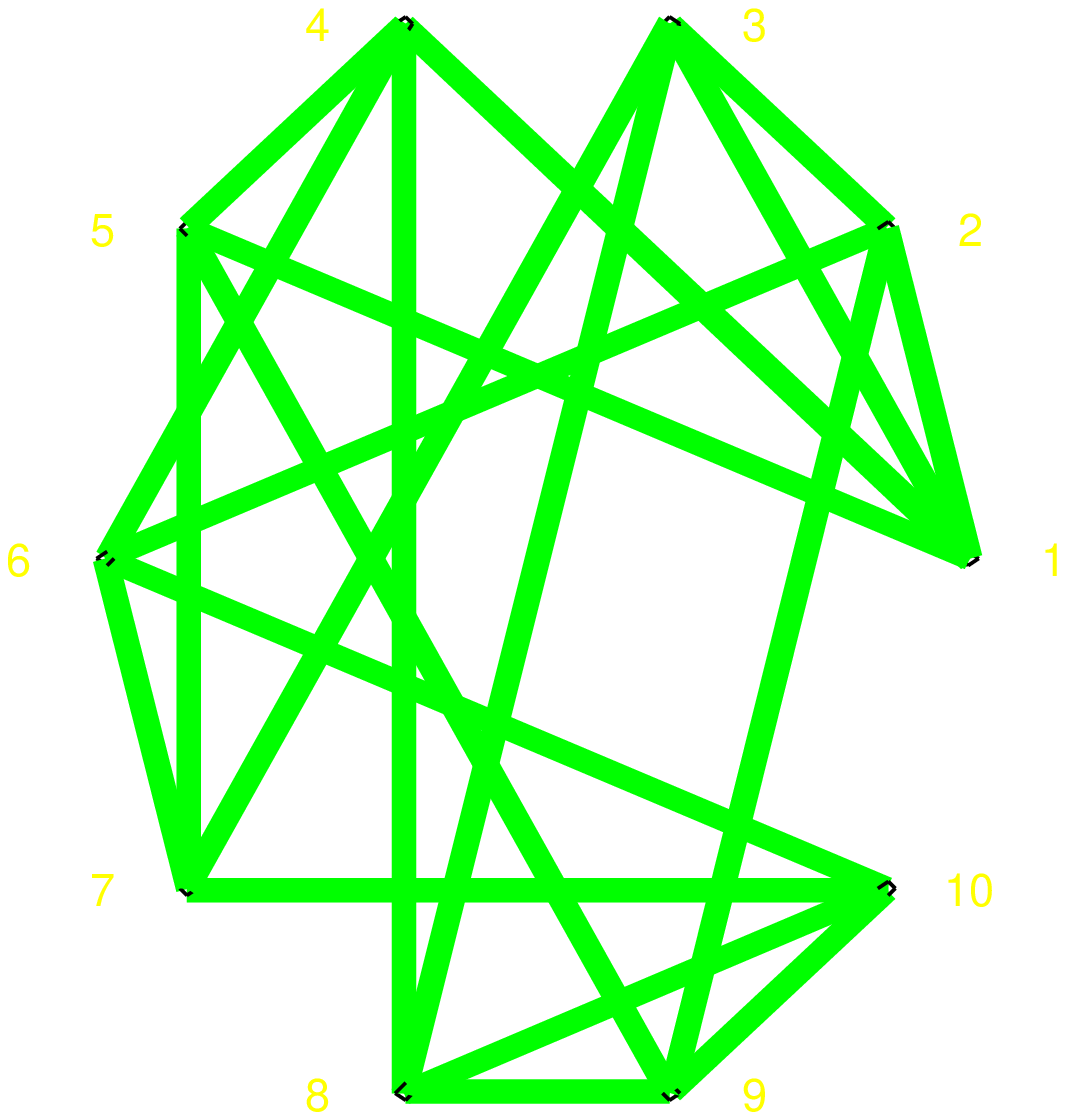}}&?&16&?&$P_{8,35}$&\\[-6mm]
13?&&\multicolumn{5}{l}{?}\\[1ex]\hline
$P_{8,36}$&\hspace*{-2mm}\raisebox{-9mm}{\includegraphics[width=12mm]{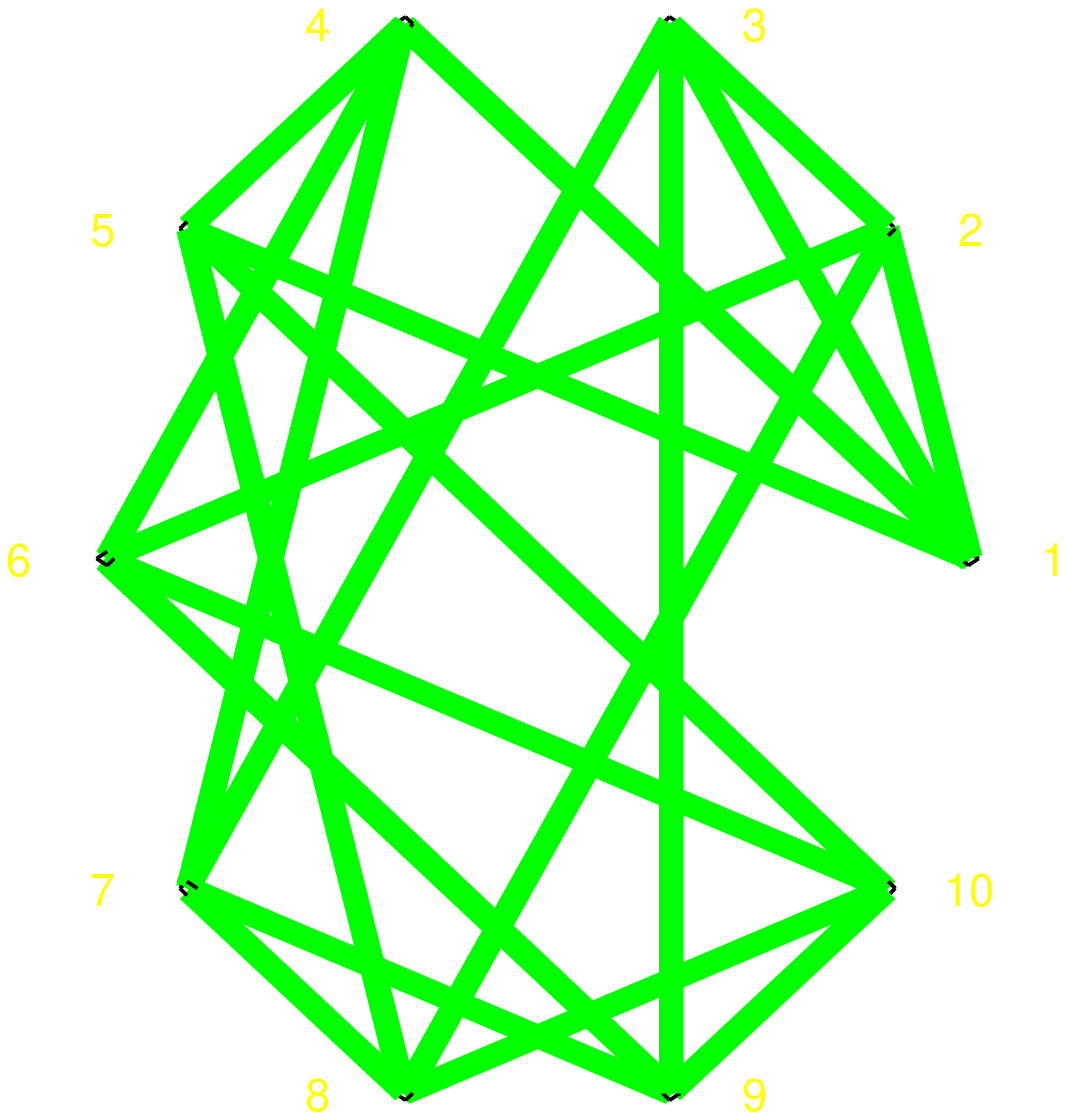}}&?&10&?&$P_{8,36}$&\\[-6mm]
13?&&\multicolumn{5}{l}{?}\\[1ex]\hline
$P_{8,37}$&\hspace*{-2mm}\raisebox{-9mm}{\includegraphics[width=12mm]{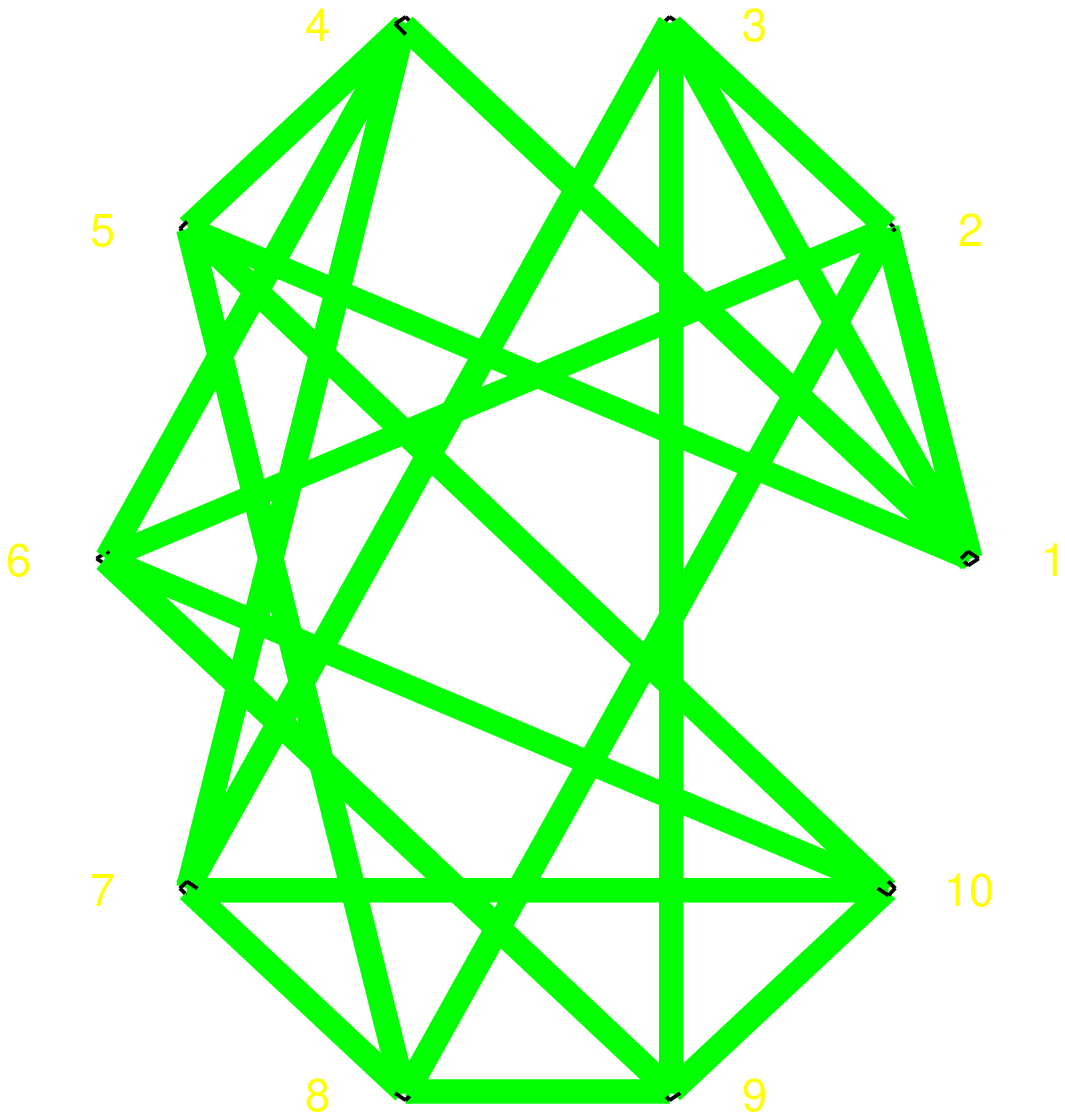}}&?&2&?&$P_{8,37}$&\\[-6mm]
?&&\multicolumn{5}{l}{?}\\[1ex]\hline
$P_{8,38}$&\hspace*{-2mm}\raisebox{-9mm}{\includegraphics[width=12mm]{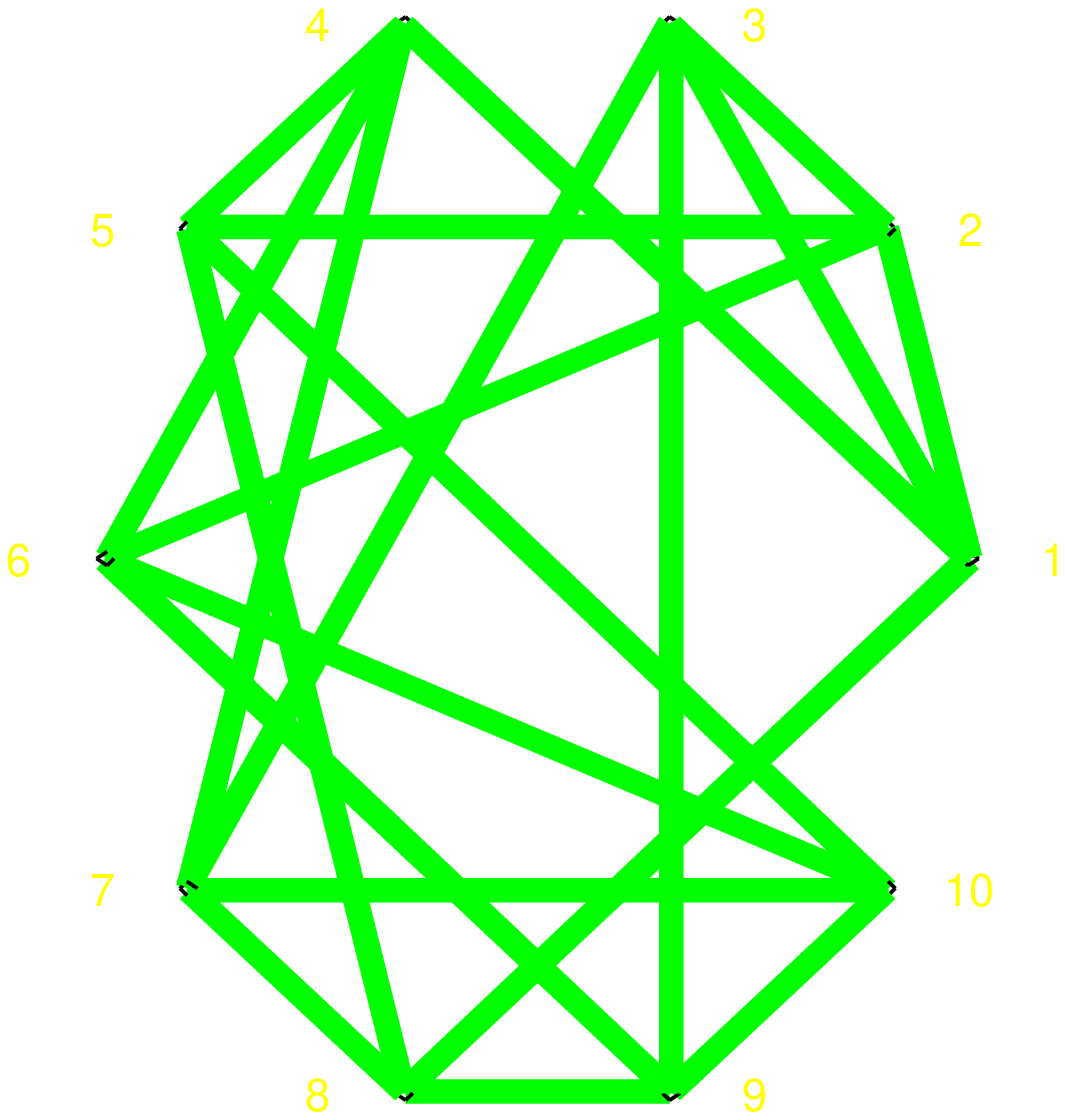}}&?&4&?&$P_{8,38}$&\\[-6mm]
?&&\multicolumn{5}{l}{?}\\[1ex]\hline
$P_{8,39}$&\hspace*{-2mm}\raisebox{-9mm}{\includegraphics[width=12mm]{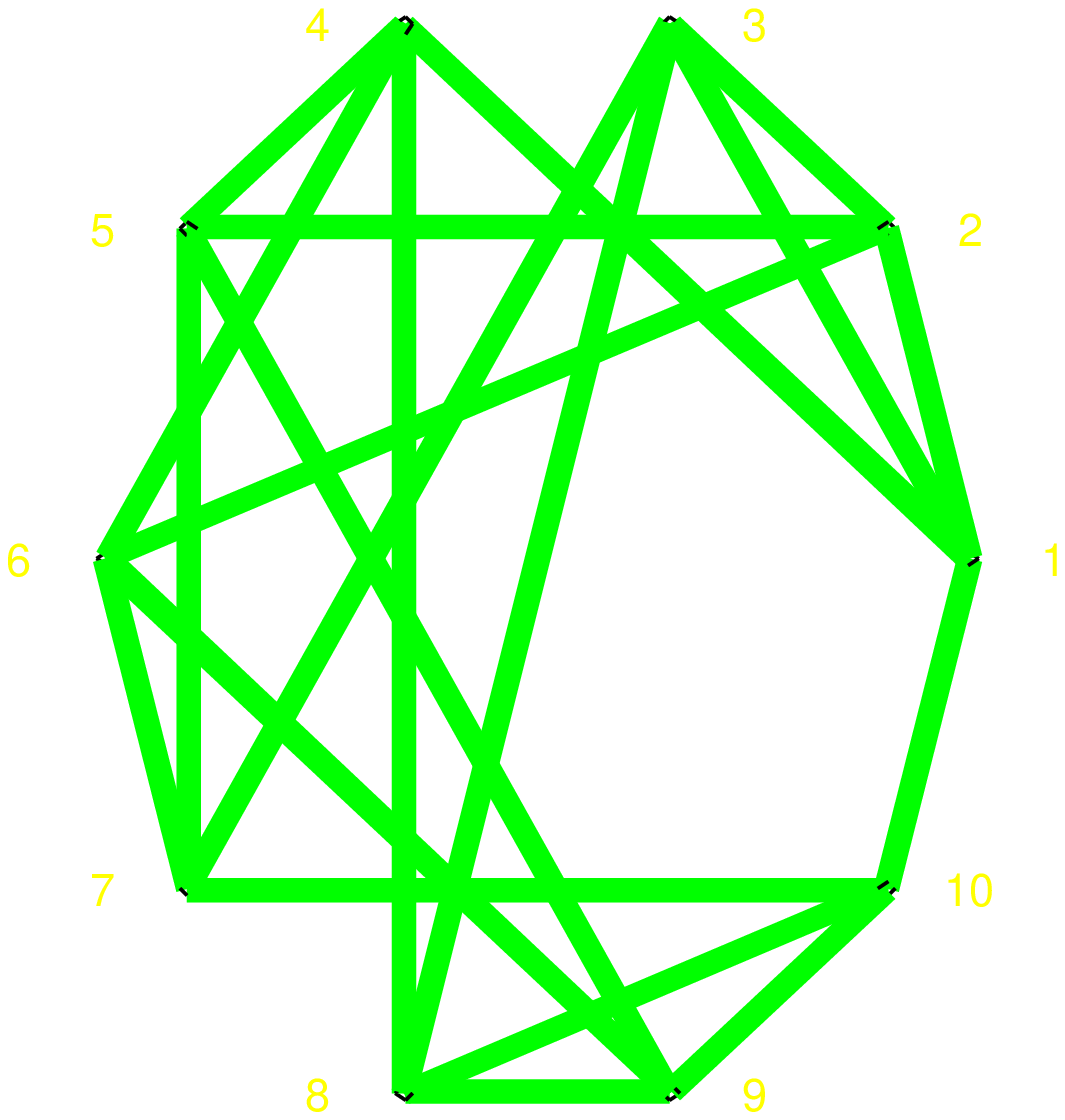}}&?&8&?&$P_{8,39}$&\\[-6mm]
?&&\multicolumn{5}{l}{?}\\[1ex]\hline
$P_{8,40}$&\hspace*{-2mm}\raisebox{-9mm}{\includegraphics[width=12mm]{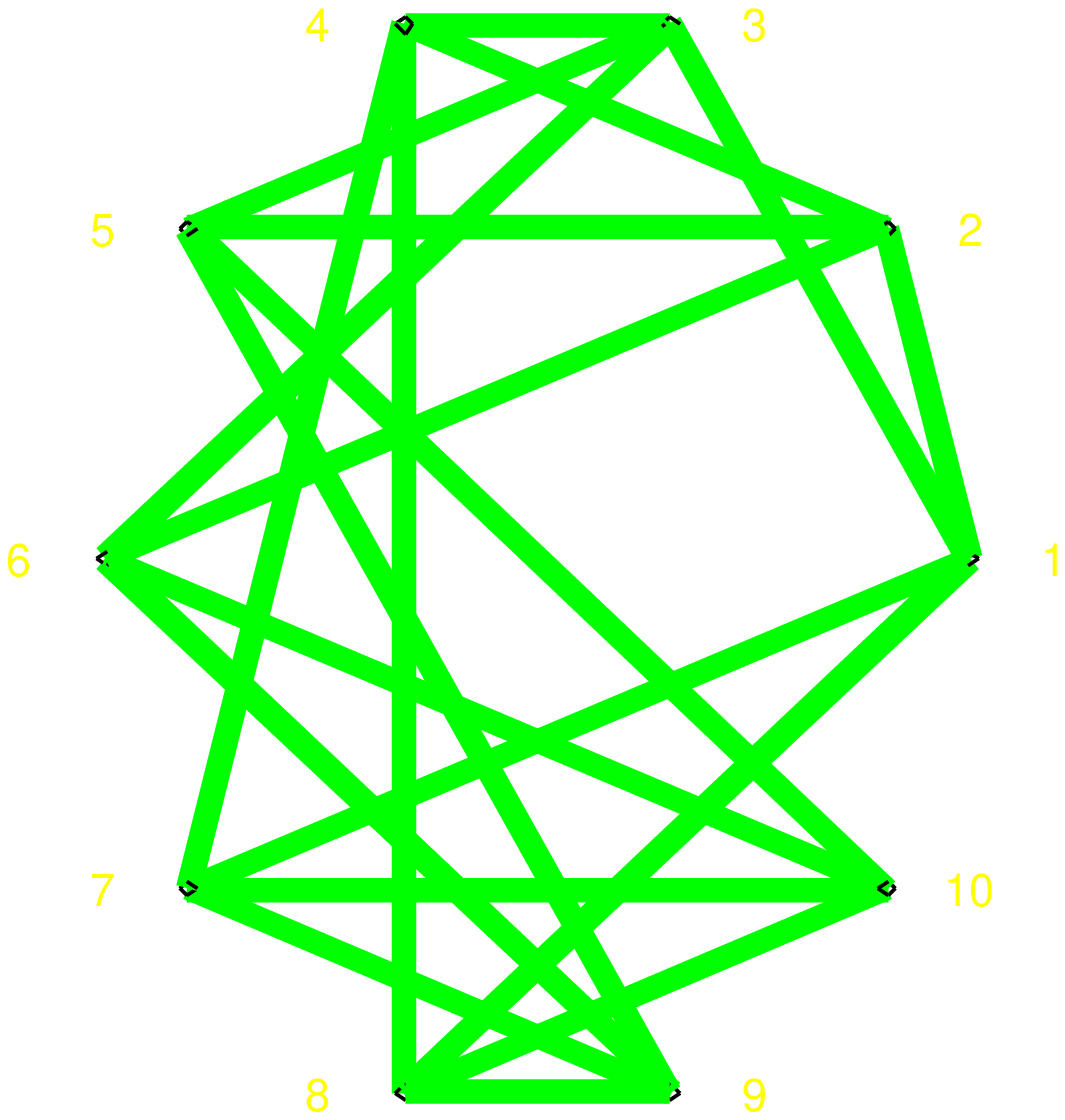}}&?&320&?&$P_{8,40}$&$C^{10}_{1,4}$\\[-6mm]
13?&&\multicolumn{5}{l}{?}\\[1ex]\hline
$P_{8,41}$&\hspace*{-2mm}\raisebox{-9mm}{\includegraphics[width=12mm]{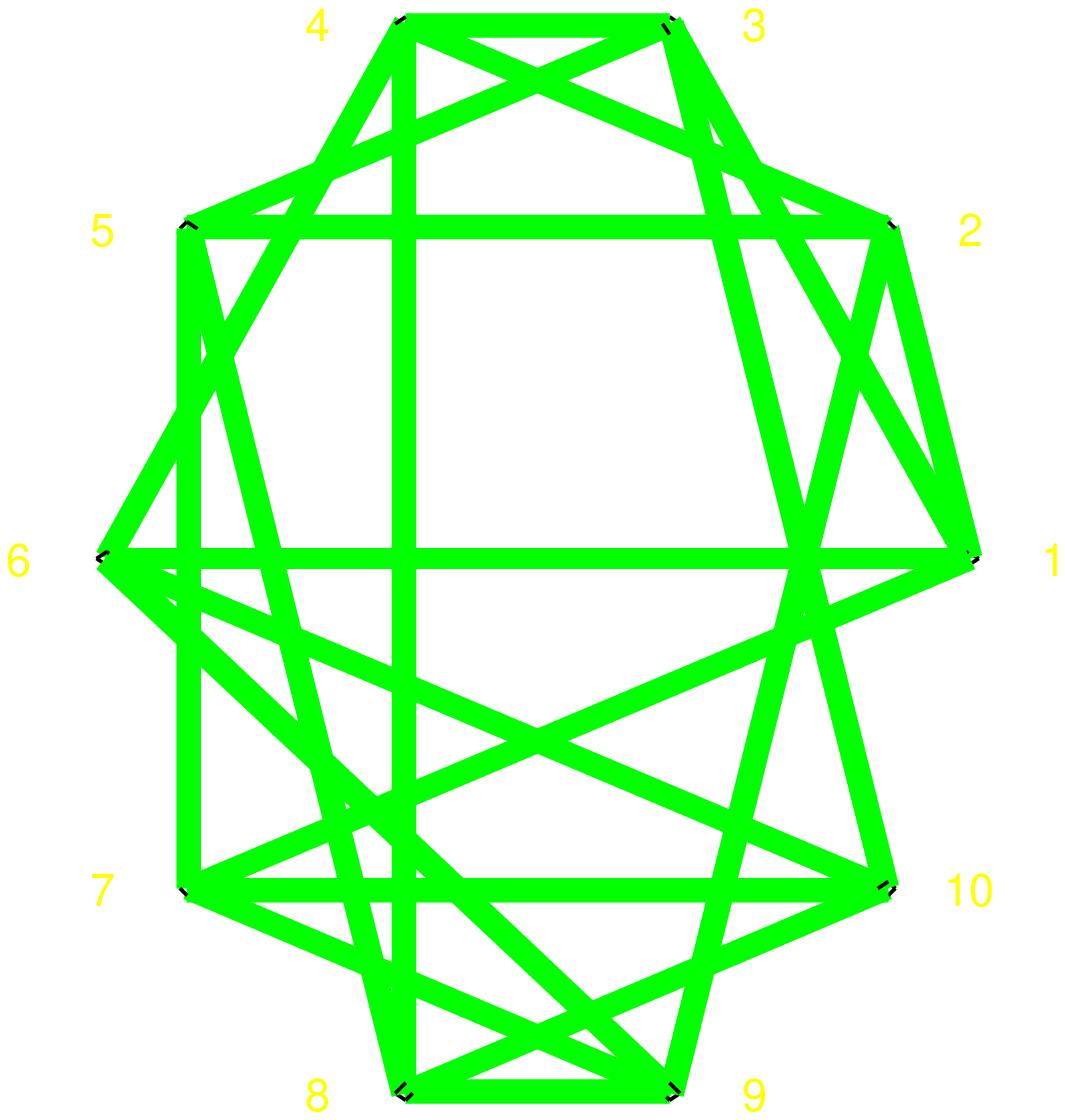}}&?&240&?&$P_{8,41}$&$C^{10}_{1,3}$\\[-6mm]
?&&\multicolumn{5}{l}{?\hspace*{101mm}}
\end{tabular}

Table 4: The census. The numbers $Q_\bullet$ are listed in Table 3.

\bibliographystyle{plain}
\renewcommand\refname{References}

\end{document}